\UseRawInputEncoding 

\pdfoutput=1

\documentclass[manuscript,11pt]{aastex}
\usepackage{epsfig}
\usepackage{soul}
\usepackage{color}

\usepackage{physymb}
\usepackage{mathrsfs}

\usepackage{lineno}
\def\oviii{O\,{\sc viii]}}
\def\ovii{O\,{\sc vii]}}

\def\fexxv{Fe\,{\sc xxv}}
\def\fexxvi{Fe\,{\sc xxvi}}

\def\coxxvii{Co\,{\sc xxvii}}

\def\nixxviii{Ni\,{\sc xxviii}}

\def\sxvi{S\,{\sc xvi}}

\def\sixiv{Si\,{\sc xiv}}

\def\mgxii{Mg\,{\sc xii}}
\def\nex{Ne\,{\sc x}}
\def\neix{Ne\,{\sc ix}}

\def\civ{C\,{\sc iv}}

\def\ovii{O\,{\sc vii}}
\def\oviii{O\,{\sc viii}}

\def\mgxii{Mg\,{\sc xii}}
\def\mgxi{Mg\,{\sc xi}}
\def\civ{C\,{\sc iv}}

\def\ovii{O\,{\sc vii}}

\def\mathB{\textbf{\em B}}

\def\mathJ{\textbf{\em J}}

\def\cm{\ifmmode {\rm cm}^{-1} \else cm$^{-1}$ \fi}
\def\s{\ifmmode {\rm s}^{-1} \else s$^{-1}$ \fi}
\def\cc{\ifmmode {\rm cm}^{-3} \else cm$^{-3}$ \fi}
\def\cs{\ifmmode {\rm cm}^{-2} \else cm$^{-2}$ \fi}
\def\g{\ifmmode \gamma \else $\gamma$\fi}
\def\G{\ifmmode \Gamma \else $\Gamma$\fi}
\def\Gs{\ifmmode \Gamma~ \else $\Gamma~$\fi}

\def\gc{\ifmmode \gamma_{\rm c} \else $\gamma_{\rm c}$ \fi}
\def\sw{Schwarzschild~}
\def\gsim{\mathrel{\raise.5ex\hbox{$>$}\mkern-14mu
             \lower0.6ex\hbox{$\sim$}}}
\def\lsim{\mathrel{\raise.3ex\hbox{$<$}\mkern-14mu
             \lower0.6ex\hbox{$\sim$}}}
\def\simless{\mathbin{\lower 3pt\hbox
     {$\rlap{\raise 5pt\hbox{$\char'074$}}\mathchar"7218$}}}   %< or of order
\def\simmore{\mathbin{\lower 3pt\hbox
     {$\rlap{\raise 5pt\hbox{$\char'076$}}\mathchar"7218$}}}   %> or of order
\def\Msun{M_\odot}                                % solar masses
\def\deg{^\circ}
\def\aa{\buildrel _{\circ} \over {\mathrm{A}}}

\def\gro1655{GRO~J1655-40}
\def\4u1630{4U1630-472}
\def\h1743{H1743-322}
\def\grs1915{GRS1915+105}

\def\xrism{{\it XRISM}}
\def\athena{{\it Athena}}
\def\nustar{{\it NuSTAR}}
\def\xmm{{\it XMM-Newton}}
\def\chandra{{\it Chandra}}

\lefthead{et al.} \righthead{\4u}

\shorttitle{Magnetic-Driving of AGN UFOs}

\shortauthors{Fukumura et al.}

\begin{document}

%\title{A Magnetic View of Warm Absorbers in NGC~3783}
%\title{Correlations with Magnetically-Driven Ultra-Fast Outflows in PDS~456}
%\title{Coronal Line Broadening by Transverse Motion of Ultra-Fast Outflows }
%\title{Modeling Magnetic Disk-Wind State Transitions in Black Hole X-ray Binaries}
%\title{Modeling modality of Persistent Disk-Wind Transitions in Black Hole X-ray Binaries }
\title{Tell-Tale Spectral Signatures of MHD-driven Ultra-Fast Outflows in AGNs}

\date{\today}

\author{\textsc{Keigo Fukumura}\altaffilmark{1},
\textsc{Mauro Dadina}\altaffilmark{2}, \textsc{Gabriele Matzeu}\altaffilmark{3}, \textsc{Francesco Tombesi}\altaffilmark{4,5,6,7}, 
\textsc{Chris Shrader}\altaffilmark{4,8},
%\textsc{Mary Ogborn}\altaffilmark{1},
\textsc{and}
\textsc{Demosthenes Kazanas}\altaffilmark{4}
}

%\textsc{and} \textsc{Ioannis Contopoulos}\altaffilmark{7} }
\altaffiltext{1}{Department of Physics and Astronomy, James Madison University,
Harrisonburg, VA 22807; fukumukx@jmu.edu}
%\altaffiltext{2}{Email: fukumukx@jmu.edu}
%\altaffiltext{3}{KITP Scholar at UC Santa Barbara (2015 - 2017)}
\altaffiltext{2}{INAF, Osservatorio di Astrofisica e Scienza dello Spazio di Bologna, via P. Gobetti 93/3, 40129 Bologna, Italy​ }
\altaffiltext{3}{Department of Physics and Astronomy, University of Bologna, Via Gobetti 93/2, 40129 Bologna, Italy​ }
\altaffiltext{4}{Astrophysics Science Division, NASA/Goddard Space Flight Center,
Greenbelt, MD 20771}
%\altaffiltext{5}{Universities Space Research Association, 7178 Columbia Gateway Dr. Columbia, MD %21046}
%
%
\altaffiltext{5}{Department of Astronomy, University of Maryland, College
Park, MD20742}
\altaffiltext{6}{Department of Physics, University of Rome ``Tor
Vergata", Via della Ricerca Scientifica 1, I-00133 Rome, Italy}
\altaffiltext{7}{INAF Astronomical Observatory of Rome, Via Frascati 33, 00078 Monteporzio Catone (Rome), Italy}
\altaffiltext{8} {Department of Physics, Catholic University of America, Washington, DC 20064}
%\altaffiltext{9}{Department of Physics, Technion, Haifa 32000, Israel}
%\altaffiltext{7}{Research Center for Astronomy, Academy of Athens, Athens 11527,
%Greece}

\begin{abstract}
\baselineskip=15pt

%Since the first discovery of near-relativistic ionized outflows often referred to as ultra-fast outflows (UFOs) observed in radio-quiet Seyfert active galactic nuclei (AGNs), it has become evident that X-ray UFOs are ubiquitously present across different AGN populations over various accretion modes. Despite an increasing number of detections in recent years, however, the launching mechanism and geometrical identification of the ionized UFOs are yet poorly understood to date. 

We aim to explore spectral signatures of the predicted multi-ion UFOs in the broadband X-ray spectra of active galactic nuclei (AGNs) by exploiting an accretion disk wind model in the context of a simple magnetohydrodynamic (MHD) framework. We are focused primarily on examining the spectral dependences on a number of key properties; (1) ionizing luminosity ratio  $\lambda_{\rm ion}$, (2) line-of-sight wind density slope $p$, (3) optical/UV-to-X-ray strength $\alpha_{\rm OX}$, (4) inclination $\theta$, (5) X-ray photon index $\Gamma$ and (6) wind density factor $f_D$. With an emphasis on radio-quiet Seyferts in sub-Eddington regime, multi-ion UFO spectra are systematically calculated as a function of these  parameters to show that MHD-driven UFOs 
imprint a unique asymmetric absorption line profile with a  pronounced blue tail structure on average. Such a characteristic line signature is  generic 
to the simplified MHD disk-wind models presented in this work due to their specific kinematics and density structure. The properties of these absorption line profiles could be utilized as a diagnostics to distinguish between different wind driving mechanisms or even the specific values of a given MHD wind parameters.
We also present high fidelity microcalorimeter simulations in anticipation of the upcoming {\it XRISM}/Resolve and {\it Athena}/X-IFU instruments to demonstrate that such a ``tell-tale" sign may be immune to a spectral contamination by the presence of additional warm absorbers and partially covering gas.

\end{abstract}


%\keywords{accretion, accretion disks --- galaxies: Seyfert ---
%shock waves --- (magnetohydrodynamics:) MHD  --- X-rays:
%galaxies}

\baselineskip=15pt

\section{Introduction}

AGNs in general are known to be fueled and energized by accretion process onto their supermassive black holes.
Such accretion is almost necessarily accompanied by outflowing gas of various physical forms; i.e. molecular gas at galactic scale (at $d \gsim$ kpc; e.g. \citealt{Faucher-Giguere12,Wagner13,Tombesi15}), UV outflows of weakly or moderately ionized gas (at $d \lsim$ pc; e.g. \citealt{Crenshaw03,Arav15}), and X-ray winds most likely originating from  the innermost circumnuclear region of active galactic nuclei (AGNs). Among the X-ray winds, roughly $\sim 50\%$ of AGN populations (most notably, radio-quiet Seyfert 1s) are historically known to exhibit a series of blueshifted absorption features due to resonant transitions from various chemical elements at different charge state; aka. warm absorbers \citep[e.g.][]{Reynolds97, Crenshaw03, Blustin05, Steenbrugge05, McKernan07, HBK07, Detmers11, Tombesi13,F18, Laha21}. While their origin and geometrical structure are still unknown to date (e.g. their launching processes, nature of the outflows either in a continuous or discrete patchy structure), the physical properties of the canonical AGN warm absorbers are often characterized by 
%
%1) hydrogent-equivalent column density $N_H$, blueshift velocity along a line of sight (LoS), $v_{out}$, and ionization parameter\footnote[1]{This is defined as $\xi \equiv L_{\rm ion} / (n r^2)$ where $L_{\rm ion}$ is ionizing (X-ray) luminosity, $n$ is plasma number density at distance $r$ from AGN.}, $\xi$. For canonical AGN warm absorbers, X-ray observations in general show 
%
i) moderate hydrogen-equivalent column densities ($10^{20} \lsim N_H \textmd{[cm$^{-2}$]} \lsim 10^{22}$), ii) a wide range of ionization parameter\footnote[1]{This is defined as $\xi \equiv L_{\rm ion} / (n r^2)$ where $L_{\rm ion}$ is ionizing (X-ray) luminosity, $n$ is plasma number density at distance $r$ from AGN.} ($-1 \lsim \log (\xi \textmd{[erg~cm~s$^{-1}$]}) \lsim 4$) and iii) slow to moderate line-of-sight (LoS) velocities ($v_{\rm out}/c \lsim 0.01$ where $c$ is the speed of light), along with turbulent velocities (e.g. $v_{\rm turb} \lsim1,000$ km~s$^{-1}$), according to a number of exploratory X-ray observations so far \citep[see, e.g.,][and reference therein]{Crenshaw03,KallmanDorodnitsyn19}. State-of-the-art dispersive spectroscopic instruments such as {\it Chandra}/HETGS and {\it XMM-Newton}/RGS with the unprecedented resolving power have greatly improved the quality of X-ray spectra for warm absorber analyses.       

Independently, a number of studies on X-ray warm absorbers from Seyfert 1s have further revealed an intriguing global nature of the ionized absorbers; i.e. a uniform column density profile as a function of ionization parameter, observationally obtained from the absorption measure distribution (AMD; e.g. \citealt{B09,HBK07,HBA10,F18}). Such a characteristic distribution of X-ray absorbers along a LoS may be indicative of a well-organized, continuous outflows rather than a patchy cloud of discrete gas. We shall address the implication of AMD more in details in \S 2.2.  

In parallel to grating observations with {\it Chandra} and {\it XMM-Newton}, intensive observations primarily carried out at CCD resolution of Seyfert AGNs (including radio-quiet, radio-loud and nearby/lensed quasars, for example) have revealed ultra-fast outflows (UFOs) exhibiting a distinct property; i.e. massive column ($N_H \lsim 10^{24}$ cm$^{-2}$) with higher ionization parameter ($\log \xi \sim 4-6$) outflowing at near-relativistic velocity ($v_{\rm out}/c \lsim 0.1-0.7$) in large contrast with the properties of conventional warm absorbers as described earlier. The most prominent UFO features detected in the X-ray spectrum is usually Fe K absorption lines  attributed to \fexxv/\fexxvi\ ions \citep[e.g.][]{Pounds03,Reeves03,Reeves09,Reeves18a,Chartas09,Tombesi10a,Tombesi10b,Tombesi11,Tombesi12,Tombesi13,Tombesi15,Nardini15,Parker17}. With an increasing number of multi-epoch observations in search for Fe K UFOs  across diverse AGN populations, it is shown that UFOs can be highly variable at least on the timescale of days \citep[e.g.][]{Reeves18b}, and there appears to be  likely correlations among the observed quantities (e.g. $N_H, v_{\rm out}, \rm{equivalent~width~(EW)}, \rm{spectral~hardness}~\Gamma$)  (e.g., see, \citealt{Chartas09, Pinto18,Parker18,Matzeu17}, but also see, e.g., \citealt{Chartas18,Boissay19}).

Some UFOs have been found in very luminous AGNs possibly accreting at a good fraction of Eddington value (e.g. PDS~456, IRAS~13224-3809, APM~08279+5255 and PG~1211+143)\footnote[2]{Ultra-luminous X-ray sources (ULXs) accreting at super-Eddington rate are also known to produce powerful X-ray UFOs, which is beyond the scope of this paper.}. However, it should be reminded that the canonical Fe K UFOs have been ubiquitously detected also in sub-Eddington AGNs such as radio-quiet Seyfert 1s \citep[e.g.][]{Tombesi10a,Tombesi11} and they appear to be present even in (much) fainter sources such as Seyfert 2s  when the accretion rate is  as low as $\sim 2\%$ of Eddington rate \citep[e.g., see][for NGC~2992]{Marinucci18}. This may be providing a deep insight into the underlying UFO driving mechanism(s) as addressed below. 

Independently, more thorough grating spectroscopic studies have revealed the extstance of UV/soft X-ray UFOs (e.g. \civ, \oviii, \neix/\nex, \sixiv, \sxvi\ and \mgxii) in addition to the canonical Fe K UFOs \citep[e.g.][]{Gupta13, Gupta15, Boissay19, Hamann18, Reeves20, Krongold21,Chartas21,Serafinelli19}. These observations of multi-ion UFOs from a number of AGNs are thus clearly pointing to the importance of studying their co-existing nature in the broadband UV/X-ray spectrum perhaps within a coherent framework.   

Although these near-relativistic outflows of ionized gas are most likely launched from a nearby {\it gas reservoir} in AGNs such as an accretion disk in their deepest part of the gravitational potential, the main driving mechanism is yet to be well constrained observationally. Currently, the most plausible launching processes of UFOs are the following; (i) {\it Radiation-pressure driving} where sufficiently high UV and soft X-ray fluxes (either via line-emitting photons or continuum radiation) are capable of accelerating weakly ionized plasma through enhanced force-multipliers \citep[e.g.][]{PSK00,Proga03,Sim08,Nomura17,Hagino15,Mizumoto21}. (ii) {\it Magnetic driving} in which the plasma is loaded onto a global poloidal magnetic field to be accelerated by Lorentz force of $\mathJ \times \mathB$ (i.e. magneto-centrifugal and magnetic pressure forces; e.g. \citealt{BP82,CL94,KK94,Ferreira97,Everett05,F10a,F10b,F15,F18,Chakravorty16,Kraemer18,Jacquemin-Ide19, Jacquemin-Ide21}). 

One of the ultimate questions to be answered in UFO physics is therefore the acceleration mechanism and morphological properties of the outflows at large scale, which might be better understood by either detailed spectroscopic modeling or perhaps, for example, variability study based on principal component analysis \citep[e.g.][]{Parker18}. 
Ideally, it is conceivable that 
%In more recent years, there have been some approach to better understand a potentially 
a certain unique spectral signature of X-ray UFOs could be exploited in those leading UFO models;
%by exploiting these leading models; 
i.e. the use of the shape of the UFO line profiles as a diagnostic proxy to distinguish the underlying launching mechanisms as described above. While this might be challenging, the difference in UFO kinematics and stratified structure (attributed to different driving processes) should in principle be imprinted in the resulting absorption signatures in such a way that one could distinctively differentiate among different scenarios (Done et al. 2021, in private communication). 

To this end, in this paper we calculate broadband X-ray spectra of multi-ion UFOs under sets of characteristic parameters for AGNs 
%(i.e. $\Gamma, \alpha_{\rm OX}, \lambda_{\rm ion}$) 
and the wind property 
%(i.e. $p,\theta, f_D$) 
in trying to qualitatively extract a unique UFO spectral feature that is primarily generic to MHD-driving. We consider, for simplicity, a pure MHD-driving process to launch UFOs in a simplified situation where the radiation field plays little role in their acceleration in an effort to study characteristic effects and dependences exclusively attributed to magnetized outflows rather than hybrid outflows \citep[e.g. MHD + line-driving;][]{Everett05}. A detailed description of the wind model is given elsewhere \citep[see, e.g.,][and reference therein] {CL94,F10a,F10b,F17,K12,K19}. Below, we list a number of fundamental properties that characterize the essence of our MHD winds. 

This paper is structured as follows; We briefly review in \S 2 the characteristic properties of our MHD-driven disk wind model that is coupled to radiative transfer calculations to obtain AMD and synthetic multi-ion UFO spectra. 
In \S 3 we will discuss how changes on the model parameters could affect the resulting spectra and show how future microcalorimeters such as {\it XRISM}/Resolve \citep[][]{XRISM20} and {\it Athena}/X-IFU \citep[][]{Barret18} will allow us to fully characterize them.
%In \S 3, calculated broadband UFO spectra are presented to discuss the effects of the model parameters, followed by  microcalorimeter simulations anticipated from {\it XRISM}/Resolve and {\it Athena}/X-IFU instruments. 
We further consider a possible contamination due to warm absorbers and partially covering gas in the spectrum to assess a plausibility of realistically extracting UFO signatures that are generic to MHD-driven fast winds. We summarize our results and discuss other implications in \S 4. 

%exclusive effects and consequences due to these five model parameters in the context of MHD scenario. 

\section{Basic Models}

\subsection{Self-Similar MHD-Driven Disk Winds}

The  essence of the MHD wind model utilized in this work is captured by a number of characteristic physical properties based on \cite{CL94} under steady-state and axisymmetric assumptions. First, the outflow is globally continuous being launched from a thin accretion disk over a large radial extent in a stratified structure \citep[e.g.][]{K12,F14,K19}. LoS wind profile is described by a self-similar prescription such that, for example, wind density and velocity are respectively given by
\begin{eqnarray}
n(r,\theta) \equiv n_o f_D \left(\frac{r}{R_{\rm in}} \right)^{-p} h(\theta) ~~~{\rm and} ~~~ v_r(r,\theta) \sim v_{\rm K,in} \left(\frac{r}{R_{\rm in}}\right)^{-1/2} g_r(\theta), \label{eq:eqn2}
\end{eqnarray}
where the wind density drops monotonically and uniformly in a self-similar form\footnote[3]{Our model is reduced to BP82 MHD wind profile for $p=3/2$.} of index $p( \ge 1)$ with distance and $n_o = 2.5 \times 10^{13}$ cm$^{-3}$ is its normalization with an arbitrary factor $f_D$, 
while the wind velocity follows the local escape velocity (i.e. Keplerian law) as a function of LoS distance. 
Both quantities are normalized at the innermost launching radius\footnote[4]{In our calculations we set $R_{\rm in} \sim R_{\rm ISCO} = 6 R_g$, the innermost stable circular orbit (ISCO) around a \sw BH, where $R_g$ is the gravitational radius \citep[e.g.][]{Bardeen72,Cunningham75}.} $R_{\rm in}$ and the angular dependences, $h(\theta)$ and $g_r(\theta)$, are calculated numerically by solving the Grad-Shafranov equation in an ideal MHD framework \citep[][]{CL94}. Thus, one of the generic features of our MHD winds is that the ionized gas is launched faster at smaller radii (closer to AGN), and the expected amount of ionic column density over a LoS distance  depends sensitively on the density gradient parameter $p$ such that
\begin{eqnarray}
N^{i}_{\rm ion}(w_i, r_{i}; \theta) &\equiv& \int_{r_{i}}^{w_i r_{i}}
n(r,\theta) dr 
\nonumber \\
&=&  n_o f_D  h(\theta)\  \times
 \left\{
\begin{array}{llr}\frac{1}{2(p-1)}
  [w_i^{2(p-1)} - 1] r_i^{2(p-1)} \propto w_i^{2(p-1)} &
~~~{\rm if}~~~ p \ne 1 ,  \\
 \ln w_i  & ~~~{\rm if}~~~ p=1 , \\
\end{array} \right. 
\label{eq:column1}
\end{eqnarray}
where the column per ion (for a given charge state) is predominantly produced over the LoS distance between $r=r_i$ and $r=w_i r_i$ ($w_i>1$) for individual ions. For example, one expects an equal amount of column (i.e. a constant $N^i_{\rm ion}$) for all ions per decade in distance  if $p=1$ (i.e. $N_{\rm ion}^i = n_o f_D h(\theta)=const$ for a given inclination $\theta$ with $w_i=10$). 
Given an ensemble of suitable objects, this can be tested observationally.

%\clearpage

%\vspace{-0in}
%------------------------------- Table~1
\begin{deluxetable}{clcc}
\tabletypesize{\small} \tablecaption{Primary Parameters in MHD UFO Model ({\tt mhdwind}) } \tablewidth{0pt}
\tablehead{ & Parameter  &  & Range of Value }
\startdata
%Power-Law Photon Index $\Gamma$  & - & $2.12_{-0.008}^{+0.01}$ $^a$  \\
%                        & $\Gamma(E < 0.5\rm{keV})=3.3^a$  & - \\
%$E_{\rm P-Cygni}$ [keV] & - & $6.43_{-0.24}^{+0.22}$ $^a$ \\ \hline
%
1 & Inclination [deg] & $\theta$ & $20\deg-50\deg$   
\\
2 & X-ray Photon Index & $\Gamma$  &  $1.5-2.5$ 
\\
3 & Optical/UV-to-X-ray Strength & $|\alpha_{\rm OX}|$ &  $1.5-1.8$ 
\\
4 & Wind Density Slope & $p$ &  $1.2-1.7$ 
\\
5 & Wind Density Factor & $f_D$ & $10^{-3}-1$  
\\
6 & Ionizing Luminosity Ratio & $\lambda_{\rm ion}$ &  $0.05-0.5$ 
\\
\enddata
%\vspace{-0.0in}
%Assuming $M_8 = 1, q=-2$ and $f_c=1.7$.
\label{tab:tab1}
\begin{flushleft}
%Unless otherwise stated, we assume the solar abundances for all elements.
%\\
%$^\dagger$ We note \windon\ state in {\bf bold} font.
%\\
%$^a$ Fixed. \\
%$^b$ Wind density normalization at the launching site in units of $10^{10}$ cm$^{-3}$.
%\\
%$^c \Delta v_{\rm line} \equiv f_{\rm sm} v_\phi$. See the text in \S 2.2.
%``$p$" denotes that values are pegged at hard limit.
\end{flushleft}
\end{deluxetable}

\subsection{Radiative Transfer with Ionizing Spectra}

Our wind simulations are incorporated to post-processed radiative transfer calculations with {\tt XSTAR} \citep[e.g.][]{Kallman01}. We track the LoS distribution of various ions of different charge state by injecting a parameterized ionizing X-ray spectral energy distribution (SED) as is commonly practiced in the literature \citep[e.g.][]{F10a,Sim10a,Sim12,Ratheesh21,Higginbottom14}. 
We consider in our spectral simulations  major ionic species over the broadband (most notably, H/He-like ions; e.g. \oviii, \nex, \mgxii, \sixiv, \sxvi, \fexxv/\fexxvi,  \coxxvii, \nixxviii, among others, from atomic data in {\tt XSTAR}) and their ionic columns $N^i_{\rm ion}$ are thus determined by the wind density $n(r,\theta)$. 
Throughout our calculations, the solar abundances are assumed for all ions (i.e. $A_{\rm ion}=A_{\rm ion, \odot}$) to reduce extra degrees of complications. 
Although the observed AGN SEDs (even within radio-quiet Seyferts and nearby quasars) are extremely diverse often exhibiting  multiple spectral components \citep[e.g.][]{Fabian00,RisalitiElvis04}, we aim to capture the broad characteristics of AGN SED, $f_E$, {\it on average}. This approach may not be sufficient to make  source-specific quantitative argument, but we feel it is justified for qualitative comparisons as a first-order study. To this end, we employ the following parameters; (1) a power-law photon index ($\Gamma$), (2) optical/UV-to-X-ray flux ratio\footnote[5]{The spectral index $\alpha_{\rm OX} \equiv
0.384 \log (f_{\rm 2keV} / f_{\rm 2500\aa})$ measures the X-ray-to-UV relative brightness where
$f_{\rm 2keV}$ and $f_{\rm 2500\aa}$ are respectively 2 keV and $2500 \aa$
flux densities \citep{Tananbaum79}} (parameterized by $\alpha_{\rm OX})$, and (3) Eddington ratio ($\lambda_{\rm ion}$) determining ionizing luminosity $L_{\rm ion}$ where $L_{\rm ion} \equiv \lambda_{\rm ion} L_{E}$ and $L_E$ is Eddington luminosity assuming $M=10^8 \Msun$ such that
\begin{eqnarray}
L_{\rm ion}  \equiv \int_{13.6 \rm{eV}}^{13.6 \rm{keV}} f_E(\Gamma,\alpha_{\rm OX}, \lambda_{\rm ion}) dE \ , 
\label{eq:xi1}
\end{eqnarray}
throughout this paper. 

Given an input SED, $f_E$, parameterized by these quantities, we compute the ionization parameter $\xi(r,\theta)$ of photoionized winds  
%%
%\begin{eqnarray}
%\xi(r,\theta) \equiv \frac{L_{\rm ion}}{n(r,\theta) r^2} \ , 
%\label{eq:xi1}
%\end{eqnarray}
%%
which determines the ionic column distribution $N^i_{\rm ion} (w_{i})$ depending on the ionization threshold  of individual ions. In order to systematically treat multi-ion UFOs, we consider AMD with the help from equations~(1) and (2) defined as 
\begin{eqnarray}
{\rm{AMD}}(p) \equiv \frac{d \left[N^i_{\rm ion}(r,\theta)\right]}{d
\left[\log \xi(r,\theta) \right]} \propto r^{1-p} \propto
\xi^{\frac{p-1}{2p-1}} \ ,
\label{eq:amd}
\end{eqnarray}
which is a useful proxy of the relative strength of multiple absorbers of different charge states from a global perspective \citep[e.g.][]{B09}. For example, one finds $\rm {AMD}=const$ with distance and ionization state if $p=1$ as previously seen in equation~(2), whereas AMD monotonically declines with increasing distance or slowly increases with increasing ionization parameter if $p>1$. In fact, a number of AGN warm absorber analyses of multiple ions have indeed favored $1 \lsim p < 1.5$ based on AMD study\footnote[6]{Similar AMD analyses for BH XRB disk-winds have independently implied similar results \citep[e.g.][for 4U~1630-472 and]{Trueba19,F21} and \citealt{F17}.} \citep[e.g.][]{B09,HBK07,HBA10,Detmers11,F18}. Since our spectral modeling considers multi-ion UFOs of a global distribution rather than being restricted to the most pronounced Fe K features, the $\rm {AMD}$ is extremely important.  

\clearpage

\begin{figure}[t]% ------------------------------------- Figure~1
\begin{center}
\includegraphics[trim=0in 0in 0in
0in,keepaspectratio=false,width=3.3in,angle=-0,clip=false]{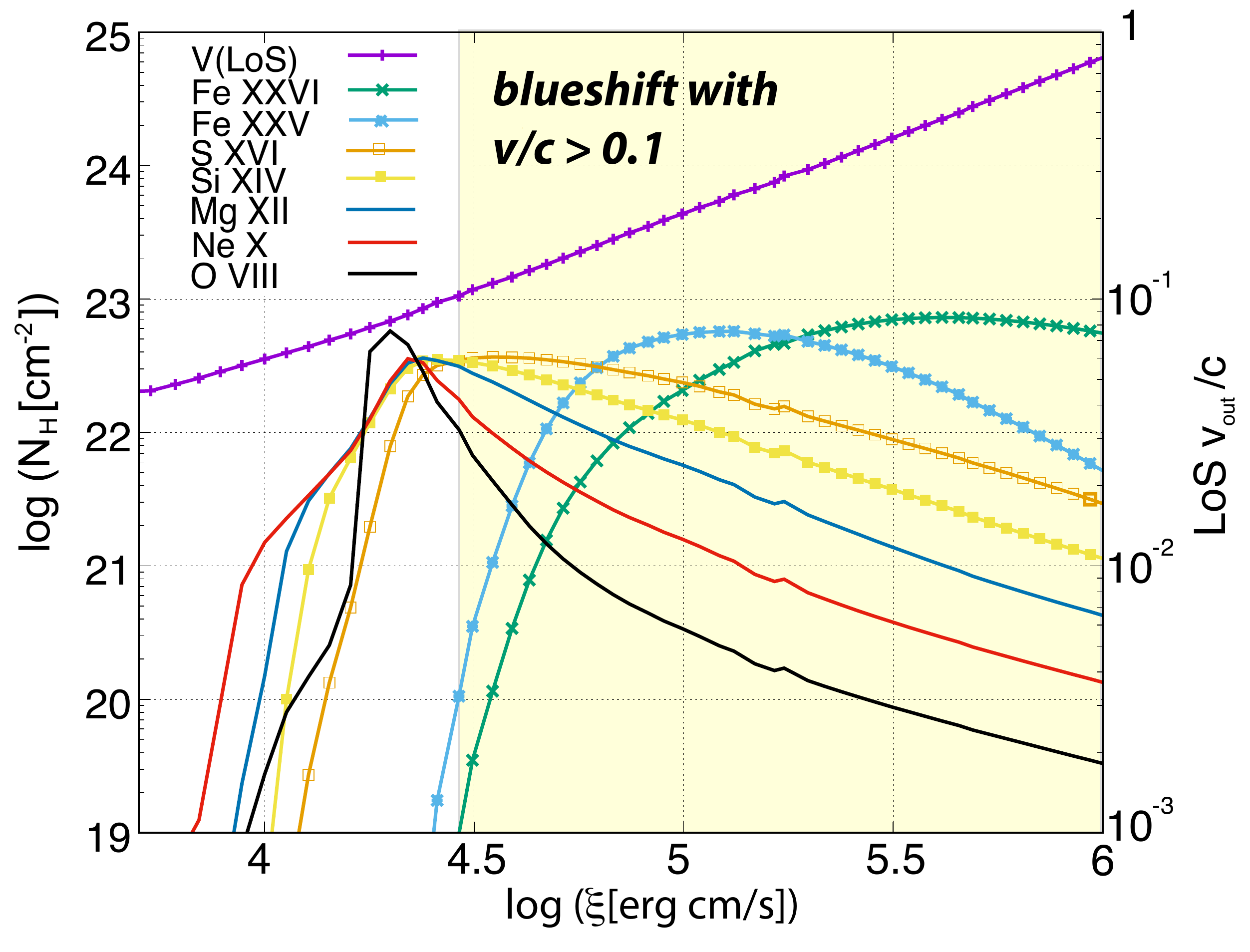}\includegraphics[trim=0in 0in 0in
0in,keepaspectratio=false,width=3.3in,angle=-0,clip=false]{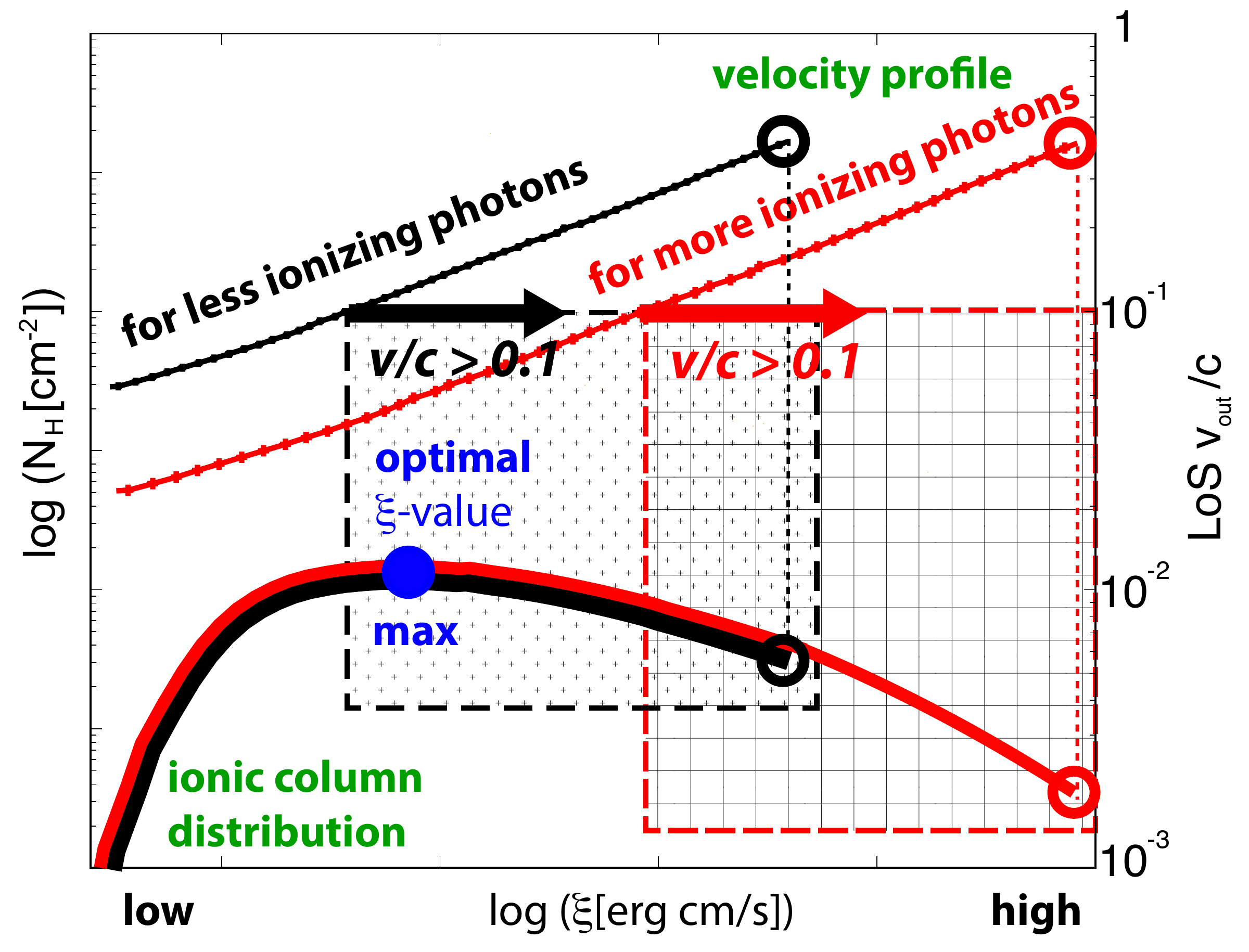}  
\end{center}
\caption{Left: An example of calculated AMD for multi-ion UFOs with MHD winds assuming $\Gamma=2, \theta=40\deg, p=1.2, f_D=1, \alpha_{\rm OX}=1.5$ and $\lambda_{\rm ion}=0.05$. Local hydrogen-equivalent columns (expressed on left ordinate) are shown for major ions; \fexxvi\ (green), \fexxv\ (cyan), \sxvi\ (orange), \sixiv\ (yellow), \mgxii\ (dark blue), \nex\ (red) and \oviii\ (dark) along with the LoS wind velocity $v_{\rm out}/c$ (purple; expressed on right ordinate). A majority of the column is highly blueshifted ($v_{\rm out}/c>0.1$ in light yellow region) for many ions. Right: Schematics of AMD with velocity profiles $v_{\rm out}(\xi)$ for different ionizing photons corresponding to the ionic column (dark for {\it less} ionizing photons and red for {\it more} ionizing photons).  Filled blue circle denotes where a local column is maximally produced. Open circle corresponds to the column produced at the innermost wind layer where $\xi$ value is the highest in each case. Note that the wind under higher ionization (in red) produces lower column of near-relativistic blueshift ($v_{\rm out}/c>0.1$ in squared domain) yielding slower UFOs in comparison with the lower ionization case (in shaded domain). }
\label{fig:amd}
\end{figure}

\clearpage

%(via $q$), the AMD is constant, i.e., independent of the radius $r$ or ionization
%parameter of the plasma $\xi$ when $q=1$, while it scales as ${\rm{AMD}} \propto r^{-1/2}
%\propto \xi$, i.e., monotonically increasing with ionization $\xi$ for the BP82 models ($q=3/4$).

Therefore, the final UFO spectrum ({\tt mhdwind}) is uniquely calculated by a set of these model parameters ($\theta, p, f_D, \Gamma, \alpha_{\rm OX}, \lambda_{\rm ion}$) for a given wind morphology. The range of the primary model parameters is listed in {\bf Table~1}.

\section{Results}

\subsection{Synthetic Multi-Ion Broadband UFO Spectra}

As stated above, we aim to globally model the multi-ion UFO. We thus need to compute the ionic column densities $N^{i}_{\rm ion}$ for various chemical elements, most notably H/He-like ions; e.g. \oviii, \nex, \mgxii, \sixiv, \sxvi, \fexxv/\fexxvi,  \coxxvii, \nixxviii, among others. Having this in mind, we handle the radiative processes between the wind material and the ionizing radiation following the prescription in our previous work \citep[e.g.][reference therein]{F10a,F17,F21}. 
%
%We follow in this work the same computational scheme adopted in our previous work \citep[e.g.][reference therein]{F10a,F17,F21} for handling radiative process between the wind material and the background ionizing radiation $f_E$ in an attempt to compute the emergent ionic column densities $N^{i}_{\rm ion}$ for various chemical elements, most notably H/He-like ions; e.g. \oviii, \nex, \mgxii, \sixiv, \sxvi, \fexxv/\fexxvi,  \coxxvii, \nixxviii, among others.
%
The photo-absorption cross-section $\sigma^i_{\rm abs}$ for a given ion is then calculated with the Voigt profile \citep[e.g.][]{Mihalas78,Kotani00,Hanke09} using the simulated wind kinematics \citep[e.g.][]{F10a}. 

In this framework, the line width (or shape) is naturally determined as a result of  the internal wind velocity gradient (or shear velocity) $v_{\rm sh}$ along a LoS from the MHD wind model, which is typically on the order of $\sim 10\%$ of the LoS velocity itself at a given distance. Hence, we obtain the velocity shear directly for the Voigt function rather than employing an arbitrary ``turbulent motion" parameterization. 

With these quantities being available, the synthetic multi-ion  UFO spectrum ({\tt mhdwind}) is calculated by computing the line optical depth $\tau^i(E)$ as a function of energy $E$ such that 
\begin{eqnarray}
\tau^i(E) \equiv \sigma^i_{\rm abs}(v_{\rm sh}; E)  N^{i}_{\rm ion}(E) \ . 
\label{eq:tau}
\end{eqnarray}
Therefore, the predicted broadband spectrum $f_{\rm obs}(E) \propto \Pi_i e^{-\tau^i(E)}$ is a function of wind-related parameters ($\theta, p, f_D$) coupled to the other parameters associated with radiation field ($\Gamma,\alpha_{\rm OX},\lambda_{\rm ion}$). This is how we simulate theoretical UFO spectra ({\tt mhdwind}) as we aim to investigate spectral signatures that could be attributed to the generic features of MHD disk-winds perhaps in terms of the wind kinematics and/or density distribution as described above. 

%Since our work is of a generic nature rather than considering 
%%not motivated for detailed case study targeted towards 
%specific objects, we fix the wind density factor $f_D=1$ as a representative value in this paper unless otherwise stated.  

To better illustrate the underlying physics responsible for characterizing broadband absorption spectra as a consequence of radiative transfer, we show an example of the simulated AMD in {\bf Figure~\ref{fig:amd} (left)} for a series of primary ions; local hydrogen-equivalent columns (value expressed on left ordinate) for \fexxvi\ (green), \fexxv\ (cyan), \sxvi\ (orange), \sixiv\ (yellow), \mgxii\ (dark blue), \nex\ (red) and \oviii\ (dark) along with the LoS wind velocity $v_{\rm out}/c$ (purple; value expressed on right ordinate) assuming a fiducial set of parameters; i.e. $\Gamma=2, \theta=40\deg, p=1.2, f_D=1, \alpha_{\rm OX}=1.5$ and $\lambda_{\rm ion}=0.05$. While the exact AMD profile  can  vary from one case to the other depending on the input SED and wind conditions, this is one of the typically obtained ionic distributions relevant for Seyfert AGN UFOs in our calculations. It should be noted that the velocity profile $v_{\rm out}(\xi)$ is always unique in the sense that the fastest part of the wind at the innermost layer always possesses the highest ionization parameter value. This is because acceleration mechanism in our MHD framework is not due to radiation, but to magnetic processes (by both magnetocentrifugal and pressure gradient forces; e.g. \citealt{BP82,CL94,F10a}). 
It is seen that the columns $N^i_{\rm ion}$ of individual ions are progressively produced over a large extent of LoS distance (or ionization parameter) as coronal X-ray  radiation traverses the wind. The predominant fraction of the LoS-integrated  column per ion is found in the higher ionization regime (i.e. in the vicinity of the central engine where the wind velocity is  near-relativistic, $v/c \gsim 0.1$, in the light yellow region of the figure) responsible for  the final absorption features we observe. The shape of a typical column profile is asymmetric in  a way that the column rapidly falls off with decreasing $\xi$ (i.e. increasing distance) past its peak point  where  the maximum local column is produced; i.e. approximately at $\log \xi \sim 4.3$ for the soft X-ray absorbers while at $\log \xi \sim 5.0-5.5$ for \fexxv/\fexxvi\ ions. This AMD hence indicates that  most of the ions (i.e. low-ionization ions such as \oviii\ and high-ionization ions such as \fexxvi) are necessarily near-relativistic (i.e. highly blueshifted) with an appreciable total column being  identified as multi-ion UFOs, especially, \fexxv/\fexxvi\ being broader and faster\footnote[7]{In general, this can depend on the specific ionizing SED and luminosity...etc.}.    

{\bf Figure~\ref{fig:amd} (right)} schematically depicts how the absorber's velocity would change for different ionizing X-ray photons with respect to AMD.  The velocity of the wind  $v_{\rm out}$ at a given $\xi$ value is lower with more ionizing photons (red) compared to that with less ionizing photons (dark) because the velocity profile $v_{\rm out}(\xi)$ systematically shifts towards higher ionization with increasing ionizing luminosity as the innermost wind is always ionized the most. 
Given that the value of the ionic column is essentially determined by the  ionization threshold, the production of local ionic column for a given element (e.g. \fexxvi) occurs roughly in the same ionization range (solid dark/red curves). %because of a fixed ionization threshold for each charge state. 
Filled blue circle denotes where the local column is maximally produced. 
The innermost wind irradiated by more (less) ionizing photons are labeled by red (dark) open circle where $\xi$ is the highest, as expected. 
In the presence of more ionizing photons in the illuminating SED (in red), the ionization parameter becomes higher by definition and the wind produces lower columns of near-relativistic blueshift (indicated by squared domain).  As a consequence, the integrated column to be observed predominantly consists of those with lower blueshift, thus ``slower" UFOs. In contrast, much of the total column is highly blueshifted if the ionizing photons in SED are less (in shaded domain), thus ``faster" UFOs. 
Therefore, the differences in blueshift and ionic column due to the amount of ionizing photons  inevitably lead to different absorption features as shall presented below in \S 3.1.1.

\clearpage

\begin{figure}[t]% ------------------------------------- Figure~2
\begin{center}
\includegraphics[trim=0in 0in 0in
0in,keepaspectratio=false,width=3.2in,angle=-0,clip=false]{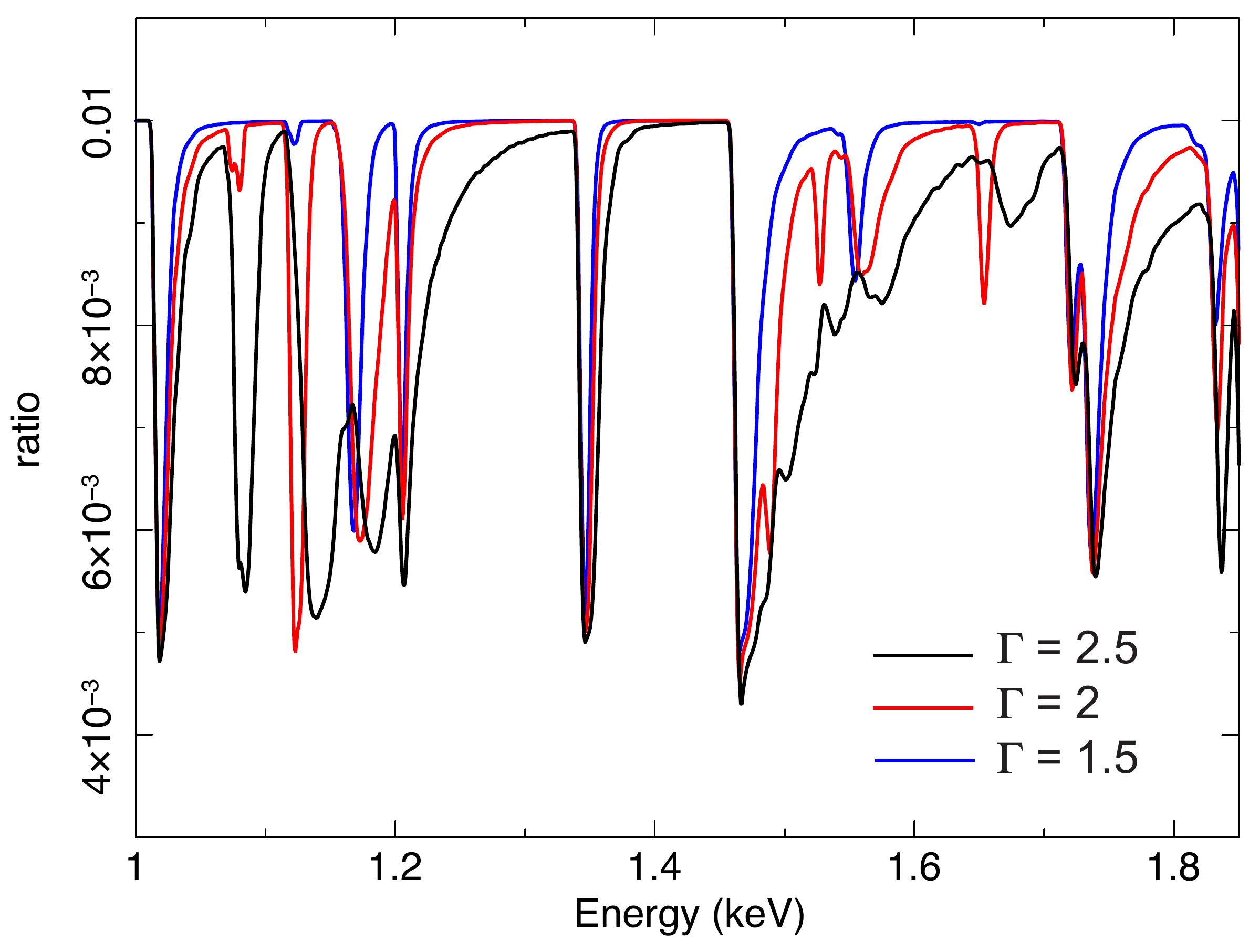} \includegraphics[trim=0in 0in 0in
0in,keepaspectratio=false,width=3.2in,angle=-0,clip=false]{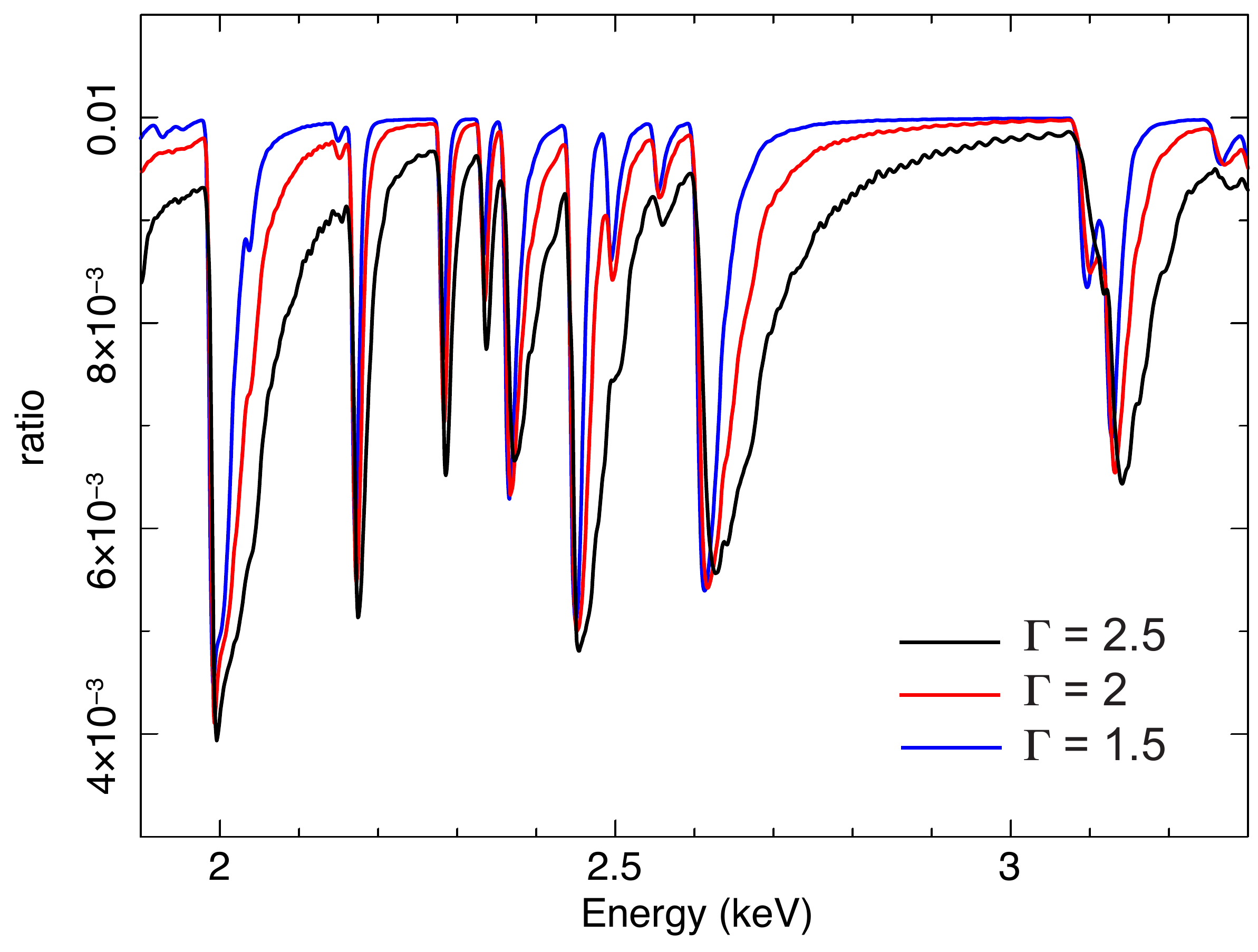} \includegraphics[trim=0in 0in 0in
0in,keepaspectratio=false,width=3.2in,angle=-0,clip=false]{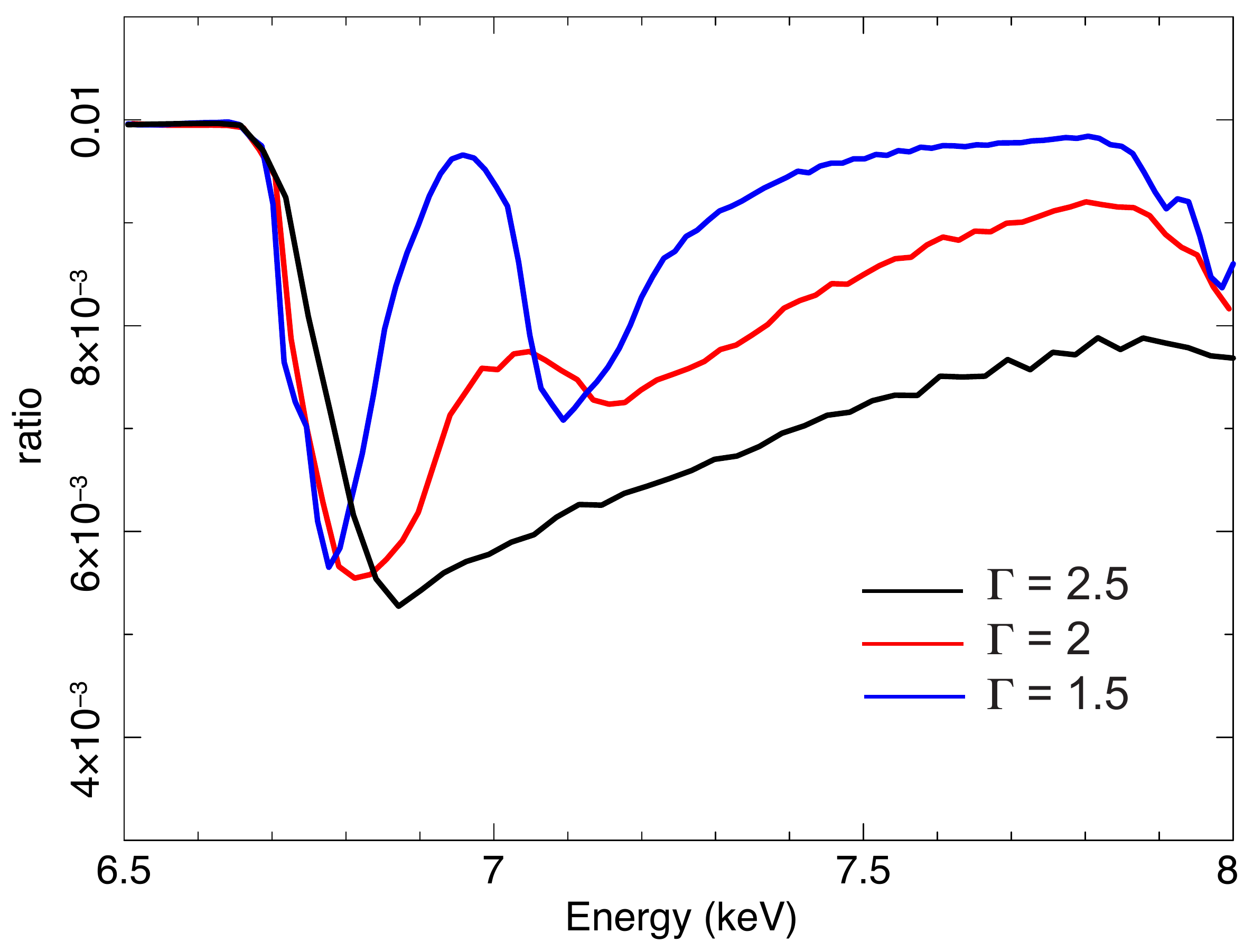}
%\includegraphics[trim=0in 0in 0in
%0in,keepaspectratio=false,width=3.2in,angle=-0,clip=false]{spec_gamma.pdf}\includegraphics[trim=0in 0in 0in
%0in,keepaspectratio=false,width=3.2in,angle=-0,clip=false]{spec_theta.pdf}
\end{center}
\caption{Theoretical multi-ion absorption spectra (ratio to the continuum) for different photon index $\Gamma$ with $\theta=30\deg, p=1.3, f_D=1, \alpha_{\rm OX}=1.5$ and $\lambda_{\rm ion}=0.1$. }
\label{fig:gamma}
\end{figure}

\subsubsection{Model Dependencies of Multi-Ion MHD Disk-Winds}

To further follow up the significance of the simulated AMD shown in {\bf Figure~\ref{fig:amd}}, we calculate a number of multi-ion UFO spectra ({\tt mhdwind}) by allowing for one parameter to be varied at a time. 
For example, the dependence of a power-law slope $\Gamma$ is shown in {\bf Figure~\ref{fig:gamma}} where we assume $\theta=30\deg, p=1.3, f_D=1, \alpha_{\rm OX}=1.5$ and $\lambda_{\rm ion}=0.1$. It is clear that the overall UFO signatures tend to be enhanced with softer (steeper) power-law component in the absence of efficient ionizing photons. Note that the trough energy in each absorption line remains almost unchanged, while the line width is dramatically affected as a result of ionization equilibrium because the corresponding velocity profile $v_{\rm out}(\xi)$ with AMD is accordingly shifted when $\Gamma$ changes, as depicted in {\bf Figure~\ref{fig:amd} (Right)}. 
When the SED is harder (i.e. smaller $\Gamma$), the inner layer (i.e. faster moving part) of the global MHD wind can be fully ionized producing little observable columns. Thus, the main absorption feature of the UFOs in that case will be governed predominantly by moderate or slower layer of the wind  where the broadening is weakened but the centroid energy remains the same. This trend is  systematically found in the broadband energy ranging from $\sim 1$ keV to Fe K complex. In particular, \fexxv/\fexxvi\ lines become so broad that the two are blended together for softer SED (e.g. $\Gamma=2.5$) as if it were a single edge feature. 

In {\bf Figures~\ref{fig:alphaox}-\ref{fig:theta}} in Appendix A, we show more synthetic multi-ion absorption spectra of the {\tt mhdwind} model to demonstrate various dependences on the model parameters. All in all, it is striking, and a common feature, that the most of the absorption features from different ions in these calculations are noticeably {\it asymmetric} such that blue tails are more extended and skewed towards higher energy. 
This is a manifestation of  the unique AMD coupled to the MHD wind kinematics shown in {\bf Figure~\ref{fig:amd}} as discussed in \S 3.1. That is, those columns produced from the wind at near-relativistic blueshift ($v/c \gsim 0.1$) largely dictate the asymmetric character in the spectrum as a ``tell-tale" signature.

\begin{figure}[t]% ------------------------------------- Figure~3
\begin{center}
\includegraphics[trim=0in 0in 0in
0in,keepaspectratio=false,width=3.5in,angle=-0,clip=false]{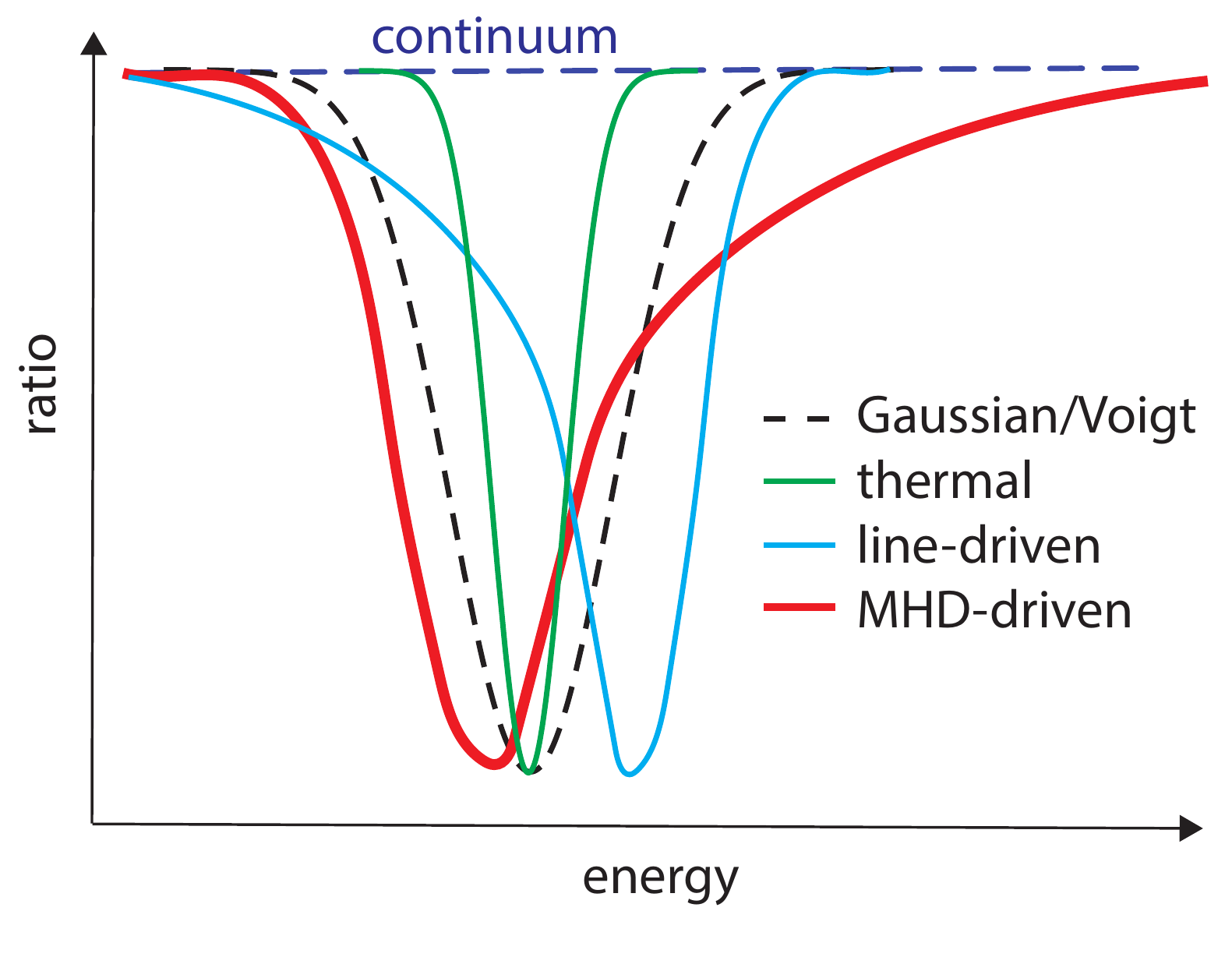}\includegraphics[trim=0in 0in 0in
0in,keepaspectratio=false,width=3.5in,angle=-0,clip=false]{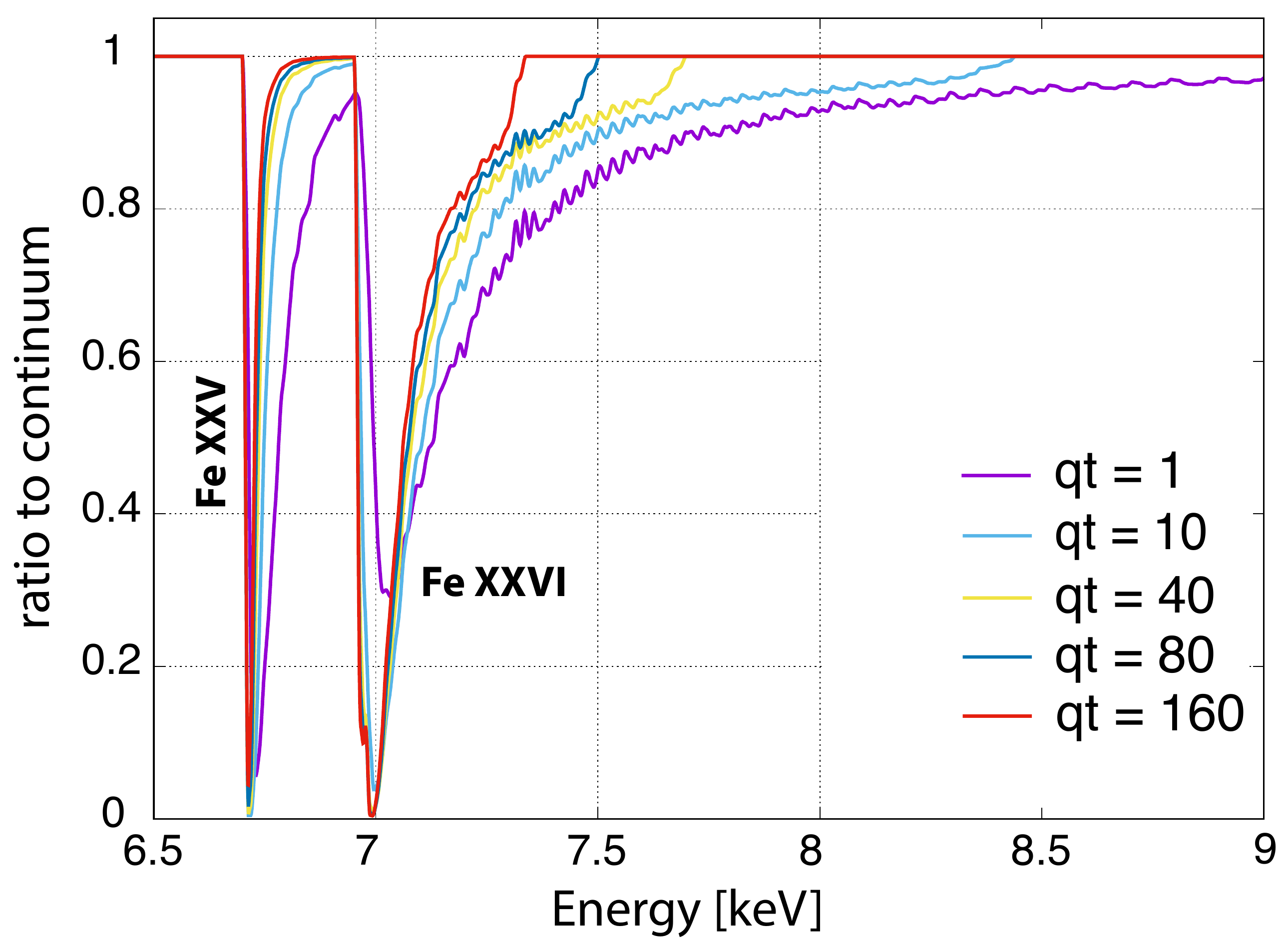}
\end{center}
\caption{Left: Schematic diagram showing a comparison of characteristic absorption profiles among various wind driving mechanisms; Gaussian/Voigt profiles (dashed), thermal-driving (green), line-driving (cyan) and MHD-driving (red). Right: Spectral change in \fexxv/\fexxvi\ lines due to disk truncation at the inner launching radius parameterized by $q_T$. }
\label{fig:schematics}
\end{figure}

As mentioned earlier, one of the common spectral features from multi-ion UFOs in these calculations is blueshifted tails 
%that is asymmetrically extended towards higher energy 
due to the generic kinematics of MHD-driven winds; i.e. the closer in to AGN, the faster the outflow. This unique characteristics of the magnetized wind can imprint a distinct absorption line profile through the shape of AMD as a consequence of photoionization balance as illustrated in {\bf Figure~\ref{fig:schematics} (Left)} (see also {\bf Figure~\ref{fig:amd}}). 
In comparison with other disk-wind models, {\it the expected shape of the line profile from MHD-driven winds (red) is generally unique and distinctive enough to be separately identified.} For example, line-driving (cyan) in general tends to produce an asymmetric line shape of an extended red wing (which is reversed to MHD-driving discussed here) due to an asymptotically increasing terminal wind velocity, whereas thermal-driving\footnote[8]{Due to its typically low velocity, thermal-driving is not really relevant for UFOs.} (green) allows for a relatively narrow line shape of little asymmetry because of a generically constant (slow) motion (C. Done and J. Reeves, in private communications). In contrast, a phenomenological Gaussian/Voigt line profiles (dashed) are symmetric.

%In comparison with other phenomenological models such as Gaussian or Voigt function, {\it the expected shape of the line profile from MHD-driven winds is generally unique and distinctive enough to be separately identified.}
%

On the other hand, it should be pointed out that the exact asymmetric line shape is very sensitive to the shape of AMD (including the velocity profile $v(\xi)$); the asymmetry could be minimized or even reversed depending on the underlying SED, wind property and other physical factors. Therefore, we emphasize that the obtained asymmetric line profiles here should be regarded as the most probable case {\it on average}. 
%rather than as a definitive characteristics in the MHD wind framework. 

Since we are focused on UFOs in the context of MHD driving, ionic columns are necessarily produced at smaller distances from AGN where the velocity is near-relativistic (see {\bf Fig.~\ref{fig:amd} (left)}). Since the broadening of lines is determined by the wind shear velocity (from the LoS velocity gradient) at a given distance as discussed in \S 3.1, the line width tends to be very large considering that $v_{\rm shear}/v_{\rm out} \sim 0.1$ along a LoS. For this reason, the UFO spectra in our model tend to be  broad such that subtle atomic line features might be blurred. For example, despite our expectation, no doublet structure is resolved for \fexxvi\ Ly$\alpha$ lines (6.973 keV and 6.952 keV) due to blending, while it should be indeed identified in BH XRB disk winds because of their much lower velocities \citep[e.g.][]{Miller15,F21}.

\subsection{The Effects of the Inner Disk Truncation}

Accretion disk winds in AGNs are often discussed in comparison with ionized outflows in BH XRBs. As an ideal laboratory owing to their much shorter time scale, sequenced X-ray observations of BH transient events can provide a valuable insight into wind's morphology, variable nature and likely correlations, for example.     
Among those BH XRBs, a handful of systems are known to exhibit  wind signatures in X-ray spectra. While there is no clear analogy for AGNs, pronounced disk winds are indeed present in BH XRBs during high/soft (bright) state, while winds appear to be extremely weak (if not absent) during low/hard (quiescent) state, as they go through a series of distinct accretion modes in the hardness-intensity diagram (aka. q-diagram) \citep[e.g.][]{Fender04, FenderBelloniGallo04,RemillardMcClintock06, McClintockRemillard06,Done07,Belloni10, Ponti12}. 
During high/soft state, the standard thin accretion disks \citep[][]{SS73,NT73} in the context of general relativity are  thought to extend all the way down to the ISCO based on X-ray observations of the thermal disk continuum \citep[e.g.][]{McClintock14,Li05,Shafee06} and interpretations of asymmetric broad Fe line profiles \citep[e.g.][]{Fabian00}. On the other hand, it has been suggested that the inner disk may actually be truncated beyond the ISCO during low/hard state (as mass accretion rate drops), considering that the inner part of accretion  becomes optically thin, hotter and radiatively inefficient (such as ADAF/ADIOS; e.g. \citealt{NY94,BB99}) forming a compact corona around a BH that is surrounded by the standard thermal disk \citep[e.g.][]{Esin97,Done07}. While such a speculation has gained much attention, the disk truncation signatures, also independently derived from reflection spectroscopy and reverberation lags,  have still remained elusive observationally  \citep[e.g.][]{Shidatsu11,Petrucci14,Plant15,Garcia15,DeMarco17}. 

To follow up a possibility of inner disk truncation in the context of Seyfert UFOs, we consider here a possible consequence of such a truncation of the inner disk. %since most of the ions responsible for UFOs are produced at smaller radius in the context of MHD driving, 
Given that the MHD-driven UFOs are primarily launched from the innermost region of the disk (where the Keplerian motion is fast and ionization state is high; see {\bf Figure~\ref{fig:amd}}), it is expected that removing the inner part of the disk should significantly reduce the amount of near-relativistic  plasma (i.e. UFOs) to be launched. As a result, the observed UFO line profile could be substantially less blueshifted depending on the truncation radius $R_T$. To probe this effect, we take one of our template runs (i.e. $\theta=40\deg, \Gamma=2, \alpha_{OX}=1.5$...) and calculate a line spectrum by parameterizing $R_T$ as $R_{T} \equiv q_T R_{\rm in} \sim q_T R_{\rm ISCO}$ where we set $q_T=1, 10, 40, 80$ and $160$. That is, the disk is not truncated for $q_T=1$, while extremely truncated at large distance for $q_T=160$. {\bf Figure~\ref{fig:schematics} (Right)} demonstrates the effect of $q_T$ on \fexxv/\fexxvi\ lines as an example. It is seen that the most blueshifted part of the feature is gradually disappearing with increasing $q_T$ as expected, whereas the centroid absorption structure (with small blueshift) remains almost unaffected as it is produced by the exterior wind  that is launched from the outer part of the disk. 
Note that the line depth is almost independent of $q_T$ since it is predominantly determined by the wind density factor $f_D$ which is fixed in this work. 

\clearpage

\begin{figure}[t]% ------------------------------------- Figure~4
\begin{center}
\includegraphics[trim=0in 0in 0in
0in,keepaspectratio=false,width=3.3in,angle=-0,clip=false]{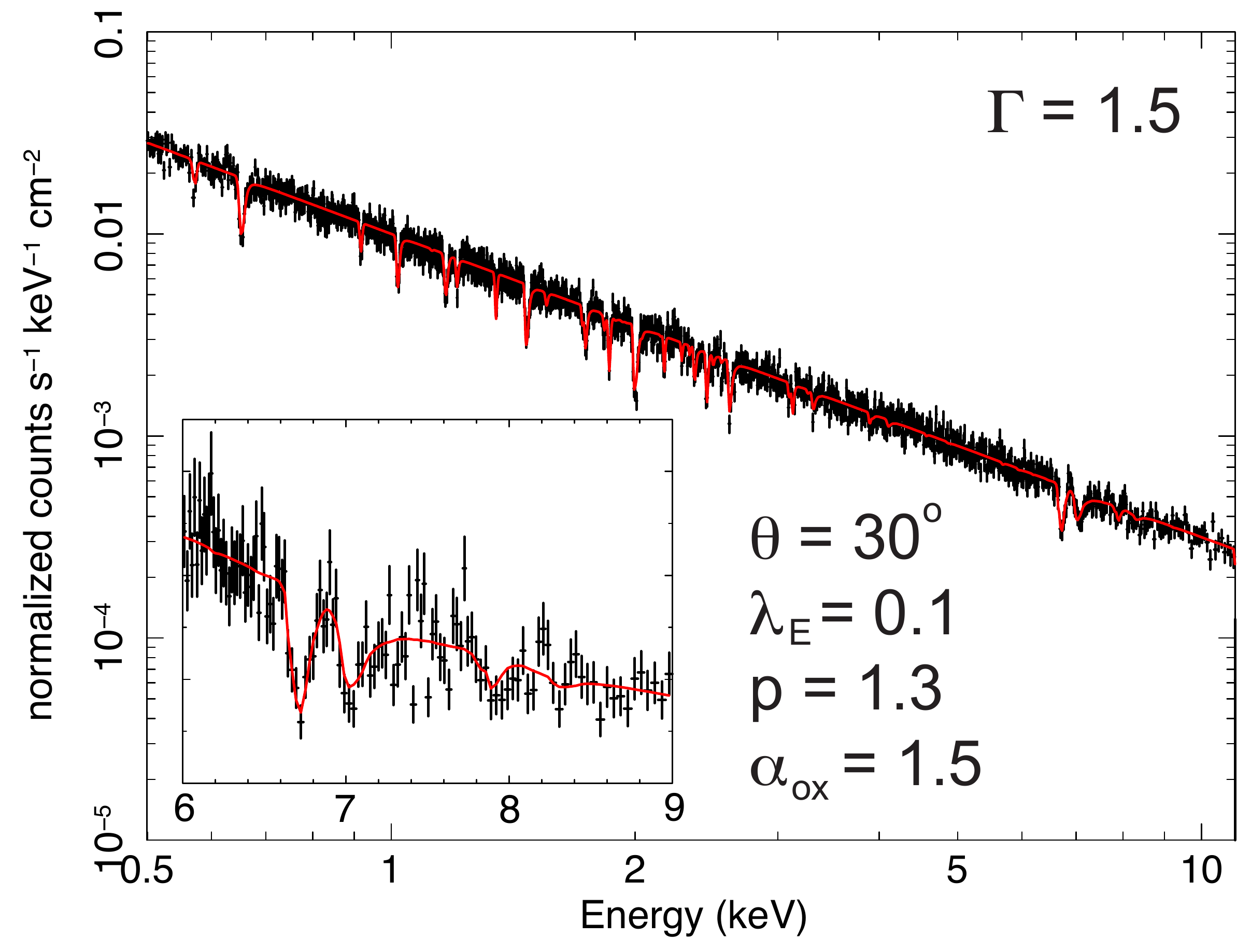}\includegraphics[trim=0in 0in 0in
0in,keepaspectratio=false,width=3.3in,angle=-0,clip=false]{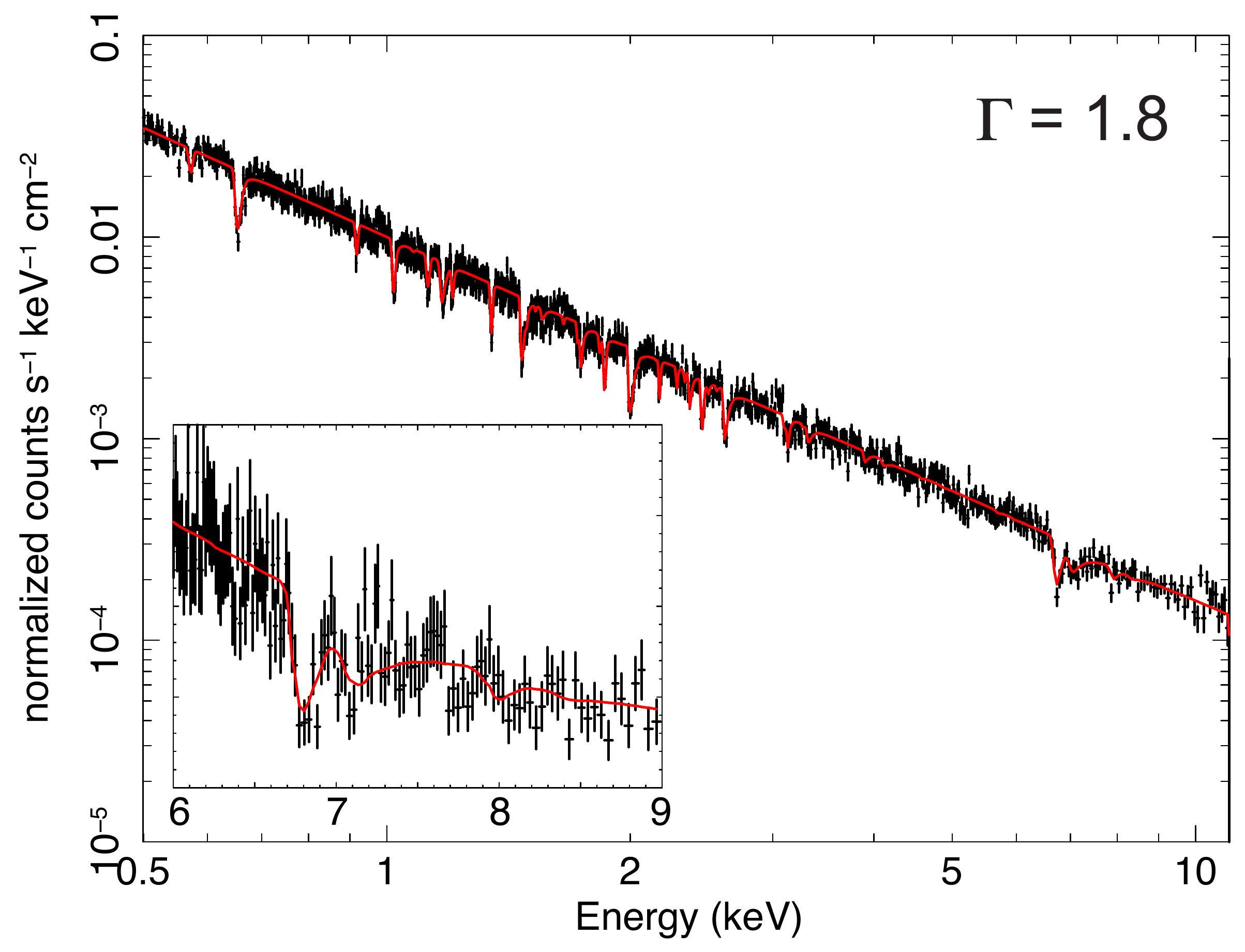}
\includegraphics[trim=0in 0in 0in
0in,keepaspectratio=false,width=3.3in,angle=-0,clip=false]{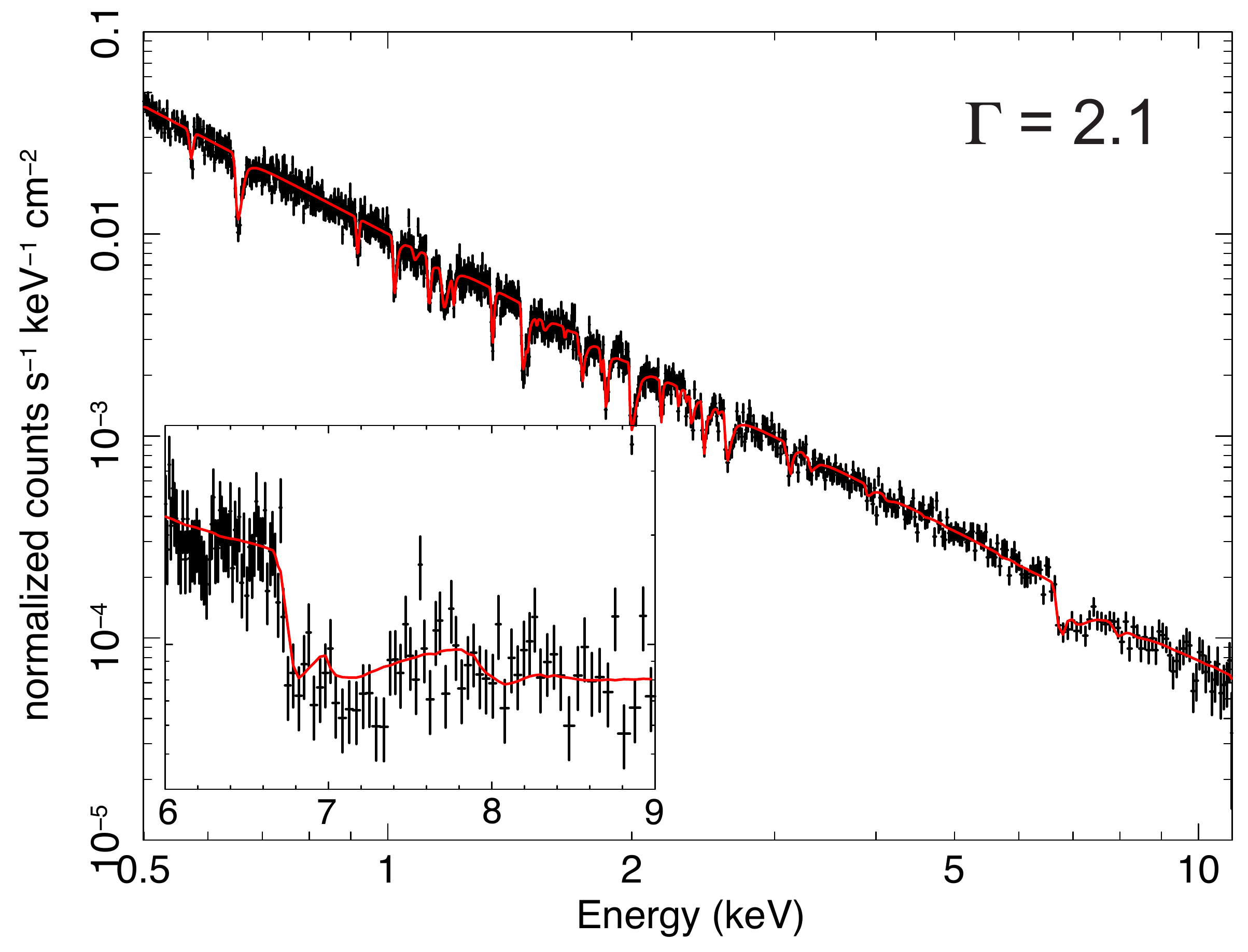}\includegraphics[trim=0in 0in 0in
0in,keepaspectratio=false,width=3.3in,angle=-0,clip=false]{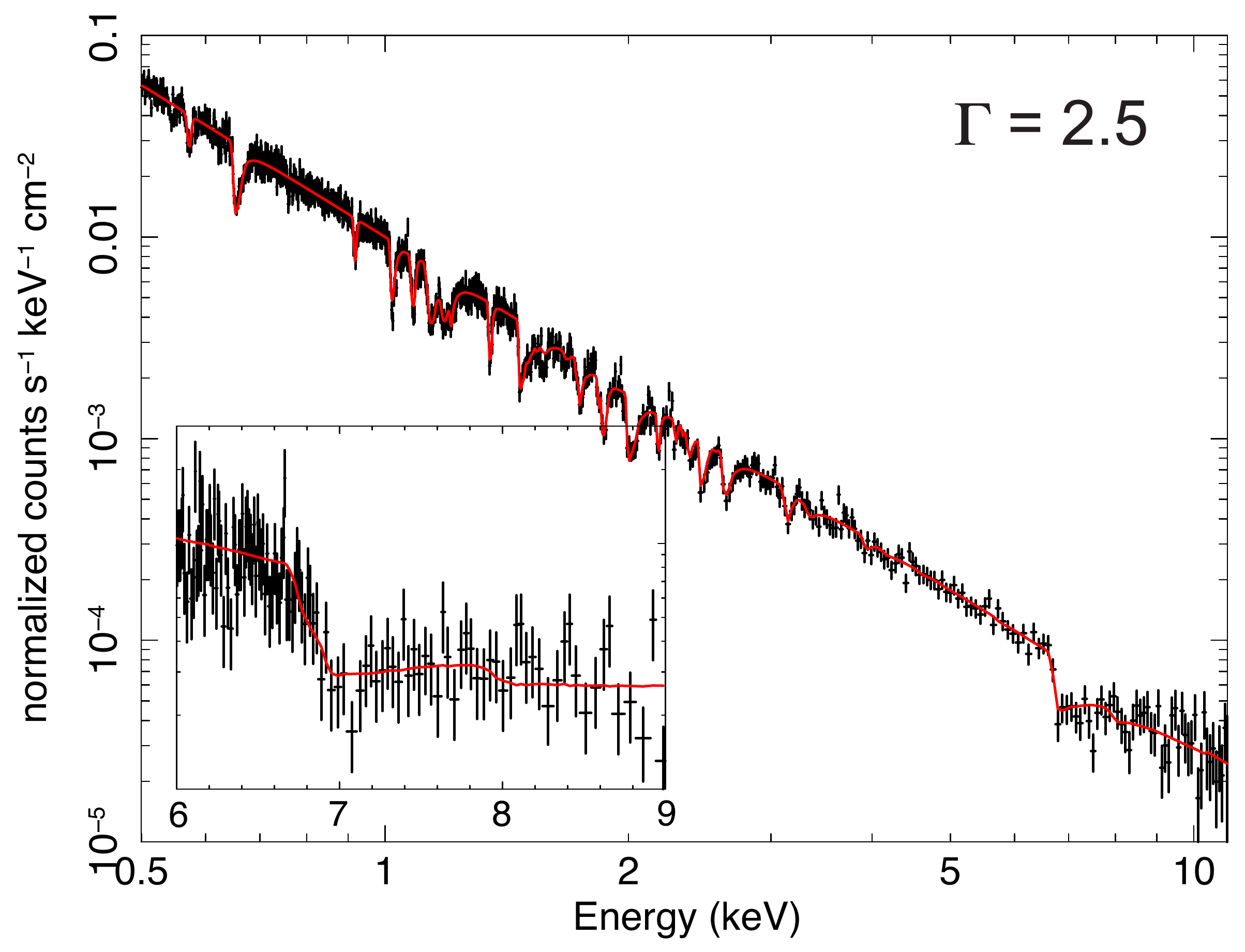}
%{sim_80ks_gamma_xrism_comp.pdf}
\end{center}
\caption{Simulated 100ks {\it XRISM}/Resolve spectra for different photon index $\Gamma$ with $\theta=30\deg, p=1.3, \alpha_{\rm OX}=1.5, f_D=1$ and $\lambda_{\rm ion}=0.1$. Inset shows a close-up of Fe K absorbers. }
\label{fig:gamma2}
\end{figure}

\clearpage

\subsection{The Effects of Soft Excess}
 
Given that a good fraction of Seyfert 1 AGNs exhibits soft X-ray excess (SE), another unique soft X-ray component, in their SED (i.e. more than 50\% to 90\%; e.g. \citealt{TurnerPounds89,Crummy06,Boissay16}), we also consider a possible effect of SE on the UFO signatures.  
Although SE is present in the energy band lower than the most of the major UFO lines (e.g. \mgxii, \sixiv, \sxvi\ and Fe K ions), a potential impact of radiative transfer on the production of these UFO columns is yet to be known. 
For an illustration purpose, we use a single blackbody of temperature $kT_{\rm SE} = 100$ eV to phenomenologically mimic the SE component in SED, which is usually a reasonable representation of it. 
For simplicity, it is assumed here that SE originates almost entirely from the same region producing the primary emission or completely within the launching region of the outflows; i.e. no ``diffuse" (larger than flow cone base) component of the SE.  
We then make a comparison, with and without SE in an assumed SED, by performing radiative transfer calculations with {\tt XSTAR}  to see the resulting UFO spectra. 

\begin{figure}[t]% ------------------------------------- Figure~5
\begin{center}
\includegraphics[trim=0in 0in 0in
0in,keepaspectratio=false,width=3.3in,angle=-0,clip=false]{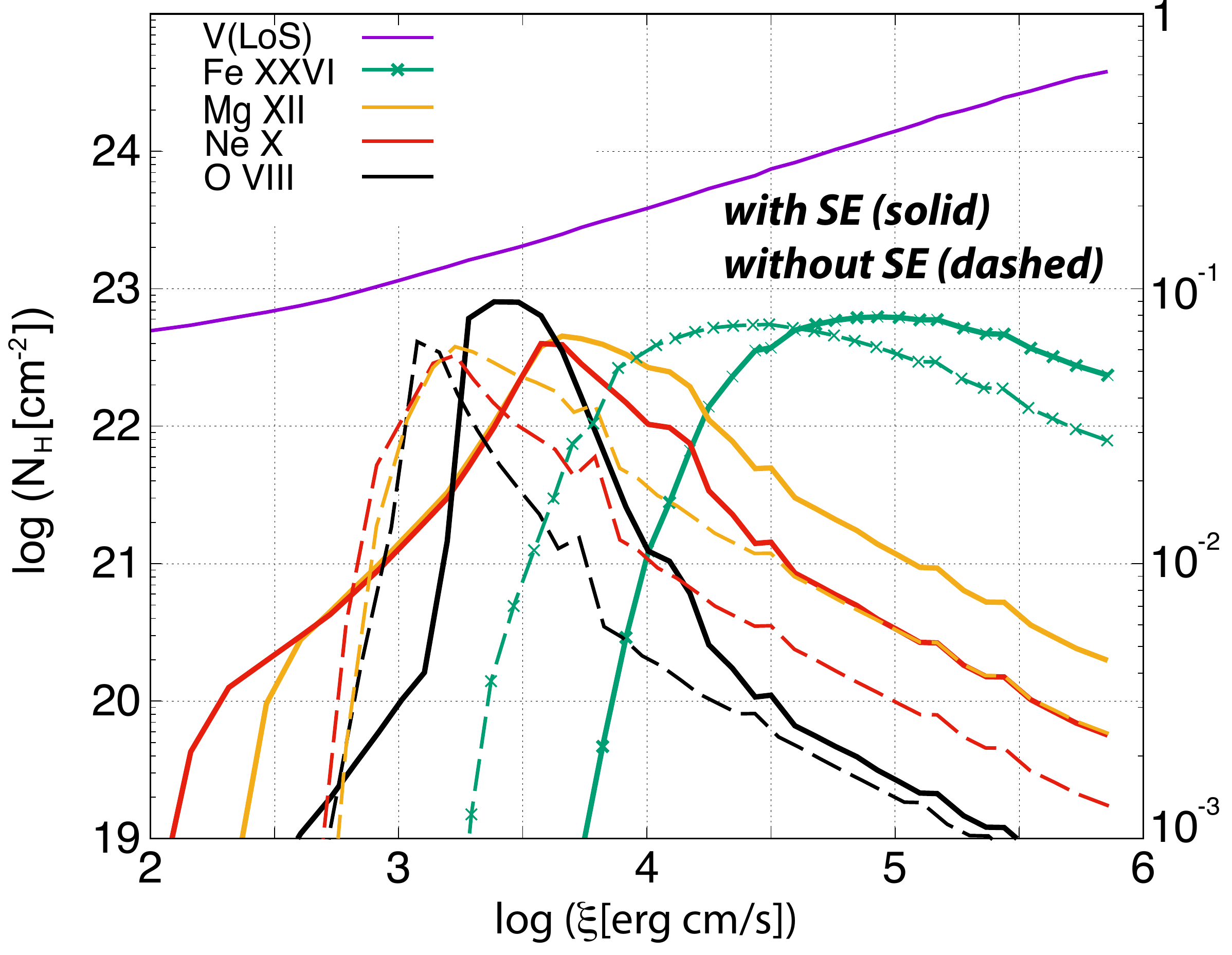}\includegraphics[trim=0in 0in 0in
0in,keepaspectratio=false,width=3.4in,angle=-0,clip=false]{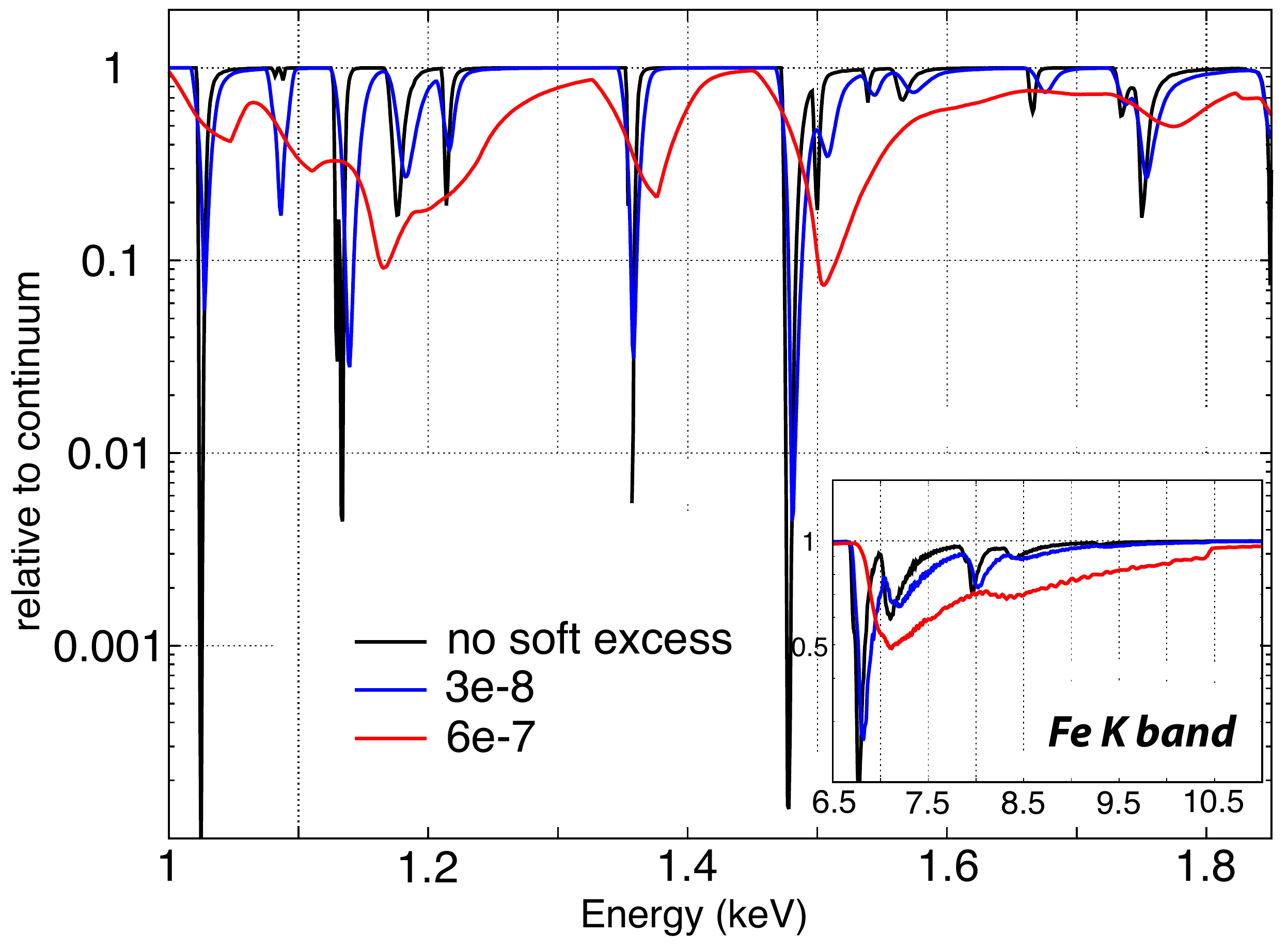}
\end{center}
\caption{Left: Sample comparison of AMD, similar to Figure~\ref{fig:amd}, with (solid) and without (dashed) SE flux in the injected SED for selected ions assuming $\theta = 40\deg, p =1.3, \alpha_{\rm OX} = 1.5, f_D = 1$ and $\lambda_{\rm ion} = 0.05$. Right: Theoretical multi-ion absorption spectra (ratio to the continuum) for different SE normalization of $K_{\rm SE}=6 \times 10^{-7}$ (strong in red) and $3 \times 10^{-8}$ (moderate in blue) in comparison with the absence of SE (in dark) assuming $\theta = 30\deg, p =1.3, \alpha_{\rm OX} = 1.5, f_D = 0.1$ and $\lambda_{\rm ion} = 0.05$. }
\label{fig:SE}
\end{figure}

Figure~\ref{fig:SE} Left shows a representative comparison of calculated AMD for selected ions with SE (solid; its normalization of  $K_{\rm SE}=3 \times 10^{-8}$) and without SE (dashed) in the injected SED of $\Gamma=2$ and $\alpha_{\rm OX}=1.5$ with the rest of the model parameters being the same in both cases. It is seen that the ionization front for various transitions are almost systematically shifted towards higher $\xi$ regime (corresponding spatially to smaller distances to the BH) due to the extra soft photons from the SE, allowing the produced UFOs to be  more blueshifted and broader in the spectrum. There is little change, though, in the calculated column. 

To confirm this view, we further calculate a broadband UFO spectrum based on a similar set of parameters by assuming different SE normalizations in Figure~\ref{fig:SE} Right; $K_{\rm SE}=3 \times 10^{-8}$ (moderate in blue) and $6 \times 10^{-7}$ (strong in red) cases. As demonstrated, the presence of SE could have a significant impact on the broadband multi-ion UFO spectra depending on its strength. In an extreme case of a strong SE flux, a large fraction of ionic column per transition line can be produced in the inner faster regions of the MHD wind in a way that the resulting absorption lines may be broad enough to be blended together (e.g. in red) in theory purely due to radiative transfer process (i.e. photoionization). 
Given that the presence of SE tends to make ionizing spectrum softer, it is consistent that the physical role of SE in UFO spectral features is qualitatively very similar to that by softer X-ray slope as shown in Figure 2 in \S 3.1.1.

%\subsection{Constraint on MHD-Wind Structure between Hard/Soft State}
\subsection{Spectral Simulations with Microcalorimeters}

By exploiting a library of theoretical spectra of multi-ion UFOs presented in \S 3.1, we further investigate observational implications here in an anticipation of upcoming microcalorimeter missions with {\it XRISM}/Resolve and {\it Athena}/X-IFU in an effort to realistically assess a plausibility of constraining  the UFO signatures. 
{\bf For subsequent simulations, we use {\tt xarm\_res\_h5ev\_20170818.rmf} for \xrism\ and  {\tt XIFU\_CC\_BASELINECONF\_2018\_10\_10\_EXTENDED\_LSF.rmf} for \athena\ through this work. }

\subsubsection{Feasibility with {\it XRISM}/Resolve Observations}

Assuming a fiducial 2-10 keV X-ray flux of $\sim 2.3 \times 10^{-11}$ erg~cm$^{-2}$~s$^{-1}$ for canonical Seyfert AGNs, we perform spectral simulations of the models considered in \S 3.1 with 100ks {\it XRISM}/Resolve exposure. These are idealized simulations considering only the theoretical UFO component with no additional spectral complexity.   

{\bf Figure~\ref{fig:gamma2}} shows the simulated spectra for different $\Gamma$ where the richness of the UFO features seem to be well captured ranging from $\sim 0.5$ keV to the Fe K complex band. 
The overall asymmetry of multi-ion absorbers are still traceable in these simulated data to the extent that the expected ``tell-tale" MHD-driving signature can be resolved especially for softer X-ray spectrum where the wind is less overionized when a fewer number of hard X-ray photons can participate in ionization.    
The other simulated {\it XRISM}/Resolve spectra for various dependences are also shown in  {\bf Figure~\ref{fig:alphaox2}-\ref{fig:theta2}} in Appendix B.

%In particular, Fe K UFOs appear to be more sensitive to X-ray hardness compared to the other low energy (and low ionization) ions. Overall, the multi-ion absorbers become more asymmetric as X-ray becomes softened (e.g. $\Gamma=2.5$) because the wind is less overionized when a fewer number of hard X-ray photons can participate in ionization.    

\subsubsection{Convolution of UFOs with Additional Absorbers in Simulated Data}

Observed X-ray spectra of a large number of radio-quiet Seyfert AGNs are generally well known to contain multiple absorption components including broadband complex atomic features; e.g. warm absorbers, typically characterized by a series of ions at different charge states \citep[e.g.][]{Crenshaw03, Blustin05, Steenbrugge05, McKernan07, HBK07, Detmers11, F18,Laha21}. Due to the atomic physics, primarily dictated by the oscillator strength and Einstein coefficient for individual resonant transitions per ion, stronger warm absorbers on average are predominantly given by H/He-like species  appearing in the soft X-ray to UV band (also occasionally accompanied by Fe L-shell transitions and unresolved transition array (UTA) due to {\it 2p-3d} transitions in Fe M-shell ions as well; e.g.  \citealt{Behar01a,Behar01b,Netzer04}). In addition, some Seyfert AGN spectra occasionally give a hint at the presence of partially covering ionized materials \citep[e.g.][]{Matzeu16}.    

To make a more realistic diagnosis for a plausible detection of MHD-driven UFO features by robustly constraining the wind parameters, we  simulate another microcalorimater spectrum based on a similar set of MHD UFO model that are now convolved with these additional  absorption components. That is, we adopt a template spectral model, {\tt mhdwind} mtable model, of $\Gamma=2, \alpha_{\rm OX}=1.5, \theta=35\deg, p=1.25, f_D=0.1, $ and $\lambda_{\rm ion}=0.08$, as a test case. For warm absorbers, we follow the conventional approach by employing a three-zone {\tt warmabs} model generated with {\tt xstar2xspec} such that 
\begin{eqnarray}
{\tt warmabs} = 
 \left\{
\begin{array}{llr}
N_H = 10^{21}, \log \xi = 2, v_{\rm out}/c = 0.001  & ~~ \textmd{(cold gas)} ,  \\
N_H = 10^{22}, \log \xi = 3, v_{\rm out}/c = 0.003  & ~~ \textmd{(warm gas)} , \\
N_H = 10^{23}, \log \xi = 4, v_{\rm out}/c = 0.01  & ~~ \textmd{(hot gas)} , 
\end{array} \right. 
\label{eq:wa}
\end{eqnarray}
%
%(1) a {\it cold} gas ($N_H = 10^{21}$ cm$^{-2}, v_{\rm out}=300$ km~s$^{-1}$, $\log \xi = 2$), (2) an {\it warm} gas ($N_H = 10^{22}$ cm$^{-2}, v_{\rm out}=500$ km~s$^{-1}$, $\log \xi = 3$) and (3) a {\it hot} gas ($N_H = 10^{23}$ cm$^{-2}, v_{\rm out}=1,000$ km~s$^{-1}$, $\log \xi = 4$), 
%
assuming a power-law continuum of  $\Gamma=2$ and a uniform turbulent velocity\footnote[9]{We have verified that the resulting {\tt warmabs} spectrum is only weakly sensitive to the choice of the turbulence velocity since the observed line width (e.g. Gaussian width $\sigma$) is typically only a few hundred km~s$^{-1}$ \citep[e.g.][for radio-quiet Seyfert 1s like NGC~3783]{Kaspi00,Netzer03,Krongold03,Blustin05,Gabel05}, which would be very distinct for those of UFOs \citep[e.g.][]{Tombesi13, Gofford15,Nardini15,Reeves18a}.} of $\sigma=500$ km~s$^{-1}$ \citep[e.g.][]{Kaspi00,Netzer03,Krongold03,Blustin05,Gabel05}. 
Independently, a partially covering ionized absorber (perhaps in a discrete patchy cloud) is represented with {\tt zxipcf} model \cite[][]{Miller06,Reeves08} by 
\begin{eqnarray}
{\tt zxipcf} = 
 \left\{
\begin{array}{llr}
N_H = 5 \times 10^{23}, \log \xi = 4, v_{\rm out}/c = 0.01  & ~~ \textmd{(partial absorbers)} ,  \\
\end{array} \right. 
\label{eq:zxipcf}
\end{eqnarray}
with a turbulent velocity of 200 km~s$^{-1}$ and the covering fraction of 0.5.
Thus, the {\tt composite} model is expressed symbolically as 
\begin{eqnarray}
{\tt composite = tbabs \times po \times mhdwind \times warmabs_{\tt C} \times warmabs_{\tt W} \times warmabs_{\tt H} \times zxipcf} 
\end{eqnarray}
where the subscript denotes each warm absorber zone in equation~(\ref{eq:wa}), which is combined with the partial covering absorber ({\tt zxipcf}) in equation~(\ref{eq:zxipcf}) while being multiplied by the underlying power-law spectrum ({\tt po}) and the Galactic neutral absorber ({\tt tbabs}) with $N_H^{\rm Gal} = 10^{20}$ cm$^{-2}$. 
We then generate a 100ks {\it XRISM}/Resolve spectrum assuming the same 2-10 keV flux of $\sim 2.3 \times 10^{-11}$ erg~cm$^{-2}$~s$^{-1}$ as is considered in \S 3.2.1. 
%

%\clearpage

\begin{figure}[t]% ------------------------------------- Figure~5
\begin{center}
\includegraphics[trim=0in 0in 0in
0in,keepaspectratio=false,width=6in,angle=-0,clip=false]{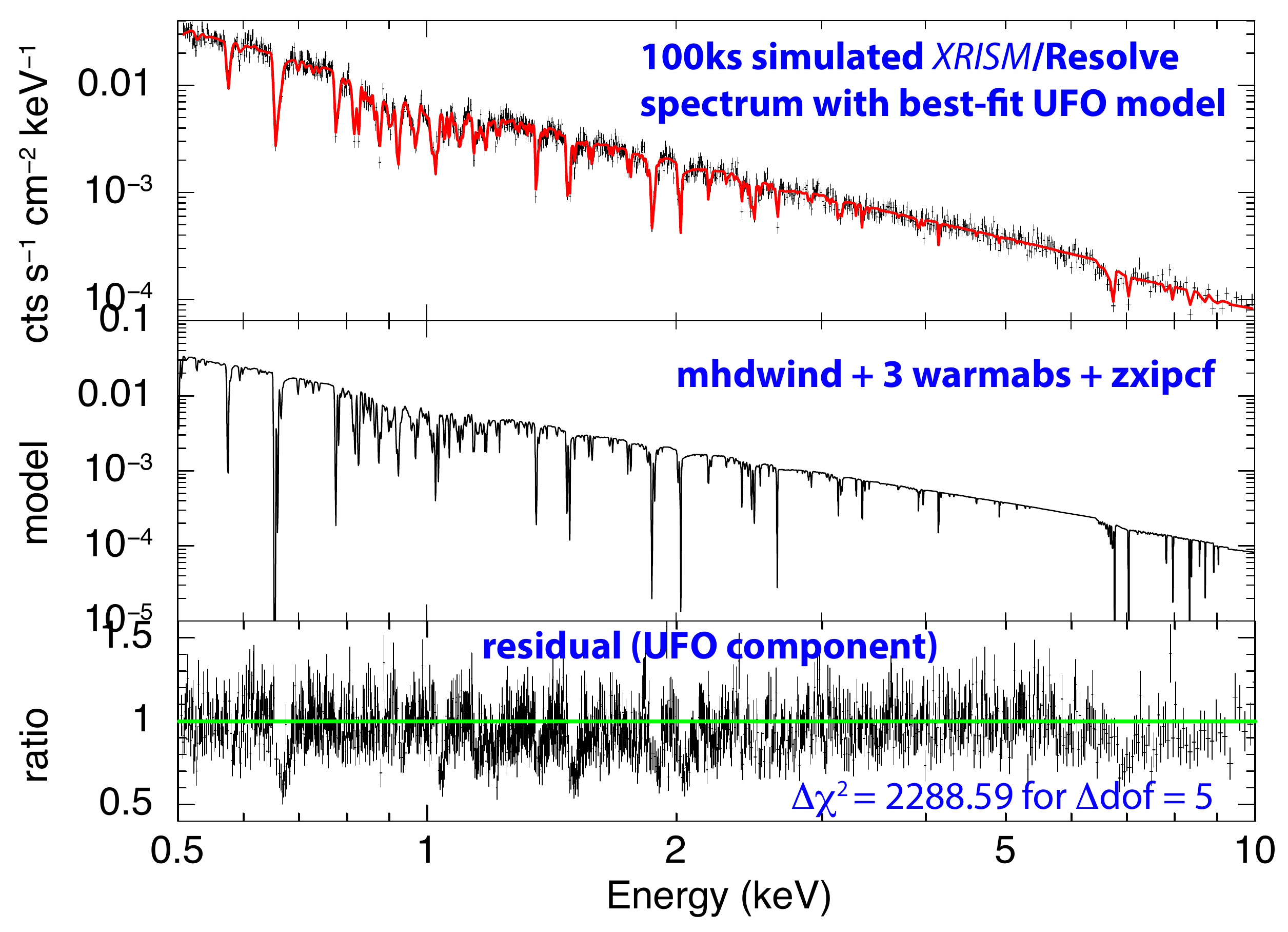}
\end{center}
\caption{Upper: Simulated 100ks {\it XRISM}/Resolve spectrum (in dark) for $\theta=35\deg, p=1.25, \Gamma=2, \alpha_{\rm OX}=1.5, f_D=0.1$ and $\lambda_{\rm ion}=0.08$ with the best-fit {\tt composite} model (in red). See {\bf Table~\ref{tab:bestfit}} for the best-fit model parameters. Middle: Unfolded best-fit model spectrum. Lower: Residuals in the form of data/model ratio in the absence of the UFO component ({\tt mhdwind}). }
\label{fig:composite}
\end{figure}

%------------------------------- Table~2
\begin{deluxetable}{ll|cc}
\tabletypesize{\small} \tablecaption{Best-Fit {\tt Composite} Model for 100 ks Simulated {\it XRISM}/Resolve Spectrum} \tablewidth{0pt}
\tablehead{Component & Parameter & Best-Fit Value & Model Value } 
\startdata
%Wind density slope $p$ & $0.7, 0.8, 0.9, 1.0, 1.15, 1.29$ \\
{\tt tbabs} & $N_H^{\rm Gal}$ [cm$^{-2}$] & $10^{20}$$^\dagger$ & $10^{20}$$^\dagger$  \\ \hline
{\tt po} & $\Gamma$  & $2.00^{+0.01}_{-0.01}$ & 2   \\ \hline
%Inclination angle $\theta$ [degrees] & $71.3^{+0.79}_{-0.69}$$^\sharp$ ~ ($67.0^{+1.07}_{-1.01}$)$^\diamond$ \\
{\tt mhdwind} & $\alpha_{\rm OX}$  & $1.51^{+0.02}_{\rm -1.51}$ & 1.5   \\
& $\lambda_{\rm ion}$   & $0.081_{-0.004}^{+0.003}$  & 0.08 \\
& $\theta$ [deg] & $34.9_{-1.0}^{+2.1}$ & 35 \\ 
& $p$  & $1.22_{-1.22}^{+0.04}$ & 1.25 \\ 
& $f_D$ & $0.074_{-0.01}^{+0.03}$ & 0.1 \\
\hline
{\tt warmabs} & $N_{\rm H,Cold}$  & $1.01_{-1.01}^{+0.33} \times 10^{21}$ & $10^{21}$ \\
& $\log \xi_{\rm Cold}$  & $1.95_{-0.07}^{+0.08}$  & $2$ \\
& $v_{\rm out,Cold}/c$ & $-8.74_{-4.5}^{+3.3}  \times 10^{-4}$ & $10^{-3}$ \\
& $N_{\rm H,Warm}$  & $1.00_{-0.06}^{+0.29} \times 10^{22}$  & $10^{22}$ \\
& $\log \xi_{\rm Warm}$  & $2.97_{-0.02}^{+0.03}$  & $3$ \\
& $v_{\rm out,Warm}/c$ & $-2.96_{-0.14}^{+0.12} \times 10^{-3}$ & $3 \times 10^{-3}$ \\
& $N_{\rm H,Hot}$  & $1.1_{-0.16}^{+0.59} \times 10^{22}$ & $10^{22}$ \\
& $\log \xi_{\rm Hot}$  & $4.00_{-0.04}^{+0.04}$ & $4$ \\ 
& $v_{\rm out,Hot}/c$ & $-9.80_{-0.23}^{+0.24} \times 10^{-3}$  & $10^{-2}$ \\ \hline
{\tt zxipcf} & $N_{\rm H,zxipcf}$  & $4.00_{-1.09}^{+1.35} \times 10^{23}$ & $5 \times 10^{23}$ \\
& $\log \xi_{\rm zxipcf}$  & $3.98_{-0.08}^{+0.07}$ & $4$ \\ 
& $v_{\rm out,zxipcf}/c$ & $-1.00_{-0.01}^{+0.01} \times 10^{-2}$ & $10^{-2}$ \\ \hline
\hline
%$r(\textmd{\fexxvi})/r_g$ & $\lesssim 25$  \\
& $\chi^2/$dof & $15930.27/18982$  ~ (18218.86/18987)$^\ddagger$ \\
%Downstream Elecetron Energy $kT_e$ & See \S 3.3 \\
%Thickness  $H \equiv h/r$  & $0.1, 0.5, 1$   \\
%Mass-Accretion Rate $\dot{m} \equiv \dot{M}/\dot{M}_{E}$ & $0.1, 0.5, 1$  \\
%Accreting Plasma  $r_{\rm sh}/r_g$  & $2?, 3?$   \\
\enddata
\vspace{0.05in}
%\begin{flushleft}
We assume $M = 10^8 \Msun$, $f_{\rm 2-10 keV}=2.3 \times 10^{-11}$ erg~cm$^{-2}$~s$^{-1}$ and $L_{\rm ion}=10^{44}$ erg~s$^{-1}$.
\\
$^\dagger$ Fixed. ~~ $^\ddagger$ In the absence of the UFO component ({\tt mhdwind}).
%$^\sharp$ Treated as a free parameter. \\
%$^\diamond$ A fixed value obtained from N15.
%\end{flushleft}
\label{tab:bestfit}
\end{deluxetable}
%

%\clearpage

The simulated {\it XRISM}/Resolve spectrum is first generated based on the {\tt composite} wind model expressed in equation~(8). We then fit the mock data with the same  model to retrieve the best-fit broadband spectrum of multi-ion UFOs by allowing {\tt mhdwind, warmabs} and {\tt zxipcf} componets to be varied simultaneously (except for the fixed turbulence velocities). The upper panel of {\bf Figure~\ref{fig:composite}} shows the simulated spectrum (in dark) along with the folded best-fit model (in red). The middle panel shows the unfolded best-fit model. 
A number of UFO signatures in this simulated spectrum seems to be well accounted for by the {\tt composite} model successfully constraining the UFO parameters at a reasonable statistical significance (yielding $\chi^2$/dof=15930.27/18982). The best-fit parameters for the simulated spectrum are listed in {\bf Table~2}. When fitted with the {\tt composite} wind model, the broadband UFO features are unambiguously resolved and the correct parameter values are successfully retained within a reasonably narrow range of uncertainties.  
It is seen that the predicted UFO signatures of relatively broad line profiles of asymmetry appear to be immune to the spectral contamination by external absorbers under this particular parameter set. The uniquely blueshifted UFO line shape of asymmetry (discussed in §3 and Figure 3) is still unmistakenly preserved and identified by simple visual inspection.

To see the significance of {\tt mhdwind} component in the spectrum, we subsequently removed the UFO component ({\tt mhdwind}) from the best-fit  spectrum ({\tt composite}). The lower panel of {\bf Figure~\ref{fig:composite}} now shows the residuals (in the form of data/model ratio) after the subtraction of the UFO component ({\tt mhdwind}) to examine whether the ``tell-tale" sign of the UFO signatures can still be observed in data.  
We find that the unique spectral features of MHD-driven UFOs of multiple ions in {\tt mhdwind} are clearly seen in the residual, thus enabling {\tt mhdwind} component to be successfully disentangled. Without {\tt mhdwind} component, the best-fit model becomes significantly worse resulting in $\chi^2$/dof=18218.86/18987 (i.e. $\Delta \chi^2=2288.59$ for $\Delta$dof = 5 by removing four {\tt mhdwind} parameters). Therefore, this simulation potentially demonstrates a realistic viability of extracting MHD-driven UFO signatures of multiple ions in the presence of contaminating non-UFO absorbers. 

From a technical perspective, it is known that bright Seyfert AGN spectra often exhibit reflection components resulting from photons reprocessed by disks \citep[e.g.][]{MagdziarzZdziarski95,RossFabian05}. It is hence conceivable that such reflection (consisting of a series of atomic features such as emission and absorption/edges as well as Compton hump) might also contaminate the underlying UFO signatures. We further investigate the impact of such reflection features by implementing both {\tt xillverCp} (for distant reflection) and {\tt relxillCp} (for relativistic reflection) models \citep[][]{Garcia13} in  \xrism/Resolve simulations. In Appendix C we present a 100 ks \xrism/Resolve spectrum with a fiducial reflection component to find that even in that case the expected MHD spectral signatures indeed remain preserved.

\begin{figure}[t]% ------------------------------------- Figure~6
\begin{center}
\includegraphics[trim=0in 0in 0in
0in,keepaspectratio=false,width=3.5in,angle=-0,clip=false]{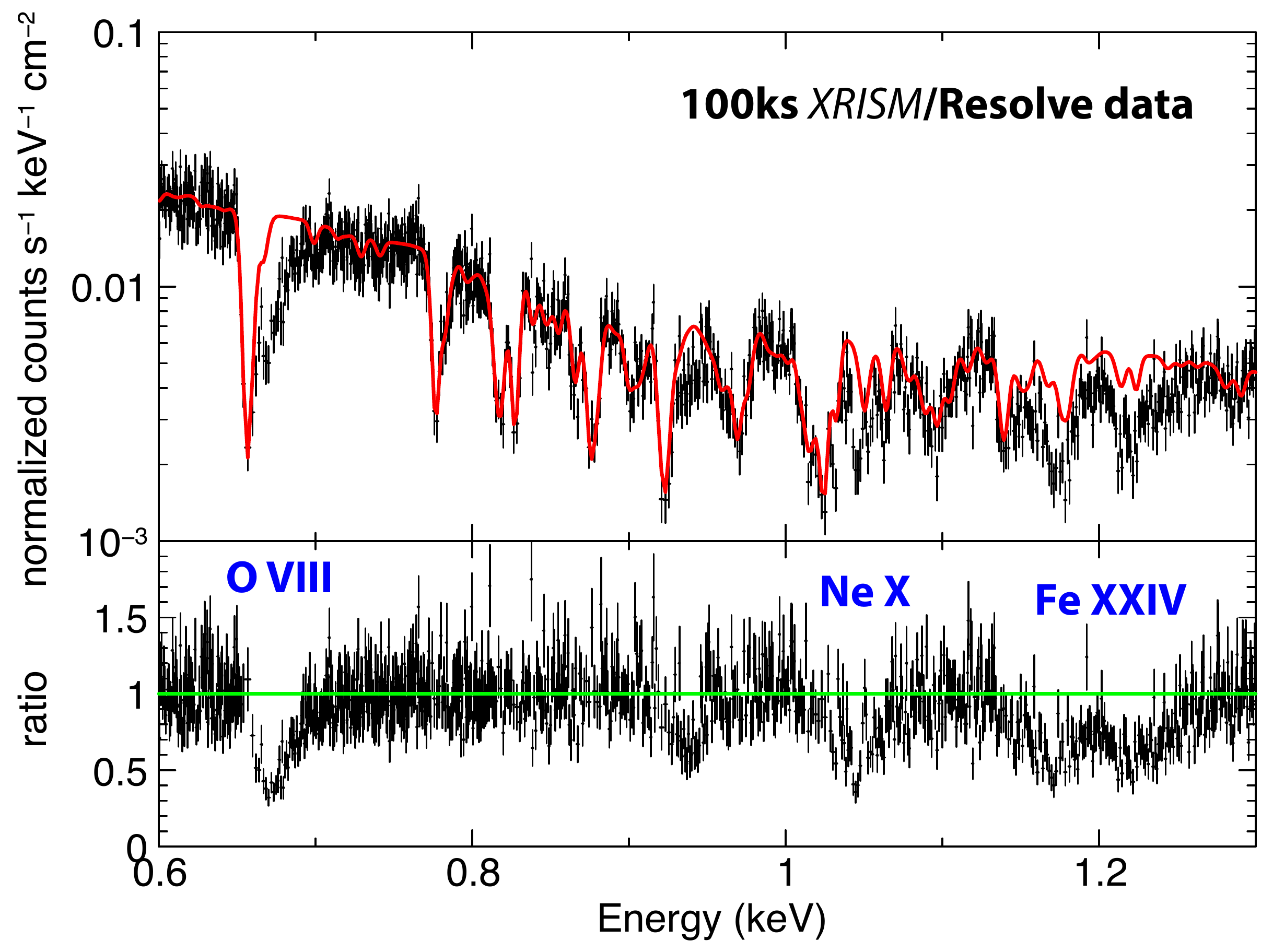}\includegraphics[trim=0in 0in 0in
0in,keepaspectratio=false,width=3.5in,angle=-0,clip=false]{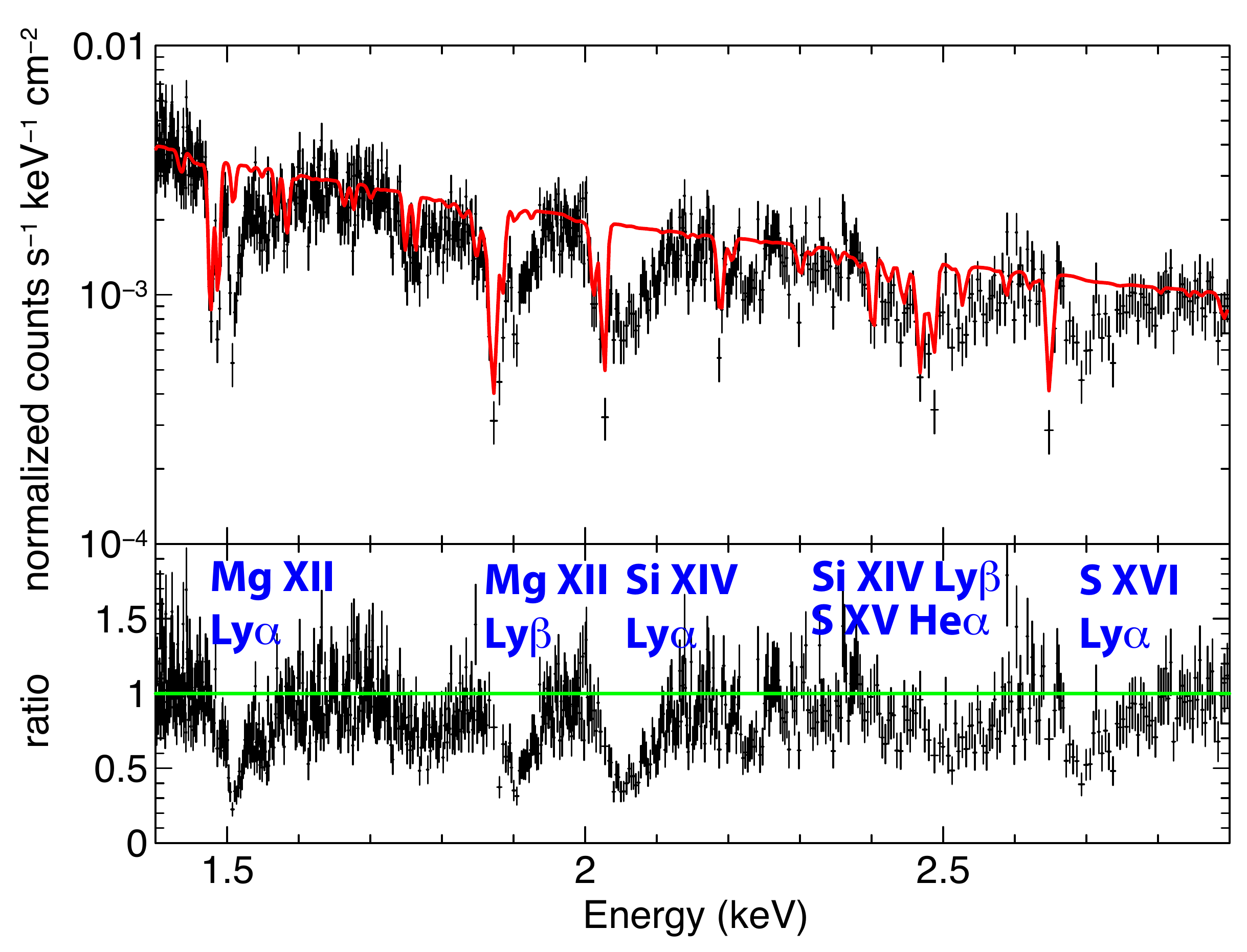} \\
\includegraphics[trim=0in 0in 0in
0in,keepaspectratio=false,width=3.5in,angle=-0,clip=false]{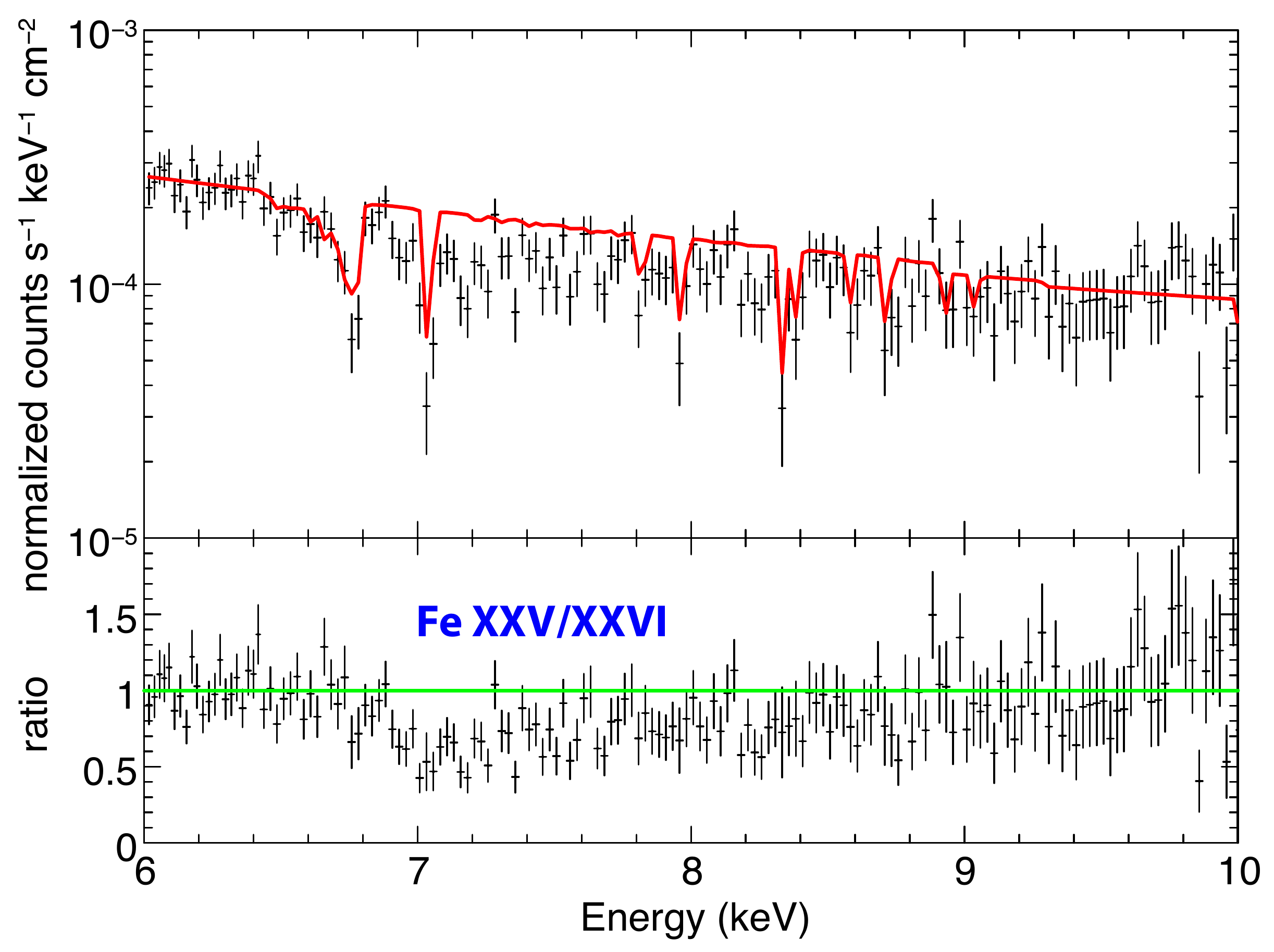}
\end{center}
\caption{Zoom-in on simulated 100 ks {\it XRISM}/Resolve spectrum for $\Gamma=2, \alpha_{\rm OX}=1.5, \lambda_{\rm ion}=0.08, p=1.3, f_D=0.1$ and $\theta=35\deg$ showing the residual (ratio) from the best-fit composite model ({\tt composite} in red) in the absence of MHD UFO component ({\tt mhdwind}). }
\label{fig:lines1}
\end{figure}

\clearpage

We further look into ion-by-ion UFO  components in a simulated broadband spectrum in order to support our claim that microcalorimeter data can in fact be exploited as a powerful diagnostic tool to realistically identify MHD-driving montages such as unique asymmetric line profiles (see {\bf Fig.~\ref{fig:schematics} left}). In {\bf Figure~\ref{fig:lines1}} we are focused on a series of major absorption lines from a 100 ks {\it XRISM}/Resolve simulation  for $\Gamma=2, \alpha_{\rm OX}=1.5, \lambda_{\rm ion}=0.08, p=1.3, f_D=0.1$ and $\theta=35\deg$ showing the residual (ratio) from the best-fit composite model ({\tt composite} in red) in the absence of MHD UFO component ({\tt mhdwind}). As demonstrated earlier, a clear UFO signature of asymmetric (broad) line is undoubtedly seen for this multi-ion MHD UFO whose shape is not necessarily matched with the phenomenological Gaussian or Voigt functions. While the exact shape and strength (i.e. line optical depth and equivalent width) of each line depends sensitively on wind parameters, the characteristic UFO features are systematically imprinted in the broadband.

\begin{figure}[t]% ------------------------------------- Figure~7
\begin{center}
\includegraphics[trim=0in 0in 0in
0in,keepaspectratio=false,width=3.5in,angle=-0,clip=false]{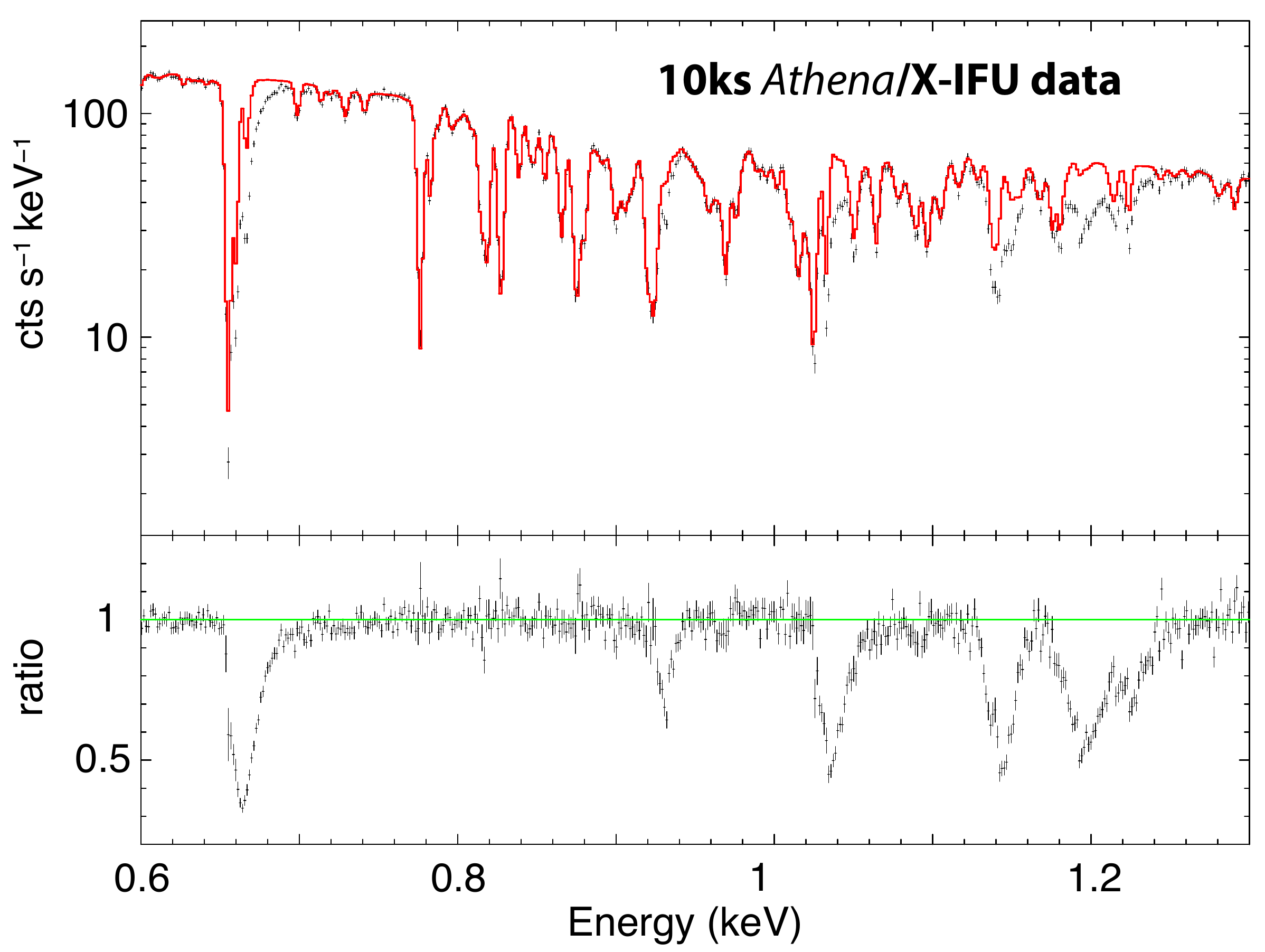}\includegraphics[trim=0in 0in 0in
0in,keepaspectratio=false,width=3.5in,angle=-0,clip=false]{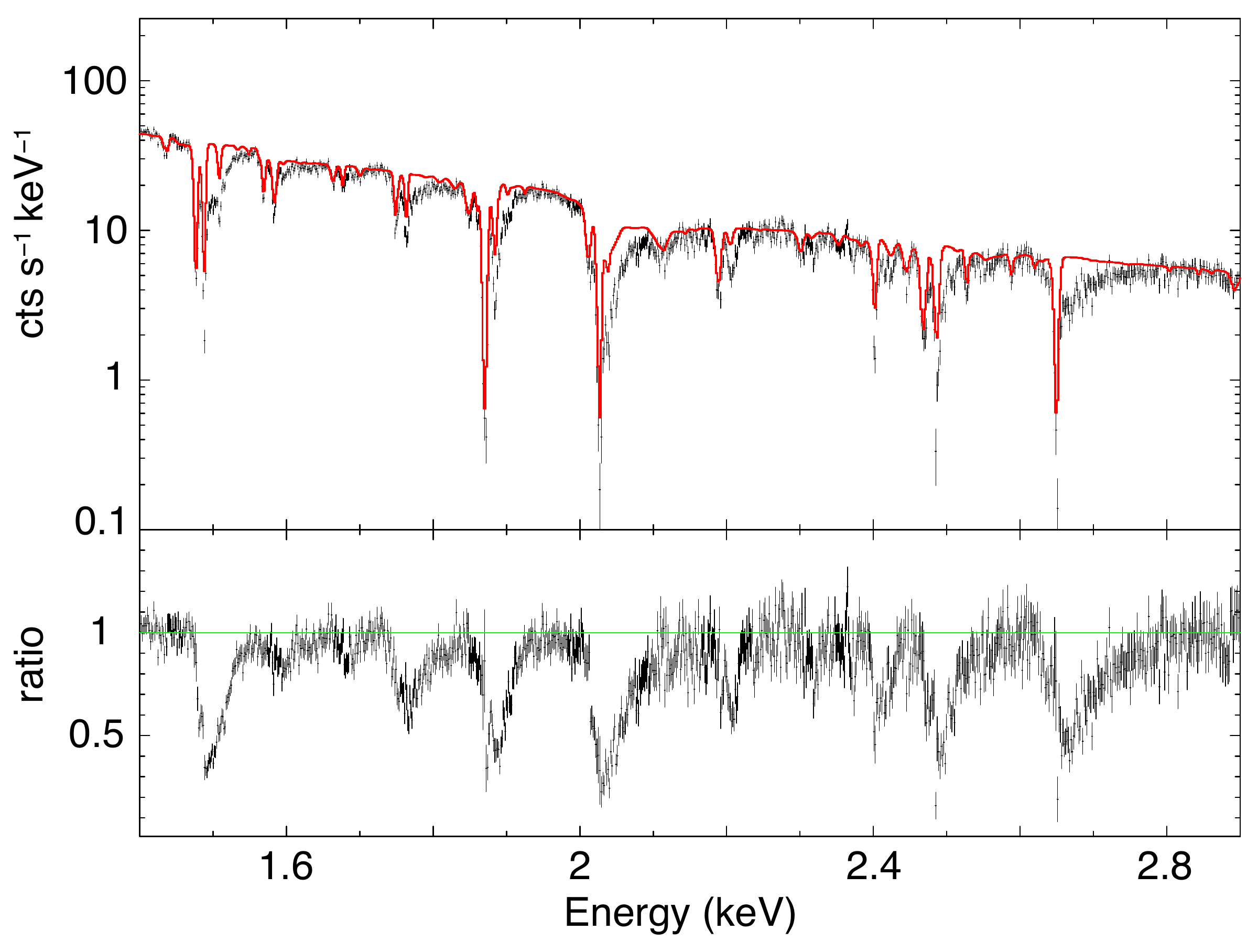} \\
\includegraphics[trim=0in 0in 0in
0in,keepaspectratio=false,width=3.5in,angle=-0,clip=false]{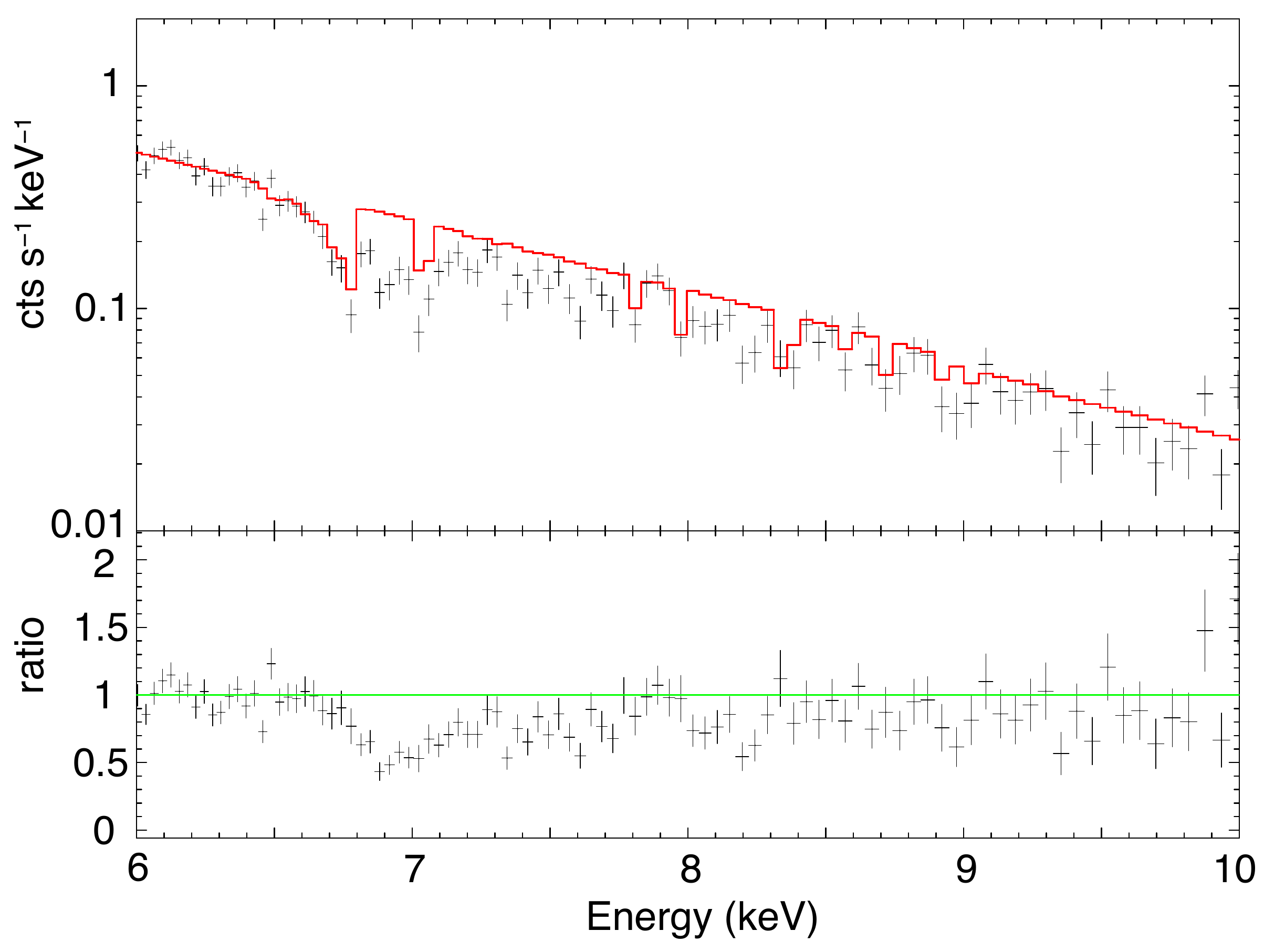}
\end{center}
\caption{Zoom-in on simulated 10 ks {\it Athena}/X-IFU spectrum for $\Gamma=2, \alpha_{\rm OX}=1.5, \lambda_{\rm ion}=0.1, p=1.2, f_D=0.1$ and $\theta=35\deg$ showing the residual (ratio) from the best-fit composite model ({\tt composite} in red) in the absence of MHD UFO component ({\tt mhdwind}). }
\label{fig:lines2}
\end{figure}

In anticipating for the next generation X-ray microcalorimeter instrument beyond {\it XRISM} mission, we also explore an observational viability of our MHD UFO model by performing a number of simulations with {\it Athena}/X-IFU instrument.  Assuming an almost identical set of fiducial model parameters as considered for {\it XRISM} simulations, broadband UFO spectra of multi-ion are simulated with 10 ks {\it Athena}/X-IFU exposure assuming $\Gamma=2, \alpha_{\rm OX}=1.5, \lambda_{\rm ion}=0.1, p=1.2$ and $\theta=35\deg$ in {\bf Figure~\ref{fig:lines2}} in comparison with the earlier {\it XRISM}/Resolve simulations in {\bf Figure~\ref{fig:lines1}}. 
It is obvious that the same simulated UFO features are further refined at more statistically significant level even with 10 ks exposure, owing to {\it Athena}'s large collecting area. In this simplified comparison, it is thus implied in principle that more than ten time-resolved observations could be made with {\it Athena} to obtain data of the same spectroscopic quality expected from a single 100 ks {\it XRISM} observation in an attempt to ambitiously conduct an outflow reverberation mapping for (highly) variable nature of the observed UFOs \citep[e.g.][]{Matzeu16,Parker18}.  

\clearpage

\clearpage

\begin{figure}[t]% ------------------------------------- Figure~8
\begin{center}
\includegraphics[trim=0in 0in 0in
0in,keepaspectratio=false,width=3.35in,angle=-0,clip=false]{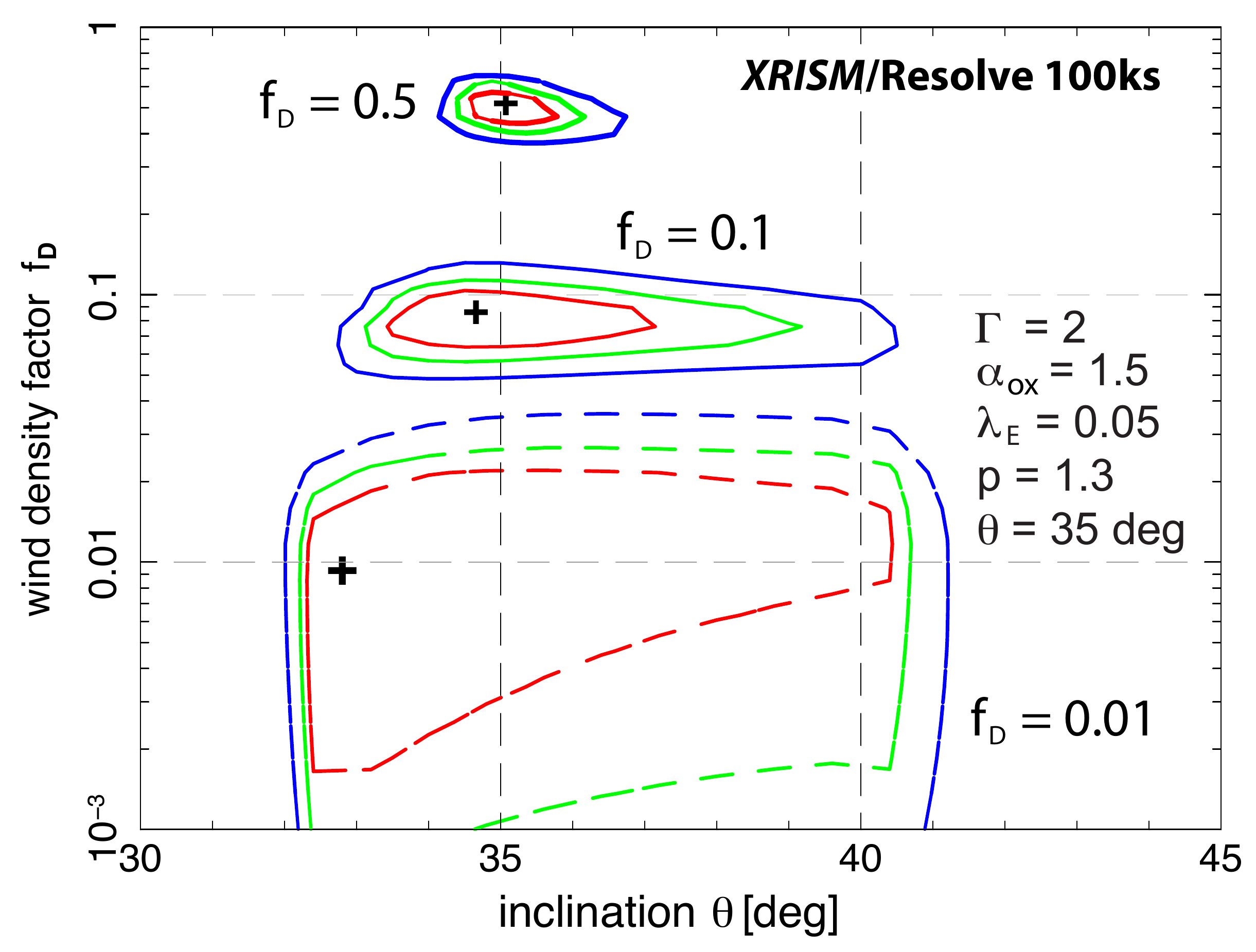}\includegraphics[trim=0in 0in 0in
0in,keepaspectratio=false,width=3.3in,angle=-0,clip=false]{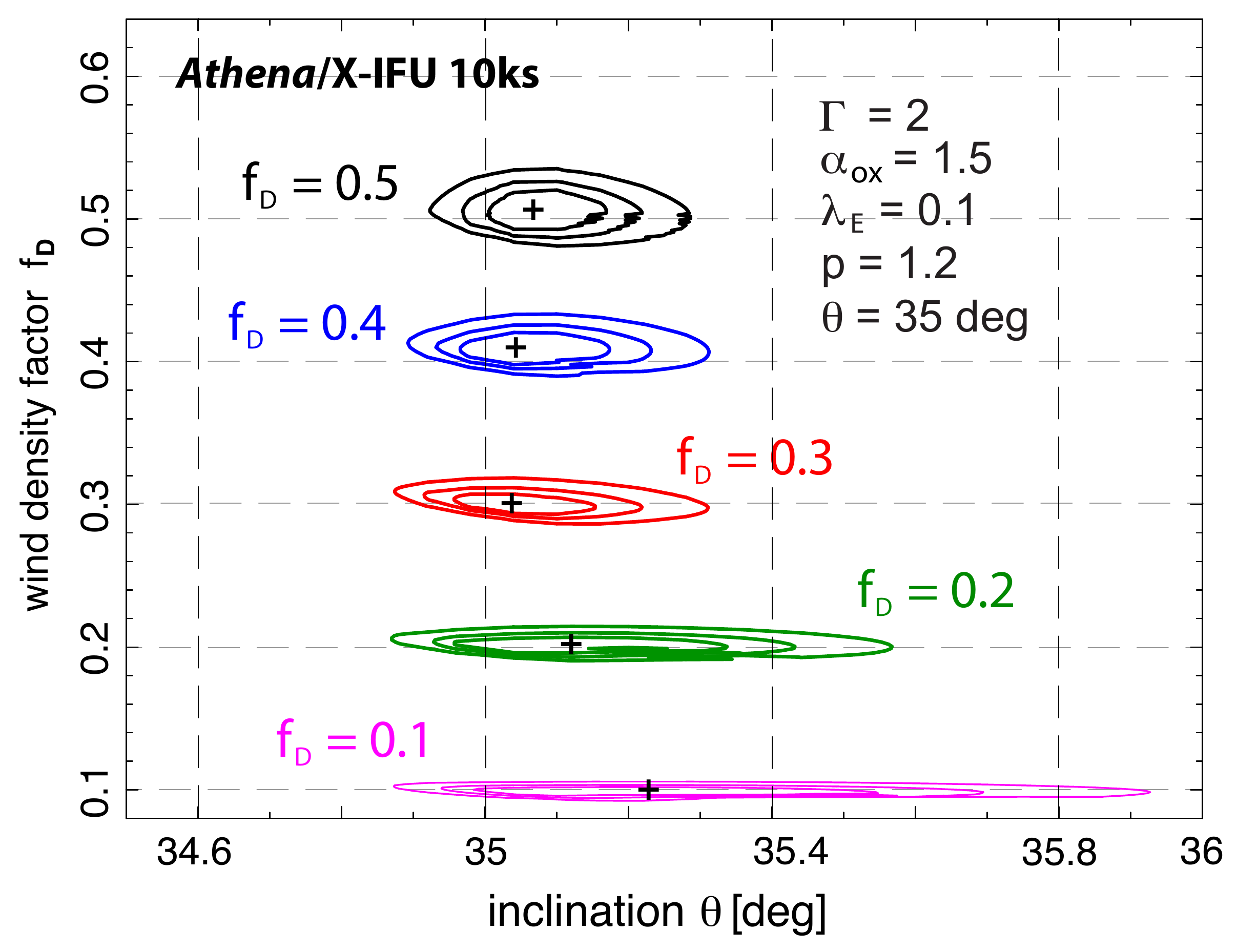}
\end{center}
\caption{Confidence contours of $(\theta, f_D)$ for different values of wind density factor $f_D$ derived from the best-fit models with 
%$\Gamma=2,  \alpha_{\rm OX}=1.5, p=1.3, 
$\theta=35\deg$ showing 68\% (red), 90\% (green) and 99\% (blue) levels. Left: {\it XRISM}/Resolve 100 ks exposure. Right: {\it Athena}/X-IFU 10 ks exposure. }
\label{fig:contour}
\end{figure}

Finally, we compare the statistical confidence in constraining, for example, the inclination $\theta$, for different wind density factor $f_D$ in {\bf Figure~\ref{fig:contour} (left)}  where their best-fit values are shown for low ($f_D=0.01$), intermediate ($f_D=0.1$) and high ($f_D=0.5$) density values assuming $p=1.3, \theta=35\deg$ and $\lambda_{\rm ion}=0.05$ for the simulated 100ks {\it XRISM}/Resolve data. The expected uncertainties in both quantities are relatively small enabling one to robustly retain the correct UFO parameters with 100 ks {\it XRISM}/Resolve exposure unless the wind density (i.e. $f_D$) is too low.  
This is because the predicted absorption lines would become very narrow and weak due to the inner part of the wind being almost fully ionized (where the wind is near-relativistic) if $f_D$ is very low. Again, a mere 10 ks exposure with {\it Athena}/X-IFU is found to drastically improve such a limitation to constraining the UFO parameters even for very weak UFO features (i.e. low $f_D$) as shown in {\bf Figure~\ref{fig:contour} (right)}, demonstrating a high efficacy in accurately recovering the correct values of $f_D$ as well as inclination $\theta$, as an example. 

\section{Summary \& Discussion}

In this work, we conduct a systematic spectral modeling of the multi-ion X-ray UFO features (i.e. $\sim 0.5-10$ keV range) relevant for radio-quiet Seyfert AGNs in the context of MHD-driven accretion disk winds. Considering some generic features of MHD-driven UFOs, our primary focus is to investigate whether or not there exist some unique spectral signatures mainly attributed to MHD driving in predicted UFO absorption lines that can  be seen in observations. By exploiting a set of {\it six} parameters in our MHD wind model, namely, X-ray photon index ($\Gamma$), X-ray to O/UV strength ($\alpha_{\rm OX}$), ionizing luminosity ratio ($\lambda_{\rm ion}$), inclination angle ($\theta$), wind density factor ($f_D$) and wind density slope ($p$), we have shown {\it on average} that many of the synthetic UFO spectra commonly exhibit an asymmetric line profile of an extended  blue tail toward higher energy manifesting the underlying MHD wind kinematics through photoionization. 
As demonstrated by the AMD in {\bf Figure~\ref{fig:amd}}, our MHD-wind model allows for simultaneous multi-ion UFOs (i.e. ranging from low to high ionization absorbers such as \oviii\ and \fexxvi) in the broadband X-ray spectrum.   
All the calculations assume the solar abundances throughout this work {\bf (i.e. $A_{\rm ion}=A_{\rm ion, \odot}$)}. 

We further perform a series of microcalorimeter simulations based on the synthetic UFO spectra in the broadband ({\tt mhdwind}) in anticipation of the future X-ray missions with {\it XRISM}/Resolve and  {\it Athena}/X-IFU instruments to better examine an observational viability of the model. Assuming a set of fiducial parameters, we find that the unique asymmetric signature of blueshift in multi-ion UFO lines can be well captured in many simulated spectra. To realistically test its plausibility of detection, additional absorption components are independently added to our simulations in the form of 3-zone warm absorbers composed of cold, warm and hot plasma ({\tt warmabs}) as well as a partially covering ionized gas ({\tt zxipcf}). Despite the presence of such a spectral contamination, it is demonstrated that the predicted UFO signatures in {\tt mhdwind} are unambiguously retained in the simulated data. If indeed imprinted in high fidelity microcalorimeter spectra, it is conceivable that the predicted UFO lines of unique features can be potentially used as a diagnostic proxy to differentiate different launching mechanisms proposed today.     

In the present formalism, a number of complications are not taken into account for simplicity. First, we treat the primary parameters ($\theta, p, f_D, \Gamma, \alpha_{\rm OX}, \lambda_{\rm ion}$) being independent of each other (see {\bf Fig.~\ref{fig:gamma2}} and other dependences in Appendices). In reality, it has been reported, for instance, that some observable AGN measurements appear to be (tightly) correlated in various samples of AGN surveys; e.g. $\Gamma$, $\alpha_{\rm OX}$, Eddington ratio and monochromatic UV luminosities \citep[e.g.][]{Steffen06,Shemmer08, Luo14,Laurenti21}. Although these  correlations are preferentially suggested in highly-accreting quasars, this may well be relevant for Seyfert 1s accreting at low Eddington rate. If this is the case, our simulation results could be less relevant in certain parameter space that we have probed in this work; e.g. X-ray bright AGNs at high Eddington ratio (i.e. higher $\lambda_{\rm ion}$ with smaller $|\alpha_{\rm OX}|$ and/or higher $\lambda_{\rm ion}$ with softer $\Gamma$). This would need to be considered for quantitative simulations specific to individual AGN/quasar UFOs.    

Additional complications associated to spectral features include SE and reflection components often seen in many Seyfert AGNs. We phenomenologically add SE component in our fiducial AGN SED to calculate its effect and find 
%that SE in principle plays 
a significant role in changing ionization structure of the wind through radiative transfer such that almost all major ions of various charge state tend to be produced at smaller distances (towards the BH) in higher ionization state where the MHD wind is faster. Stronger SE causes multi-ion UFO lines to be more blueshifted and broader in a way similar to the role of photon index $\Gamma$, both of which involves more soft photons being available for photoionization.      
On the other hand, the presence of reflection features (i.e. emission and absorption/edges) is also considered as another source of spectral contamination against the expected MHD-driven UFO signatures. By implementing {\tt xillver} and {\tt relxill} models, we perform a similar 100ks microcalorimeter simulation for \xrism/Resolve. It is demonstrated that even a relatively pronounced reflection flux (e.g. refl\_frac=2 and $A_{\rm Fe}/A_{\rm Fe, \odot}=5$) does not significantly distort the tell-tale MHD signature of the blueshifted  UFO absorbers.

We emphasize that our choice of defining the Eddington ratio, $\lambda_{\rm ion}$, is simply a convenient way to control the X-ray luminosity $L_{\rm ion}$ responsible for photoionization in this model (see \S 2.2). More subtle factors related to $L_{\rm ion}$ (e.g. radiative efficiency in accretion power, black hole spin, coronal geometry and disk structure) have been ignored to avoid further complications and extra degrees of freedom in this problem. It is also expected that X-ray luminosity should be intimately tied to wind density (i.e. $n_{o}$ in our model) determined by the disk density that essentially signifies accretion rate. In our calculations, a fixed value of the wind density factor $f_D$ is naively decoupled from our Eddington ratio $\lambda_{\rm ion}$. 
As addressed earlier, the mass accretion rate (thus the wind density normalization parameterized by $f_D$) should in reality be physically coupled to the X-ray luminosity and perhaps $\alpha_{\rm OX}$, for example, as suggested in the past studies \citep[e.g.][]{Steffen06,Shemmer08, Luo14,Laurenti21}. 
For this reason, it is not our primary interest to compare the {\it absolute} line properties (i.e. line optical depth and EW), which is subject to the value of $f_D$. Instead, our main focus is the {\it relative} comparison of line shape of multi-ion UFOs from one case to the other so one can learn a systematic trend as a whole rather than a specific quantitative change.  Another prescription of X-ray luminosity such as $L_{\rm ion}  \propto f_D$, for example, would hence be more physically motivated as is often employed in our past spectral analyses \citep[e.g.][]{F17,F18,F21}, which, however, is beyond the scope of the present work.    

%*$\Gamma = 1.99_{-0.0079}^{+0.0082}, \theta=30.1_{-1.1}^{+0.43}, \lambda_{\rm ion}=0.098_{-0.0036}^{+0.012}, \alpha_{\rm OX}=1.5_{\rm peg}^{+0.012}, p=1.29_{-0.009}^{+0.003}$...

Our microcalorimeter simulations clearly show a high efficacy of retaining the expected multi-ion UFO components in the composite broadband spectrum for a given set of model parameters, indicating a persistent immunity of the UFO features against the external contaminations due to other forms of ionized absorbers. This immunity, however, is not always warranted depending on a number of conditions. We simply assume a typical 2-10 keV flux of $\sim 2.3 \times 10^{-11}$ erg~cm$^{-2}$~s$^{-1}$ with certain exposures as a test case for Seyfert 1 AGNs, but disentangling the predicted MHD-driven UFO features (i.e. near-relativistic blueshift of asymmetry to higher energy) could obviously become more challenging and very enigmatic in some cases (e.g. lower flux and shorter exposure). Therefore, given the model limitation due to the fixed value of $f_D$ in our simulations, the absolute values of the model parameters that set the appearance and disappearance of multi-ion UFOs (shown in our calculated spectra) should not be regarded as the generalized values, although a qualitative trend can still be learned from those calculations.

With the photon statistics achievable  with the current spectroscopic instruments, it is not easy  to conduct time-resolved analysis of rapid UFO variability.
%With the achievable photon statistics with the current spectroscopic instruments, it is not practically trivial to conduct time-resolved analysis of rapid UFO variability. 
As demonstrated in  {\bf Figure~\ref{fig:contour}} for a given UFO parameter set, only a 10 ks exposure with {\it Athena}/X-IFU appears to be sufficient to robustly resolve these Seyfert UFOs owing to its large collecting area. This would revolutionize the physics of UFOs since it can enable us to take a series of snapshot spectra, each of which on a much shorter timescale that is currently possible. By being able to study a temporal response of individual UFO absorbers ranging from soft X-ray band to Fe K lines, spectroscopic information can be further cast into timing property of various absorbing ions, from which one could potentially make a geometrical (spatial) diagnosis of multiple ions in the UFOs as the continuum varies in time. This is another reason why multi-ion UFO modeling, and not just Fe K absorbers, is important for understanding a global picture of the UFOs. Such a reverberation mapping of multi-ion UFOs has  been partially hinted at with the use of principal component analysis (PCA) of {\it XMM-Newton}/EPIC data in order to study in detail the correlations between the continuum source  and the multiple ions in the observed UFOs \citep[e.g.][]{Parker18}.

Concerning the expected asymmetric line profiles unique to MHD-driving process, there have been some observational hints of the expected asymmetric Fe K UFO features in \nustar\ and \xmm\ data (e.g. a nearby Seyfert 2 AGN, MCG-03-58-007 in \citealt{Braito22}; a lensed broad absorption line quasar, APM~08279+5255 in \citealt[][]{Chartas09}; a complex Fe K trough feature seen in narrow-line Seyfert, 1H~0707-495 in \citealt{Hagino17}, among others). 
Also, it might be relevant for other types of disk-winds such as those of BH XRBs occasionally detected, for example, with \chandra, in Galactic transients \citep[e.g.][]{Miller15,Chakraborty21} as well as AGN warm absorbers \citep[e.g.][]{Kaspi00,Kallman09,F17}. The observed velocities of the ionized outflows in these systems are much lower than those in UFOs; i.e. a few hundred to at most a thousand km~s$^{-1}$. The expected blueshift is thus only up to $v/c \lsim 0.01$ meaning that the line width is at most $\Delta E/E \lsim 0.001$. This implies that the characteristic asymmetry would be observationally more intangible even if indeed magnetically launched in those systems. While it might still be plausible to differentiate between MHD-driving (with such an asymmetry of smaller magnitude) and, for example, another promising process such as thermal-driving (C. Done, in private communications),  a robust disentanglement could be perhaps less trivial while not impossible. Again, analyzing multiple ions rather than a single Fe K feature would be a great complement in order to probe a wide range of ionization structure in the wind.  

In this work, we do not incorporate emission lines also expected to originate from the same outflowing gas or separate regions. For example, Monte Carlo approach has been often employed to calculate self-consistent emission and scattering components (e.g. \citealt{Hagino15,Hagino16,Hagino17} with {\tt MONACO} and \citealt{Sim08,Sim10a,Sim10b} also including special relativistic effects), while {\tt XCORT} has utilized {\tt XSTAR} runs \citep[][]{SchurchDone07}. More recently,  \citet{Luminari18} has proposed {\tt WINE} model with a special focus on  detailed special relativistic dimming for UFOs \citep[see also][]{Luminari20,Luminari21}. Such emission process might be important to accurately assess the physical properties of ionized absorbers because they can ``fill in" the intrinsic absorption features. For example, some very luminous quasars accreting at near-Eddington rate such as PDS~456 is found to show a broad P-Cygni profile in Fe K band associated with its prototype UFO  where the wind may well be Compton-thick \citep[][]{Nardini15}. On the other hand, one of the well-studied Seyfert 1 AGN showing canonical X-ray warm absorbers, NGC~3783, is also known to exhibit a handful emission lines in UV/soft X-ray band mostly from the He-like triplets (resonance, intercombination, and forbidden lines) of \ovii, \neix, and \mgxi\ as well as Ly$\alpha$ lines from the H-like species of these elements with a very small systematic velocity shift ($130\pm290$ km~s$^{-1}$; \citealt[e.g.][]{Kaspi00,Gabel05, Kallman09}). However, given a moderate accretion rate in Seyfert 1 population with typically very weak (if not none) P-Cygni profiles in X-ray spectra, it is reasonable to speculate that X-ray absorbing gas is most likely optically thin. Scattering within the absorbers might then be less efficient. 

While we further investigate in this paper a possible effect of inner disk truncation on UFO spectra (see \S 3.2), a slight change in UFO kinematics and/or wind density could also naturally change UFO line profiles without invoking a geometric change in disk configuration (i.e. truncation). Hence, such a degeneracy may make it physically challenging to uniquely attribute variable nature of UFOs to only disk truncation in observations especially in AGNs where phenomenology of accretion flows still remains very enigmatic \citep[e.g.][]{Antonucci13}. It is also hard to identify ``state transition" of accretion mode due to their much longer time scales. On the other hand, we have similarly considered a potential impact of disk truncation in the context of AGN warm absorbers and BH XRB disk-winds in the past \citep[][]{F21}, and we would like to note, in the context of MHD scenario, that the inner disk truncation in those slow/moderate disk winds would not significantly change the absorption features in contrast with UFOs. This is because the ionized ions responsible for non-UFOs are produced in a part of the wind that is predominantly launched from much larger disk radii from the BH, thus the inner truncation would play little role in reshaping the absorption lines. 
The inner disk truncation therefore results in a distinct consequence in absorption spectra between UFOs and non-UFOs.

%We find it a fundamentally distinct response of winds to inner disk geometry.        

Considering the elusive origin of the UFO and its launching mechanism, this work is originally motivated by the notion that a detailed UFO spectral feature may potentially shed light on the underlying physics of UFOs. As one of the promising scenarios, we exploit the MHD-driving process by calculating synthetic UFO spectra of multiple ions in the broadband X-ray. As illustrated in {\bf Figure~\ref{fig:schematics} (left)}, the predicted UFO line shape is generally found to be uniquely asymmetric with an extended blue tail attributed to the generic kinematics of MHD disk-winds; i.e. the closer in to the BH, the faster the wind. While this is indeed a generic spectral signature in most of the calculations made so far, such a claim should be understood with care. That is, it is conceivable that the degree of such characteristic asymmetric signature may be made very small to the extent that the expected asymmetry could become virtually unnoticeable in observations depending on the wind condition and or AGN properties. This point is  intimately related to AMD and velocity distribution addressed in \S 3.1 (see {\bf Figure~\ref{fig:amd}}). On the other hand, another leading scenario, line-driving process tends to imprint a different tell-tale sign in the UFO line shape that is generally reversed\footnote[10]{Note, however, that even some line-driven absorption spectra may look similar to those from MHD-driven outflows under certain wind kinematics \citep[e.g.][]{GiustiniProga12}.} to MHD-driven UFO spectra because of its different underlying wind kinematics; i.e. the gas accelerates along a LoS by radiation pressure until it  asymptotically reaches a finite terminal  velocity (thus coasting) at large distance \citep[e.g.][]{SV93,KWD95,LongKniggle02,Sim08,Hagino15}. Such a distinct velocity field of line-driven UFOs is generally expected to produce a broad asymmetric UFO line spectrum with an extended red tail being fundamentally distinct from that by MHD-winds \citep[e.g.][also Reeves J. N. in private communication]{LongKniggle02,Sim05,Sim10a,Sim10b,Hagino15,Mizumoto21}. Nonetheless, it is possible again that the apparent difference between two models could be made either too small or too subtle in some specific cases to observationally enable us to robustly differentiate. Hence, we emphasize that our claim for obtaining the unique spectral features due to MHD-driving should be viewed as an {\it average} feature. With this caution in mind, we believe that our modeling and simulations in this work can help improve our fundamental understanding of the physics of X-ray UFOs that could otherwise be  observationally inaccessible.

\acknowledgments
%We are grateful to James Reeves for providing us  with the {\it
%XMM-Newton/NuSTAR} spectrum and KF thanks to Alex Sanner for his insightful help on
%wind visualization calculations.
%
K.F. is grateful to James Reeves, Daniel Proga, Maria Diaz Trigo and Chris Done, for a number of inspiring comments and questions that have partly motivated this project. 
We appreciate a number of constructive comments and questions by an anonymous referee to improve the manuscript. 
This work is supported in part by NASA grants, NNH18ZDA001N-ADAP and NNH20ZDA001N-ADAP, through Astrophysics Data Analysis Program. 
M.D. and G.A.M. acknowledges the financial support from Attivit\`{a} di Studio per la comunit\`{a} scientifica di Astrofisica delle Alte Energie e
Fisica Astroparticellare: Accordo Attuativo ASI-INAF n. 2017-14-H.0.

\section*{Appendix A: Model Dependences of Synthetic Multi-Ion Broadband UFO Spectra}

\begin{figure}[t]% ------------------------------------- Figure~9
\begin{center}
\includegraphics[trim=0in 0in 0in
0in,keepaspectratio=false,width=3.2in,angle=-0,clip=false]{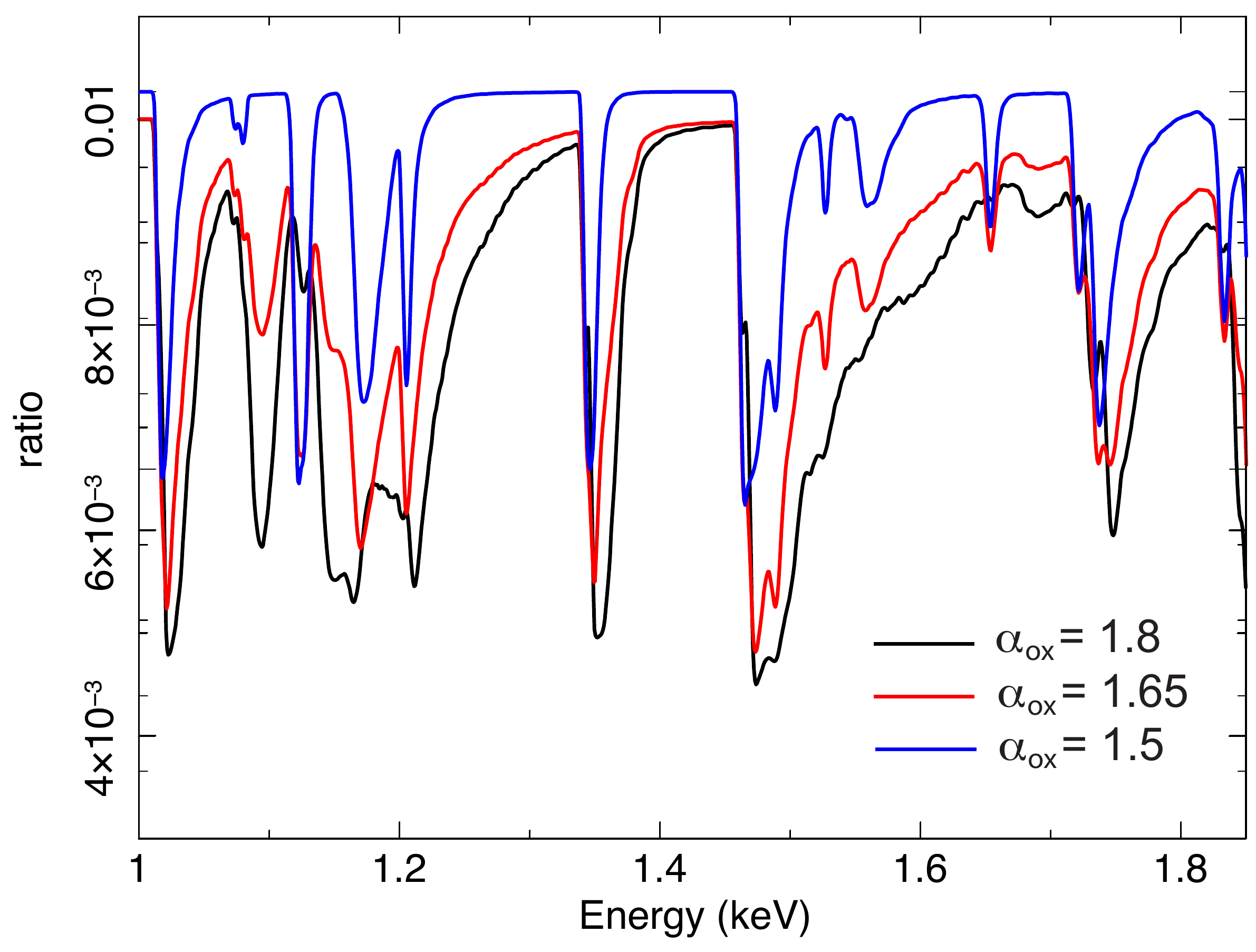} \includegraphics[trim=0in 0in 0in
0in,keepaspectratio=false,width=3.2in,angle=-0,clip=false]{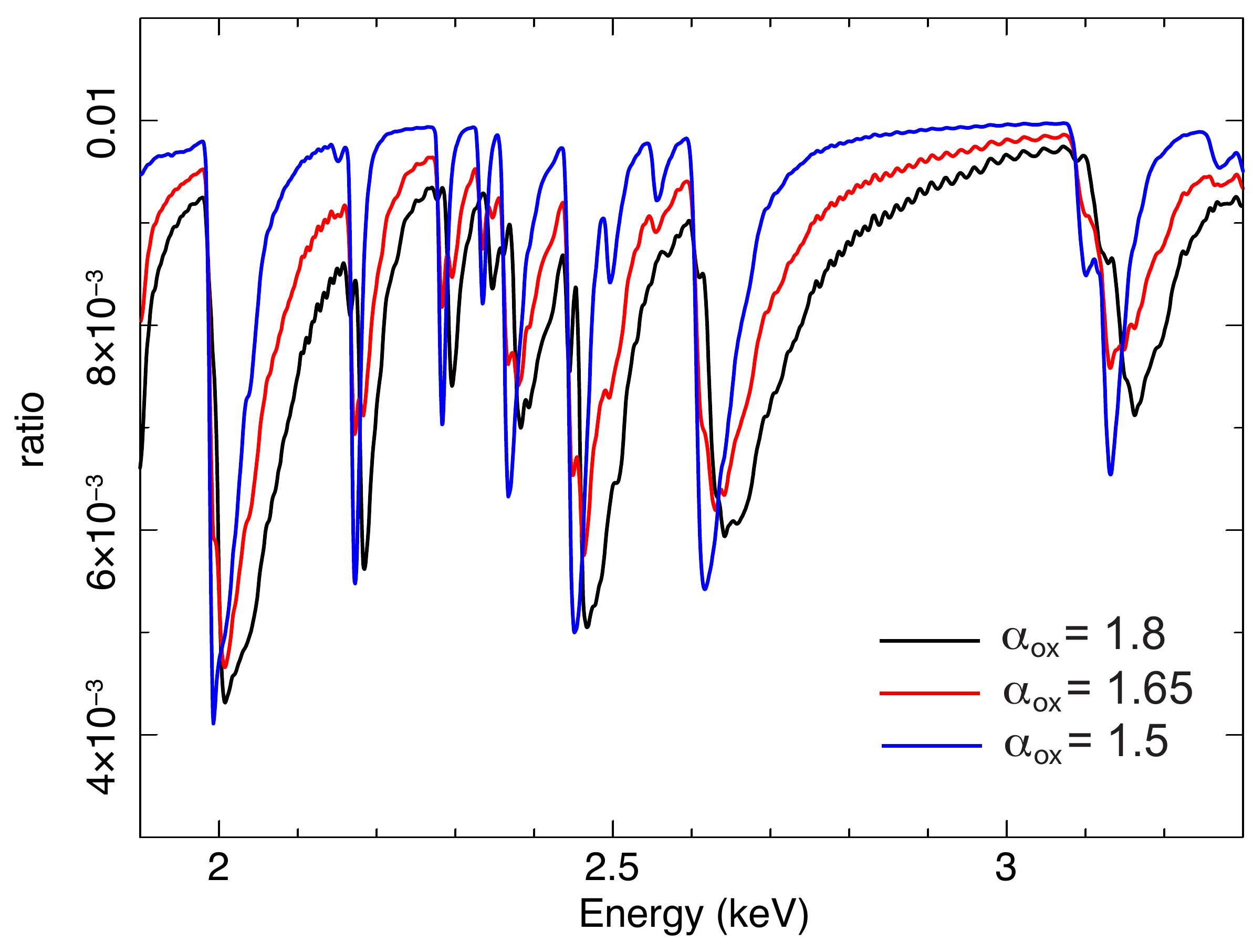} \includegraphics[trim=0in 0in 0in
0in,keepaspectratio=false,width=3.2in,angle=-0,clip=false]{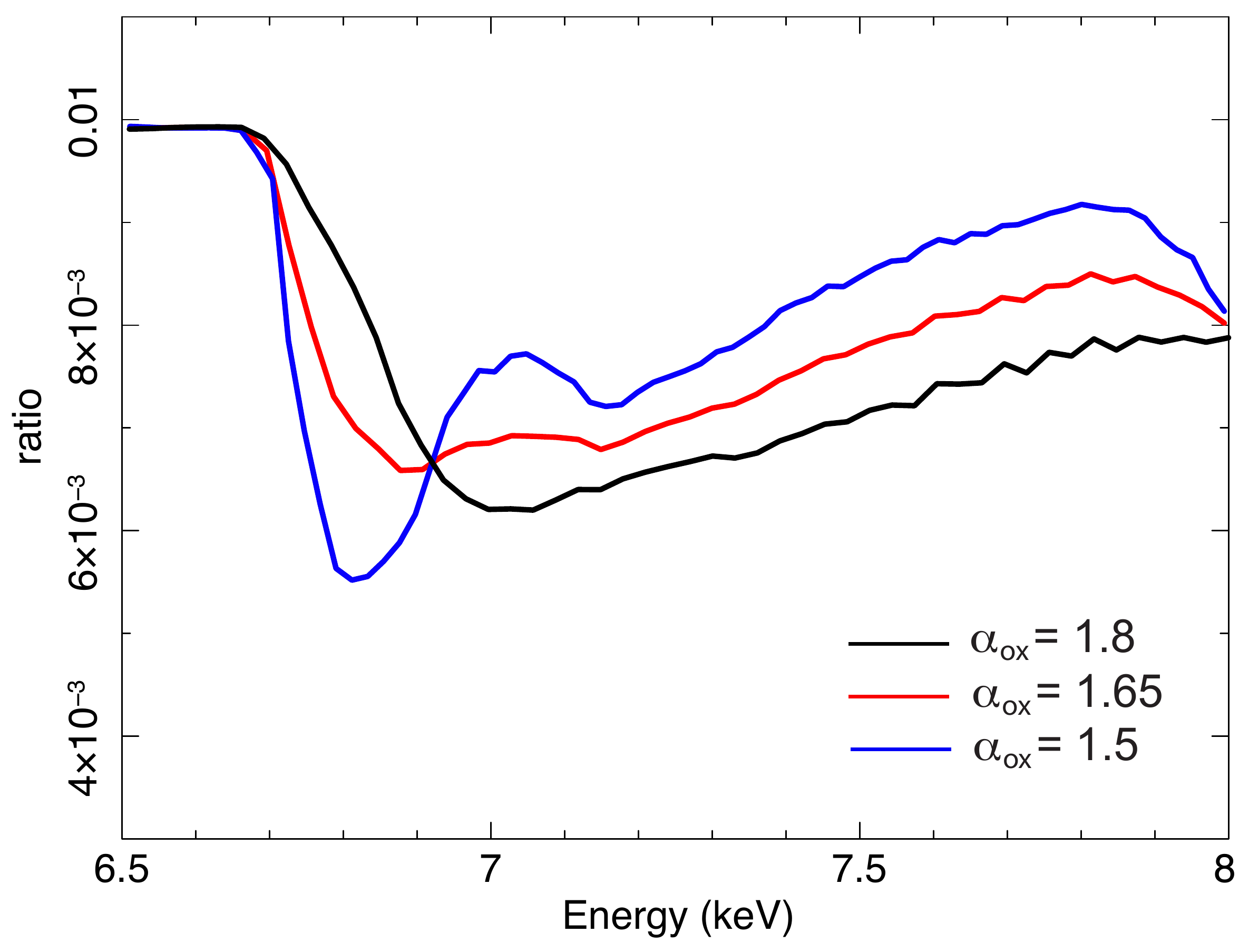}
%\includegraphics[trim=0in 0in 0in
%0in,keepaspectratio=false,width=3.2in,angle=-0,clip=false]{spec_gamma.pdf}\includegraphics[trim=0in 0in 0in
%0in,keepaspectratio=false,width=3.2in,angle=-0,clip=false]{spec_theta.pdf}
\end{center}
\caption{Theoretical multi-ion absorption spectra (ratio to the continuum) for different X-ray strength $\alpha_{\rm OX}$ with $\theta=30\deg, \Gamma=2, p=1.3, f_D=1$ and $\lambda_{\rm ion}=0.1$. }
\label{fig:alphaox}
\end{figure}

As part of investigating the model parameter dependences, we explore the effect of X-ray hardness via $\Gamma$ on the UFO line profiles in \S 3.1. In the Appendix A, the effects of the other model parameters (see Table 1) are  considered and demonstrated. 
We examine the dependence of X-ray strength relative to optical/UV flux parameterized by $\alpha_{\rm OX}$ in {\bf Figure~\ref{fig:alphaox}} with $\theta=30\deg, \Gamma=2, p=1.3, f_D=1$ and $\lambda_{\rm ion}=0.1$. Although the selected range of $\alpha_{\rm OX}$ in this work is limited, we clearly see its strong effect on the UFO spectra. By suppressing more X-ray photons relative to UV photons (i.e. larger $|\alpha_{\rm OX}|$ value), for example, the resulting velocity curve $v(\xi)$ tends to shift towards lower $\xi$ domain with AMD (see {\bf Figure~\ref{fig:amd}b}) producing more blueshifted UFO lines with larger width. As similarly seen in {\bf Figure~\ref{fig:gamma}}, soft X-ray UFO features are more sensitive to the change in X-ray strength compared to Fe K UFOs. Trough energies in individual lines are more noticeably blueshifted with increasing $|\alpha_{\rm OX}|$, while there is little change due to $\Gamma$  in {\bf Figure~\ref{fig:gamma}}.  

\begin{figure}[t]% ------------------------------------- Figure~10
\begin{center}
\includegraphics[trim=0in 0in 0in
0in,keepaspectratio=false,width=3.2in,angle=-0,clip=false]{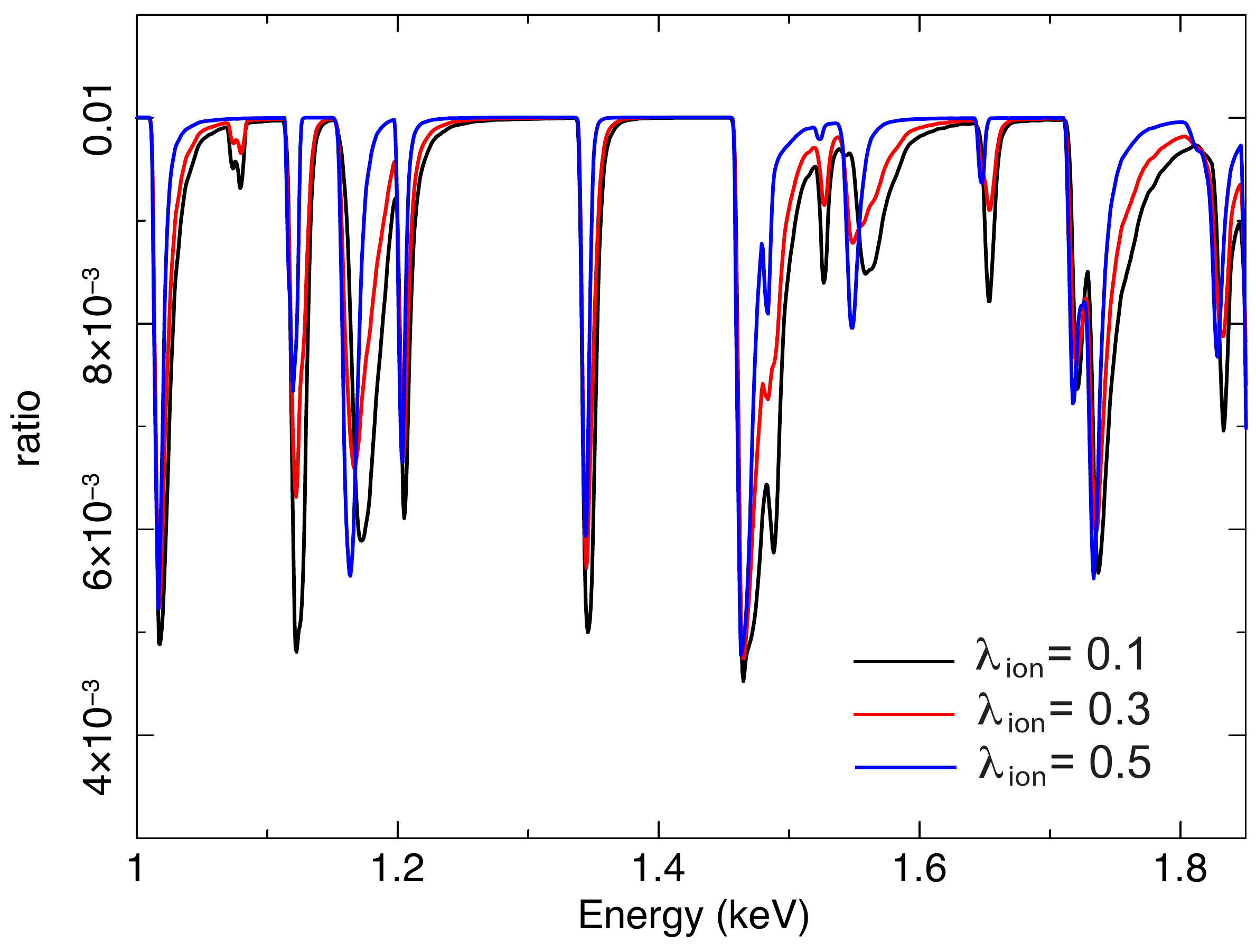} \includegraphics[trim=0in 0in 0in
0in,keepaspectratio=false,width=3.2in,angle=-0,clip=false]{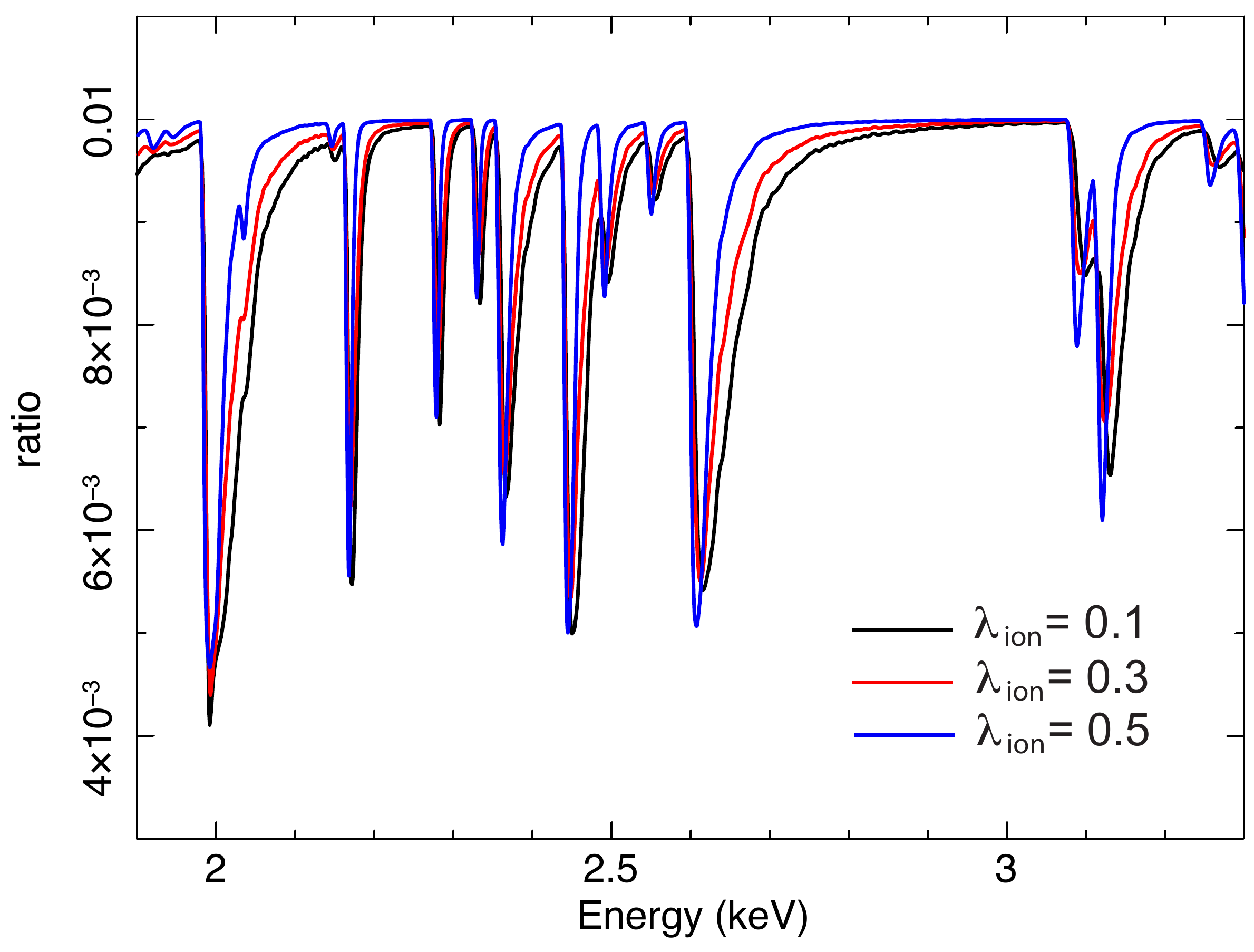} \includegraphics[trim=0in 0in 0in
0in,keepaspectratio=false,width=3.2in,angle=-0,clip=false]{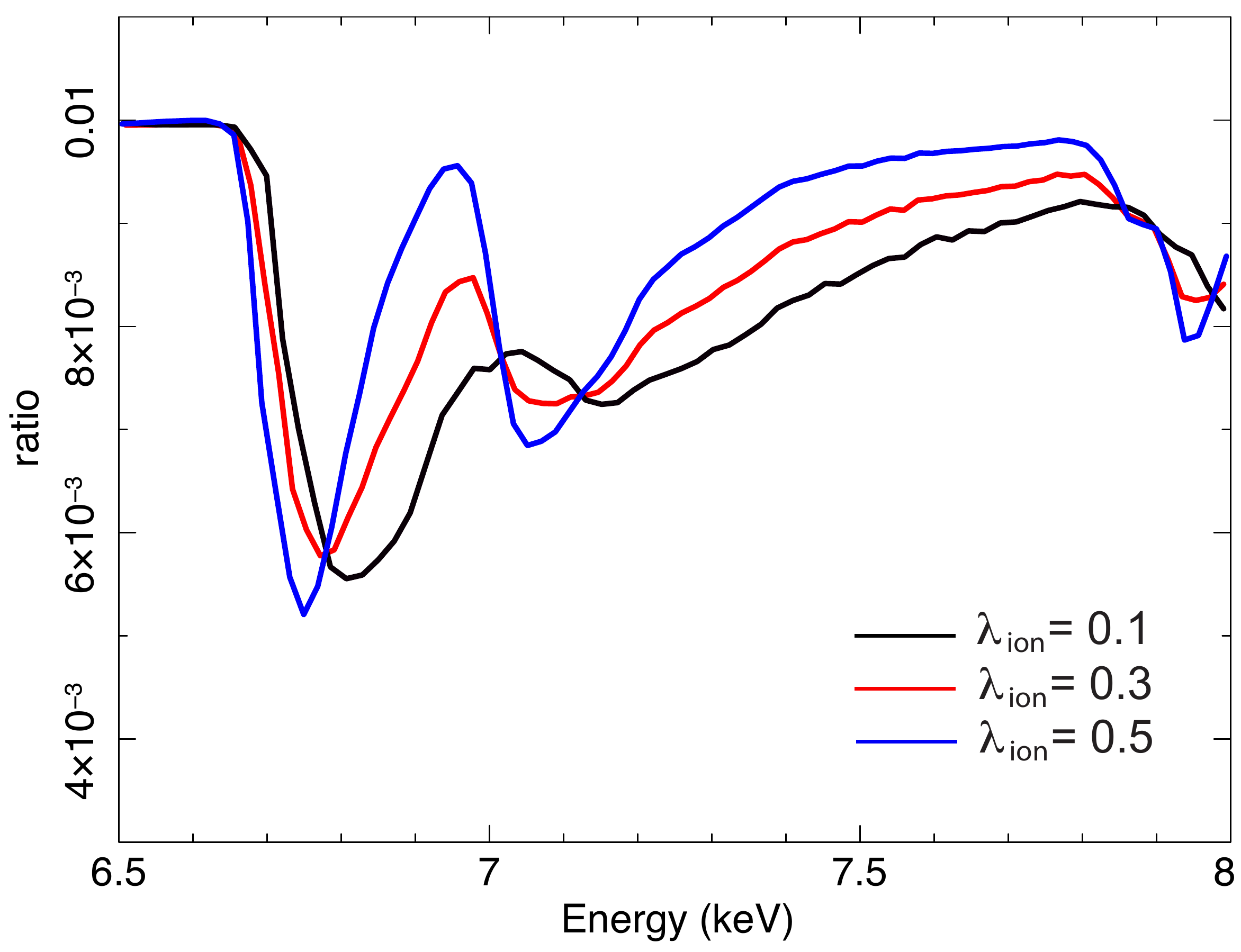}
%\includegraphics[trim=0in 0in 0in
%0in,keepaspectratio=false,width=3.2in,angle=-0,clip=false]{spec_gamma.pdf}\includegraphics[trim=0in 0in 0in
%0in,keepaspectratio=false,width=3.2in,angle=-0,clip=false]{spec_theta.pdf}
\end{center}
\caption{Theoretical multi-ion absorption spectra (ratio to the continuum) for different ionizing luminosity ratio $\lambda_{\rm ion}$ with $\theta=30\deg, \Gamma=2, p=1.3, f_D=1$ and $\alpha_{\rm OX}=1.5$. }
\label{fig:lambda}
\end{figure}

The effect of ionizing luminosity in units of Eddington luminosity $\lambda_{\rm ion}$ is explored in {\bf Figure~\ref{fig:lambda}} with $\theta=30\deg, \Gamma=2, p=1.3, f_D=1$ and $\alpha_{\rm OX}=1.5$. In a simplified situation where only ionizing luminosity is varied while being decoupled from spectral shape (i.e. $\Gamma$ and $\alpha_{\rm OX}$), we find that the UFO signatures are weakened in general with increasing $\lambda_{\rm ion}$ as naturally expected. In comparison, the spectral dependence is less sensitive to $\lambda_{\rm ion}$ than it is to  spectral indices, especially for soft X-ray UFOs. Understanding of this trend, however, needs some caution because the variation of X-ray luminosity in AGNs is often accompanied by the change in spectral shape as well, which is not currently taken into account in these calculations for simplicity. We shall discuss more on this point later in \S 4.  

\clearpage

\begin{figure}[t]% ------------------------------------- Figure~11
\begin{center}
\includegraphics[trim=0in 0in 0in
0in,keepaspectratio=false,width=3.2in,angle=-0,clip=false]{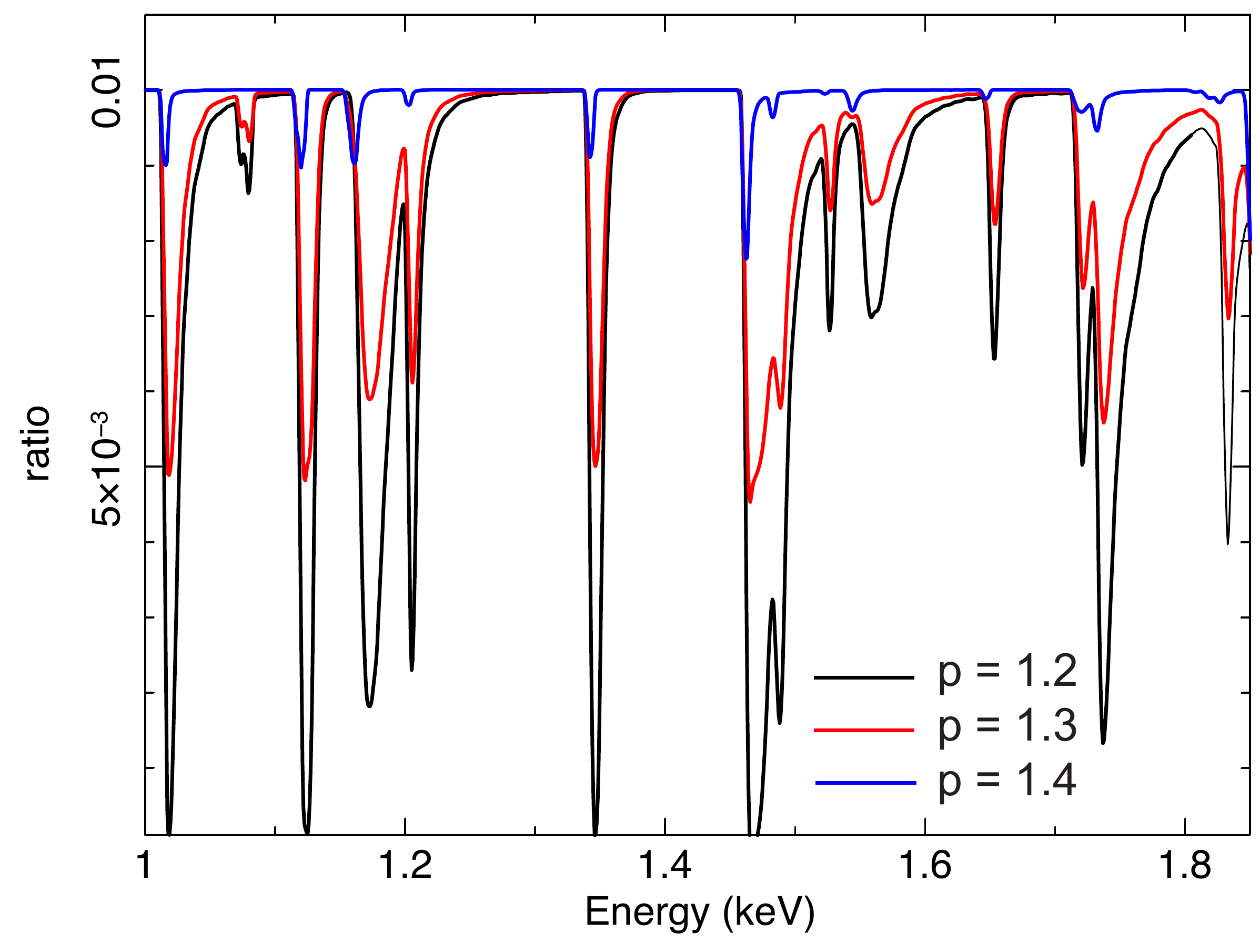} \includegraphics[trim=0in 0in 0in
0in,keepaspectratio=false,width=3.2in,angle=-0,clip=false]{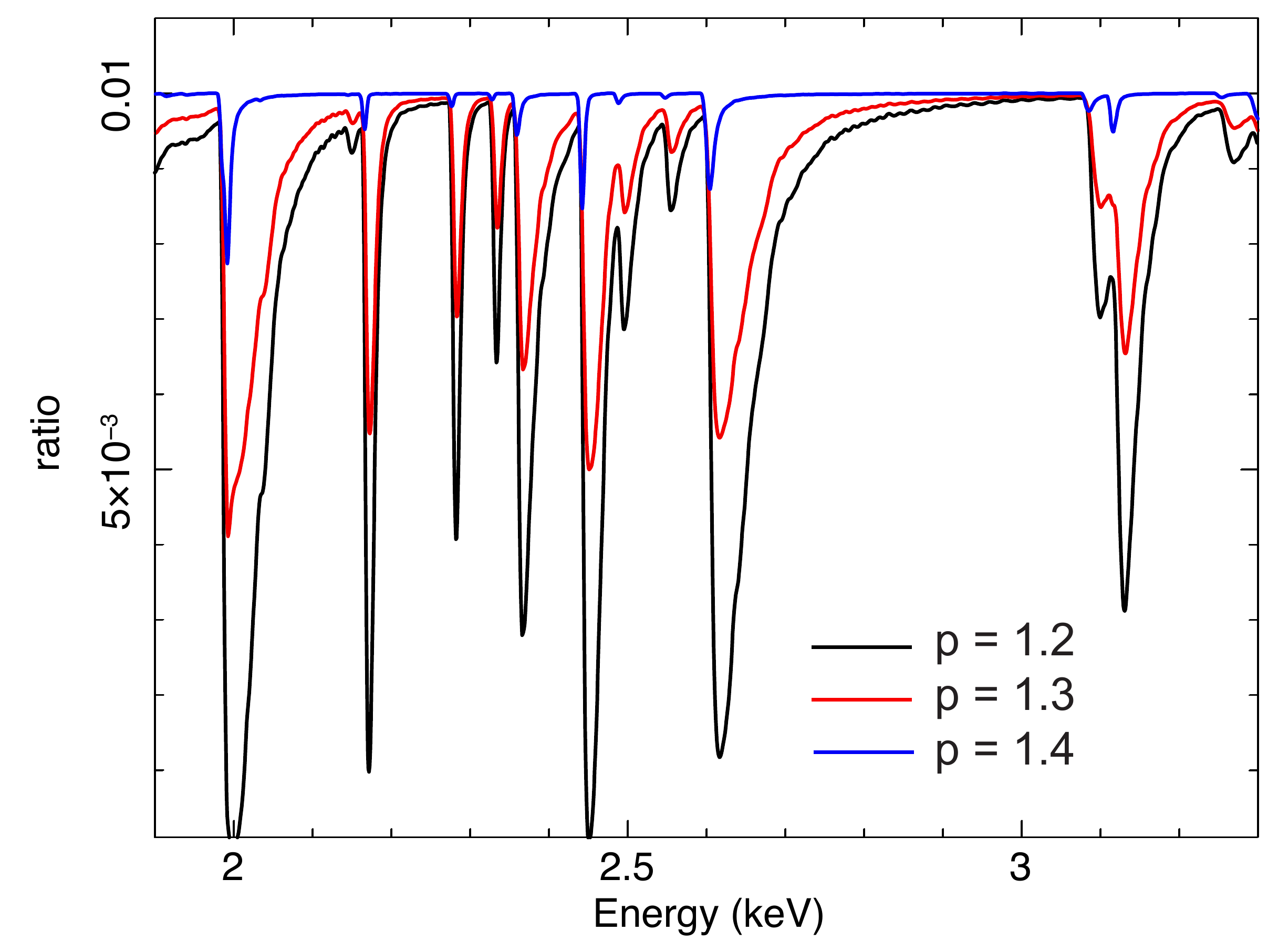} \includegraphics[trim=0in 0in 0in
0in,keepaspectratio=false,width=3.2in,angle=-0,clip=false]{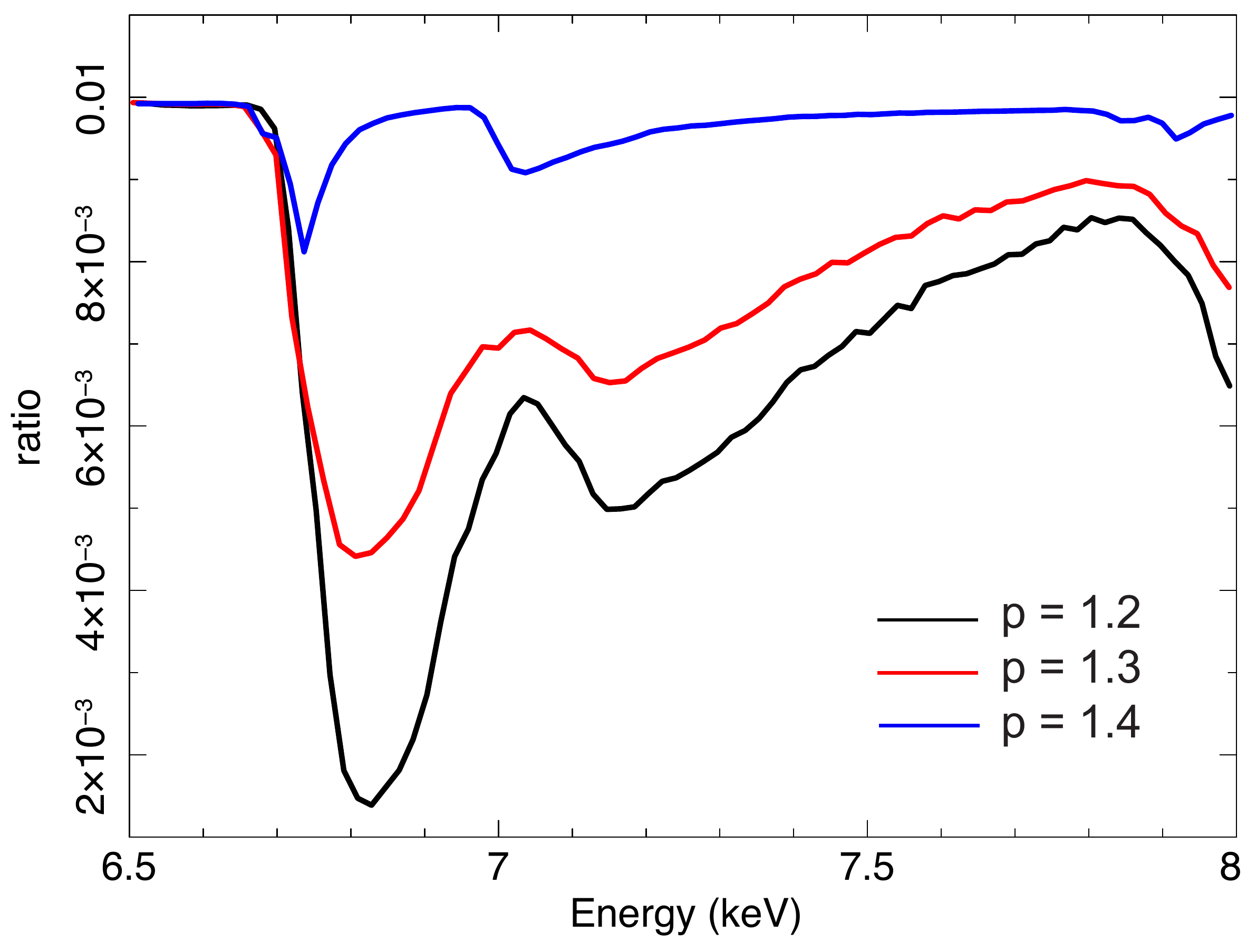}
%\includegraphics[trim=0in 0in 0in
%0in,keepaspectratio=false,width=3.2in,angle=-0,clip=false]{spec_gamma.pdf}\includegraphics[trim=0in 0in 0in
%0in,keepaspectratio=false,width=3.2in,angle=-0,clip=false]{spec_theta.pdf}
\end{center}
\caption{Theoretical multi-ion absorption spectra (ratio to the continuum) for different density slope $p$ with $\theta=30\deg, \Gamma=2, \alpha_{\rm OX}=1.5, f_D=1$ and $\lambda_{\rm ion}=0.1$. }
\label{fig:slope}
\end{figure}

\begin{figure}[t]% ------------------------------------- Figure~12
\begin{center}
\includegraphics[trim=0in 0in 0in
0in,keepaspectratio=false,width=3.2in,angle=-0,clip=false]{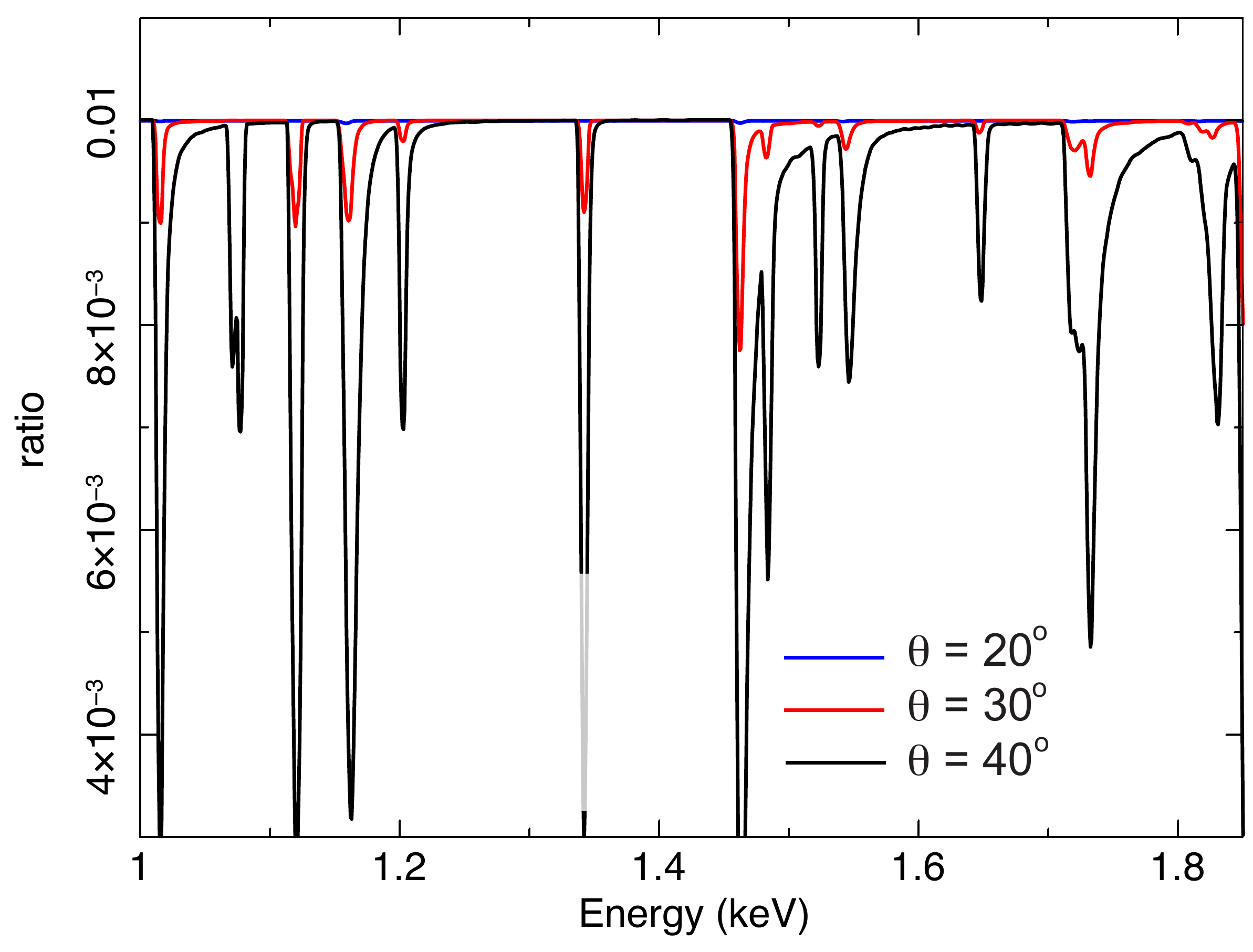} \includegraphics[trim=0in 0in 0in
0in,keepaspectratio=false,width=3.2in,angle=-0,clip=false]{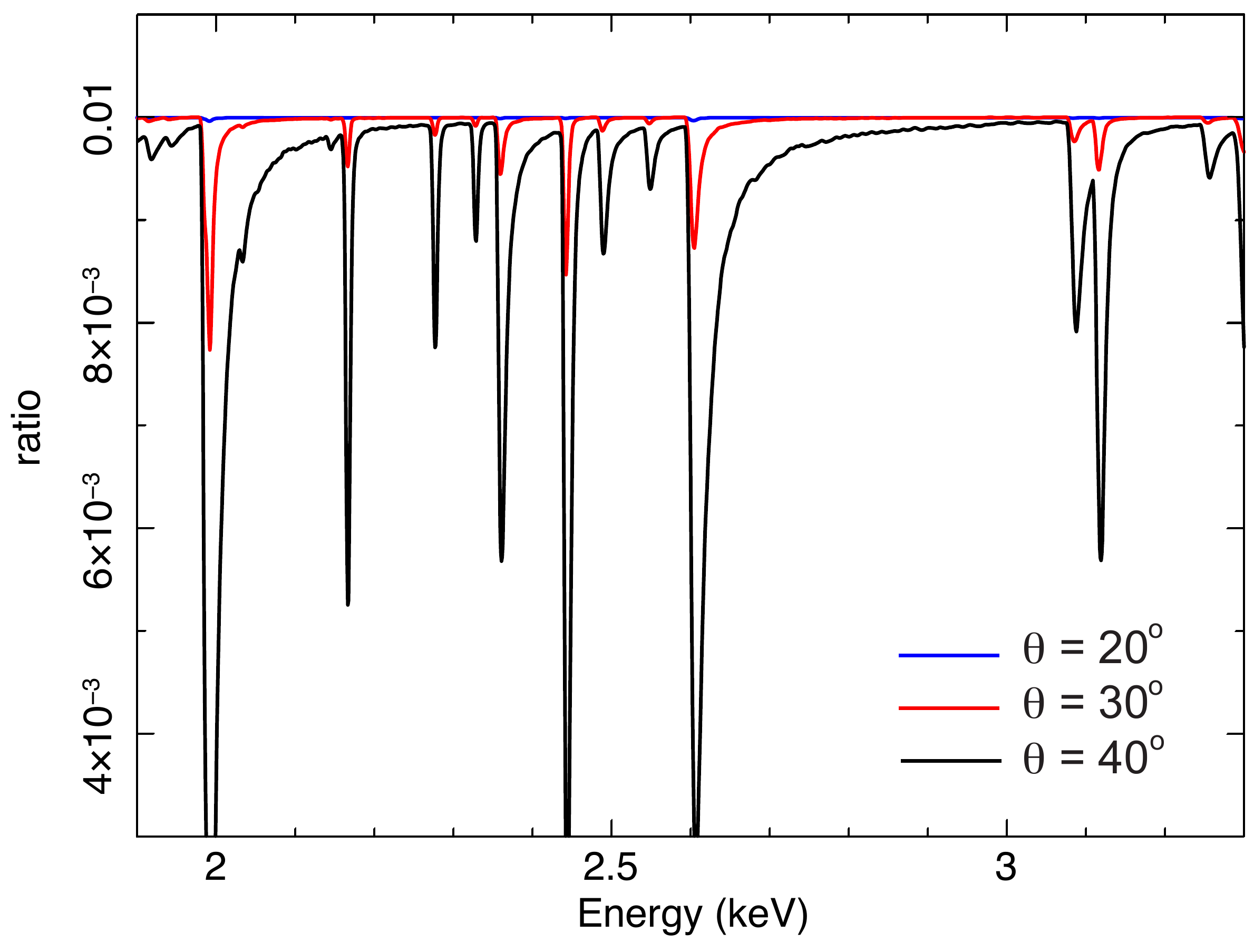} \includegraphics[trim=0in 0in 0in
0in,keepaspectratio=false,width=3.2in,angle=-0,clip=false]{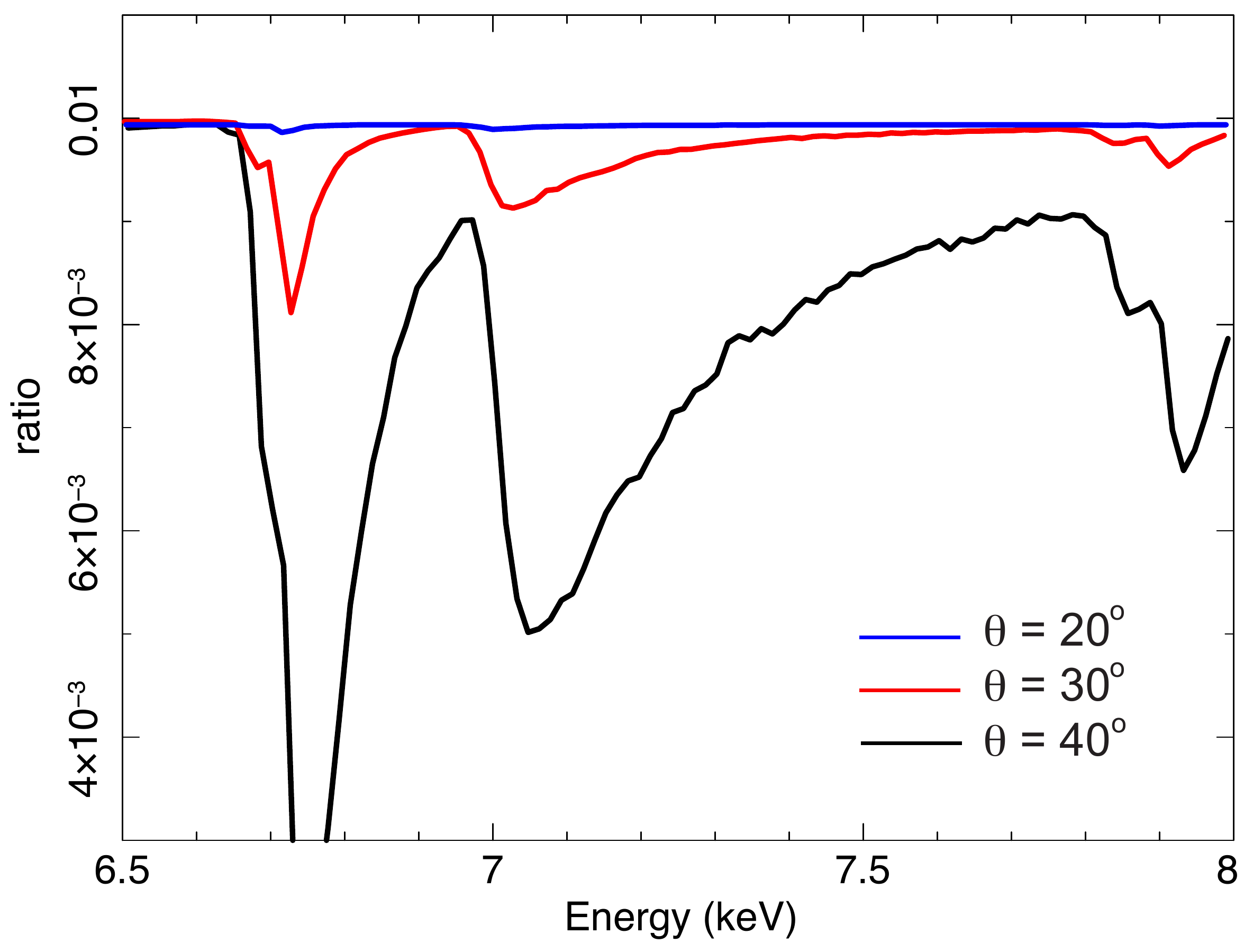}
%\includegraphics[trim=0in 0in 0in
%0in,keepaspectratio=false,width=3.2in,angle=-0,clip=false]{spec_gamma.pdf}\includegraphics[trim=0in 0in 0in
%0in,keepaspectratio=false,width=3.2in,angle=-0,clip=false]{spec_theta.pdf}
\end{center}
\caption{Theoretical multi-ion absorption spectra (ratio to the continuum) for different inclination $\theta$ with $\Gamma=2, \alpha_{\rm OX}=1.5, p=1.3, f_D=1$ and $\lambda_{\rm ion}=0.1$. }
\label{fig:theta}
\end{figure}

In addition to the effect of photoionization due to differences in the radiation field, we further study the dependence on the wind conditions. In {\bf Figure~\ref{fig:slope}} we show the UFO spectra for various wind density gradients with $p$ assuming $\theta=30\deg, \Gamma=2, \alpha_{\rm OX}=1.5, f_D=1$ and $\lambda_{\rm ion}=0.1$. In this exercise, soft X-ray UFOs (located at larger distances from AGN) have lower density with increasing $p$, whereas Fe K UFOs are impacted the least, as the innermost wind at the launching radius  remains fixed (at $r \sim R_{\rm in} = 6R_g$). As a result, soft X-ray UFO signatures in the spectrum are significantly weakened, for example, for $p=1.4$, as expected. With $p =1.2$ wind density structure, on the other hand, the broadband UFO features are clearly present being attributed to an almost equal column $N^i_{\rm ion}$ for all ions as expected from AMD (see \S 2.2).

Finally, we consider spectral dependence on inclination $\theta$ in {\bf Figure~\ref{fig:theta}} assuming $\Gamma=2, \alpha_{\rm OX}=1.5, p=1.3, f_D=1$ and $\lambda_{\rm ion}=0.1$. The obtained dependence is two-fold; i.e. a combination of velocity change and wind density variation with $\theta$. This is because the LoS wind velocity (i.e. velocity component projected onto LoS direction) is dependent on the relative angle between the LoS orientation and a tangential direction of magnetic field lines. At the same time, the wind density is lower by many orders of magnitude near the funnel region close to the symmetry axis \citep[e.g.][]{BP82,CL94,F14,Chakravorty16}. These factors combined together generally cause the absorption lines to be dramatically weaker and less blueshifted with decreasing $\theta$.

\section*{Appendix B: Model Dependences of Simulated Multi-Ion Broadband UFO Spectra}

\begin{figure}[t]% ------------------------------------- Figure~13
\begin{center}
\includegraphics[trim=0in 0in 0in
0in,keepaspectratio=false,width=3.3in,angle=-0,clip=false]{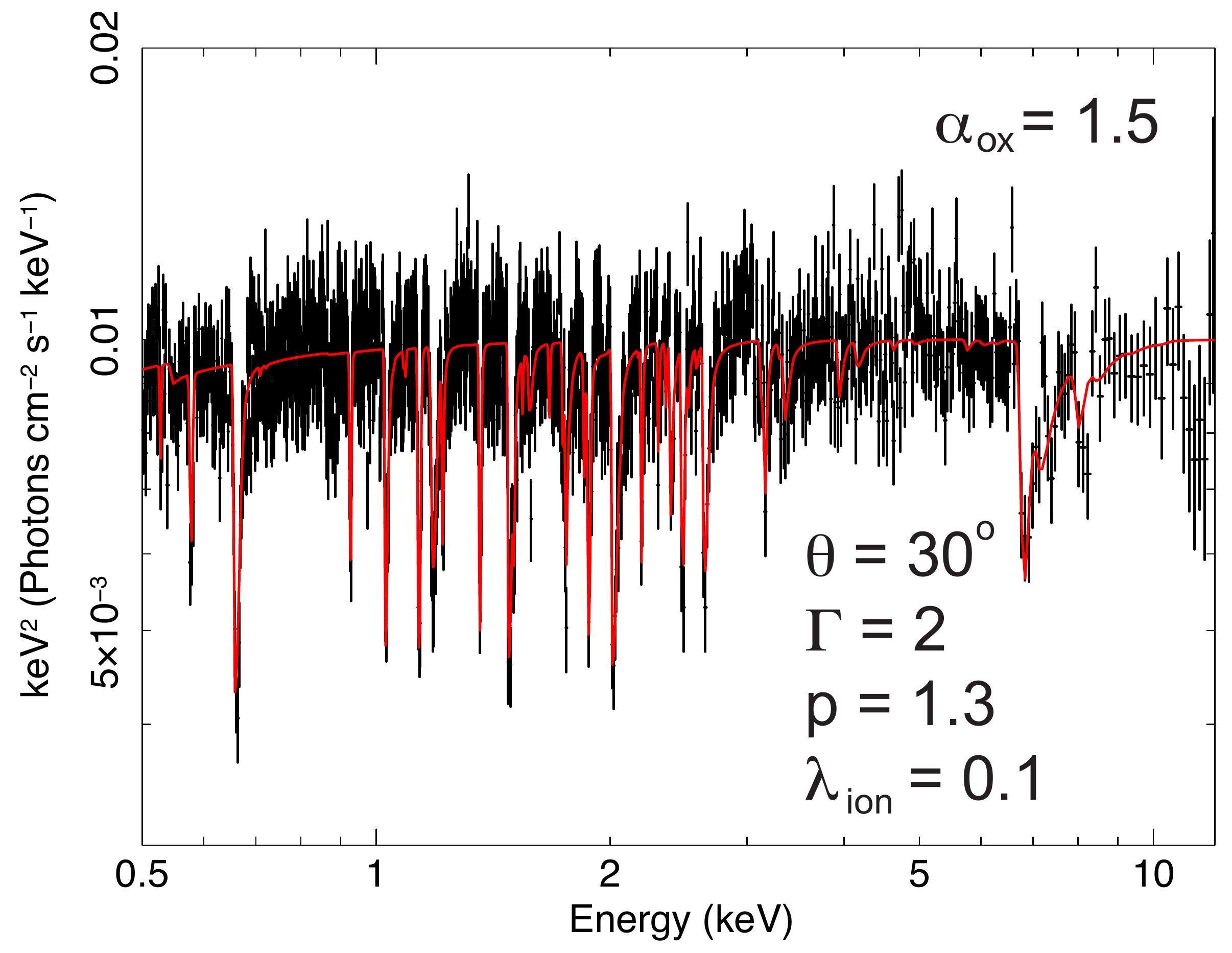}\includegraphics[trim=0in 0in 0in
0in,keepaspectratio=false,width=3.3in,angle=-0,clip=false]{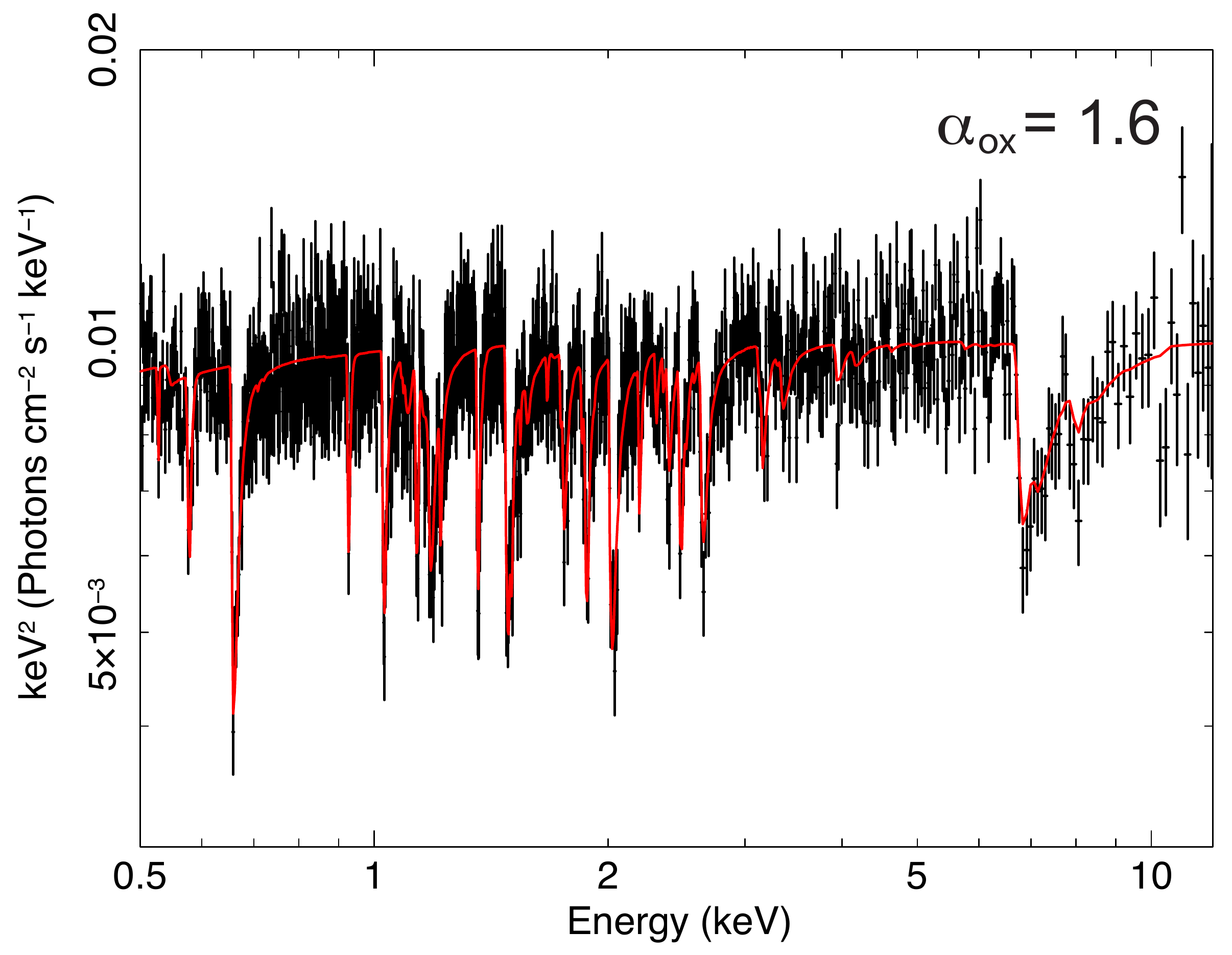}
\includegraphics[trim=0in 0in 0in
0in,keepaspectratio=false,width=3.3in,angle=-0,clip=false]{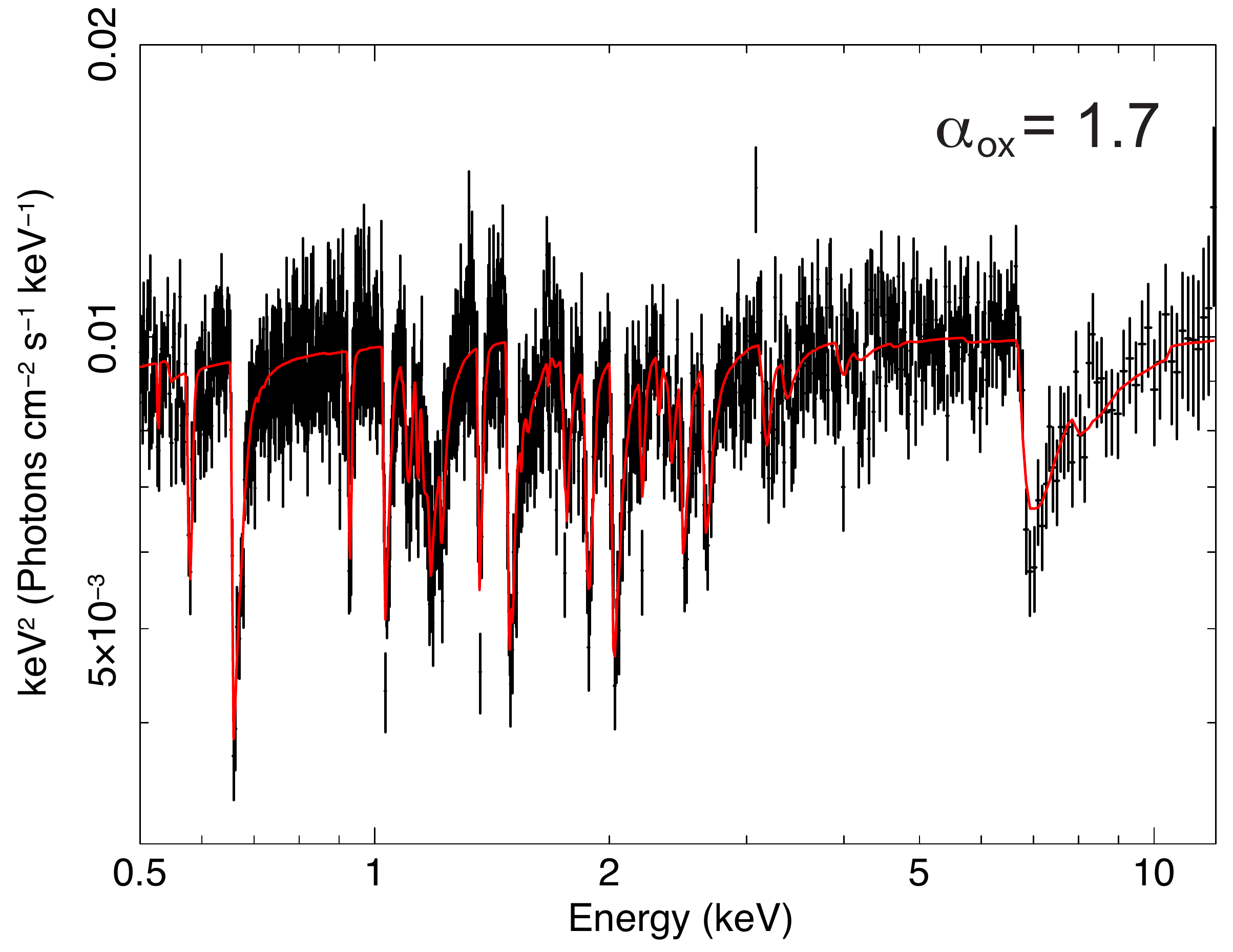}\includegraphics[trim=0in 0in 0in
0in,keepaspectratio=false,width=3.3in,angle=-0,clip=false]{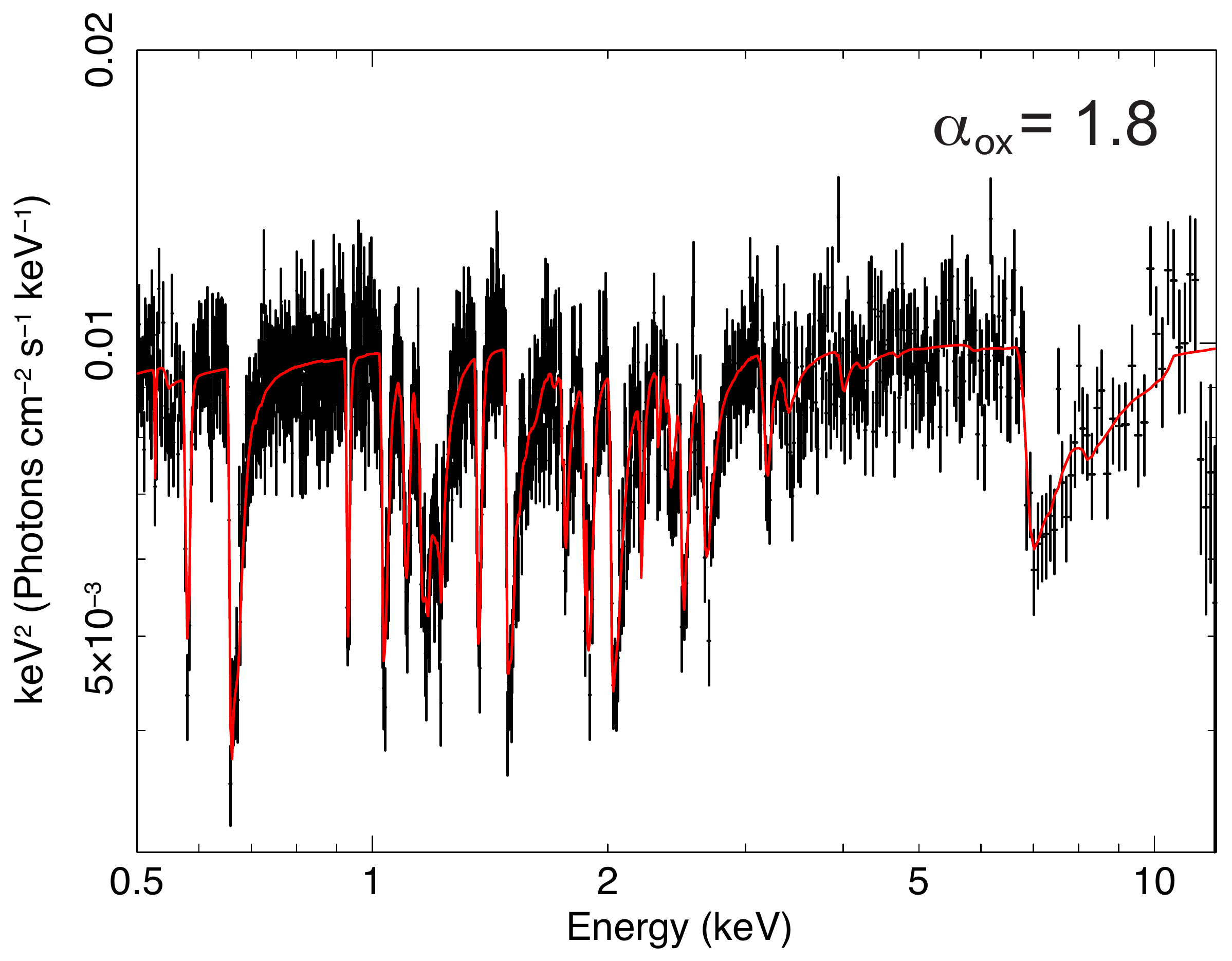}
\end{center}
\caption{Simulated 100ks {\it XRISM}/Resolve spectra for different X-ray strength $\alpha_{\rm OX}$ with $\theta=30\deg, p=1.3, f_D=1, \Gamma=2$ and $\lambda_{\rm ion}=0.1$. }
\label{fig:alphaox2}
\end{figure}

\begin{figure}[t]% ------------------------------------- Figure~14
\begin{center}
\includegraphics[trim=0in 0in 0in
0in,keepaspectratio=false,width=3.3in,angle=-0,clip=false]{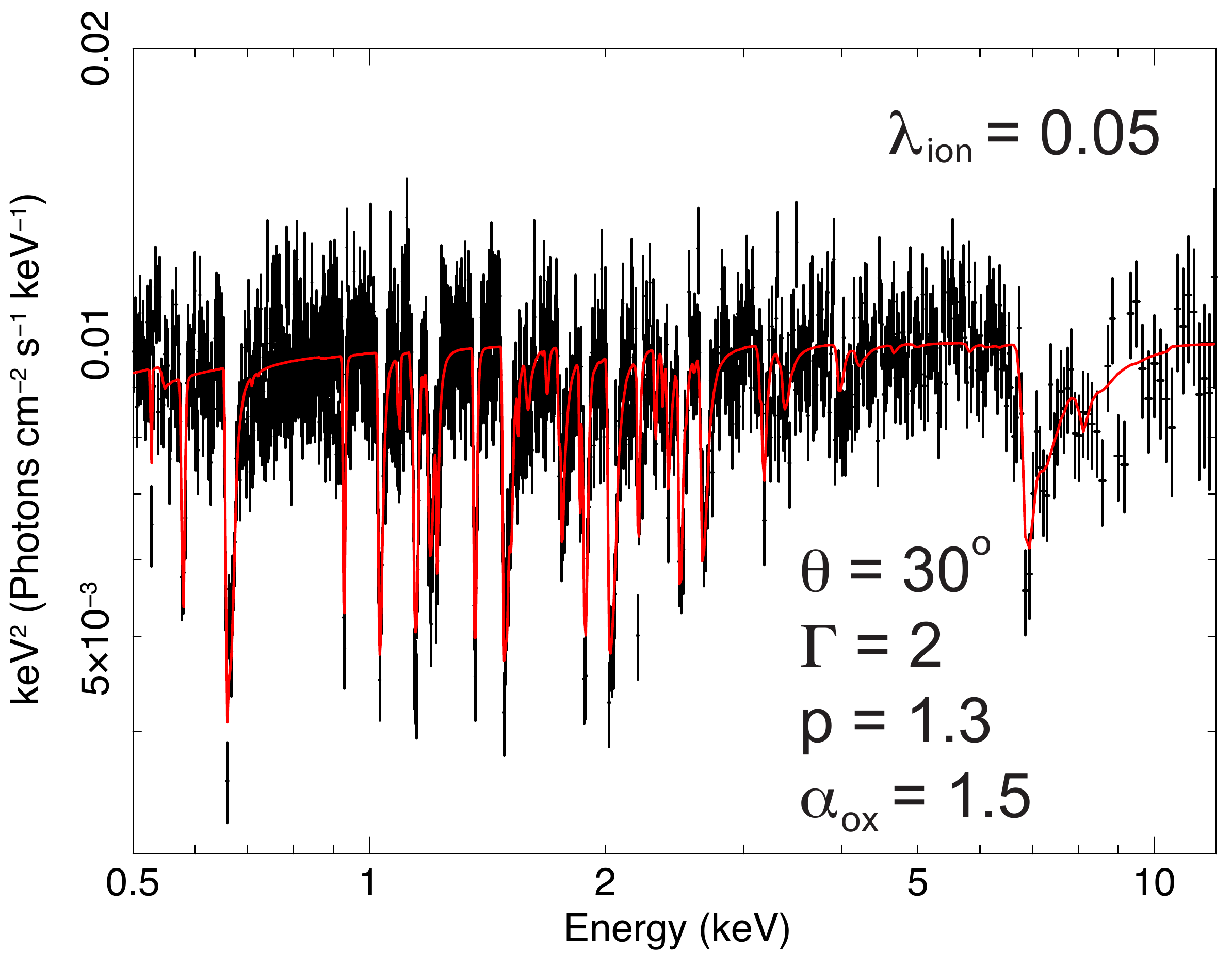}\includegraphics[trim=0in 0in 0in
0in,keepaspectratio=false,width=3.3in,angle=-0,clip=false]{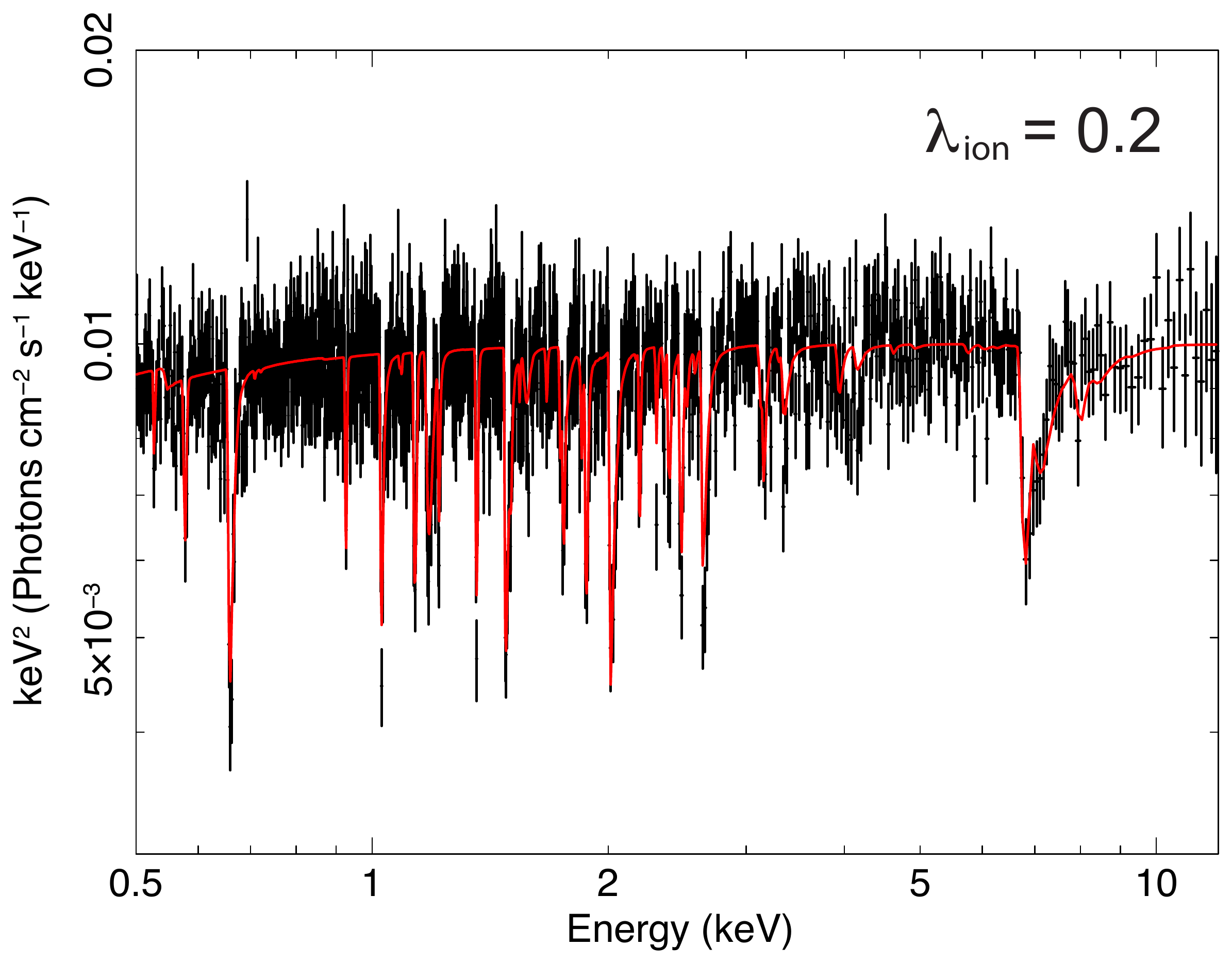}
\includegraphics[trim=0in 0in 0in
0in,keepaspectratio=false,width=3.3in,angle=-0,clip=false]{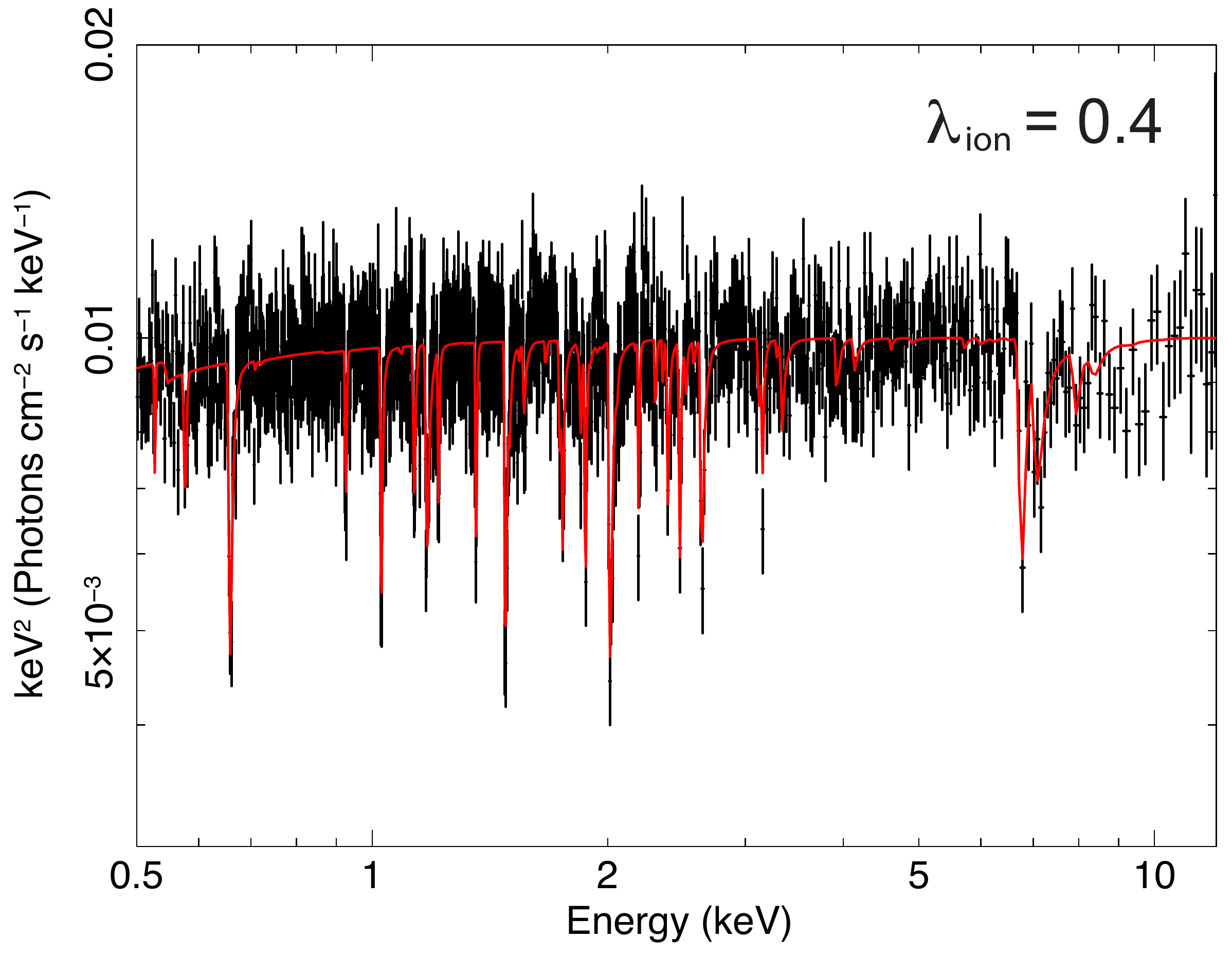}\includegraphics[trim=0in 0in 0in
0in,keepaspectratio=false,width=3.3in,angle=-0,clip=false]{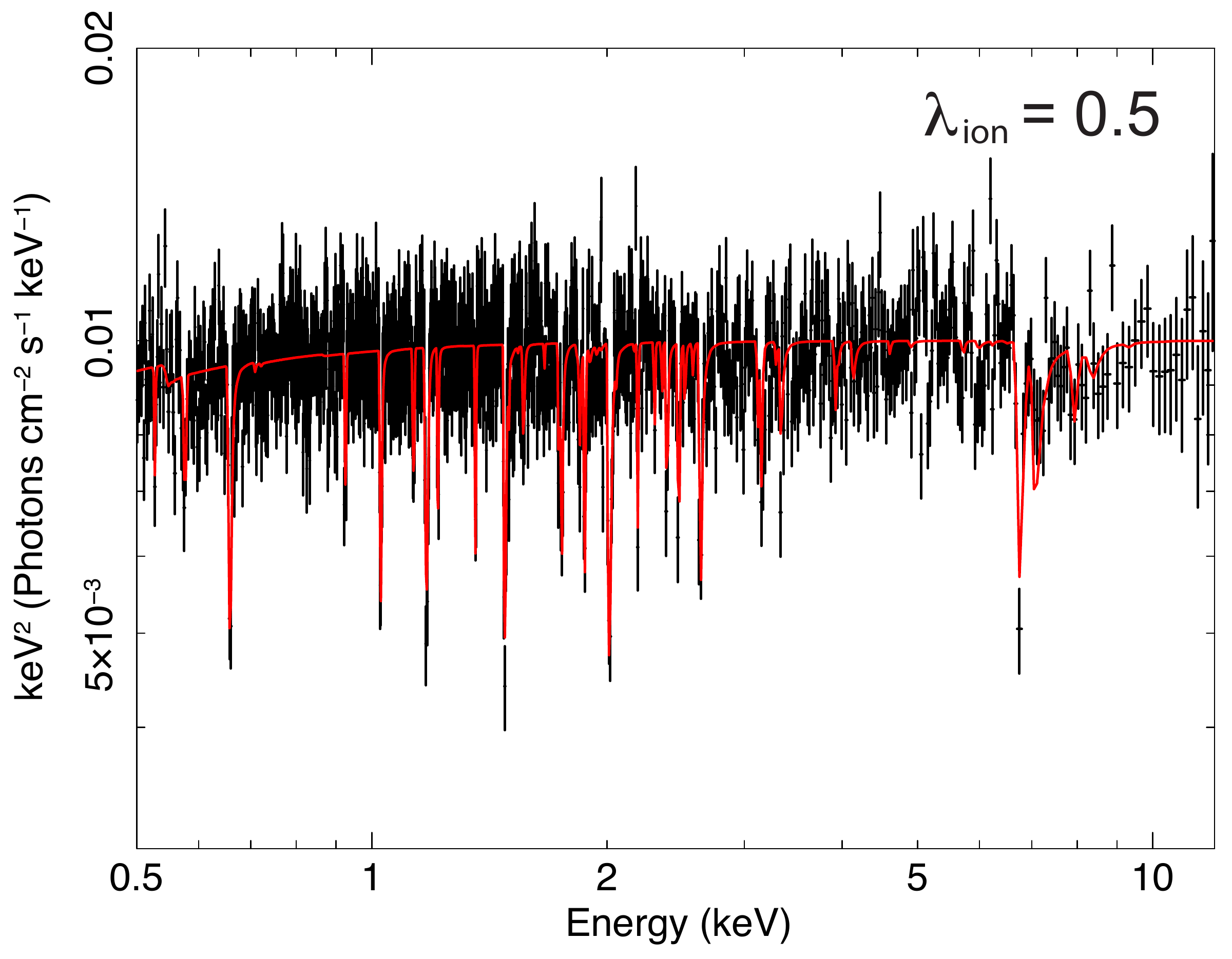}
\end{center}
\caption{Simulated 100ks {\it XRISM}/Resolve spectra for different ionizing luminosity ratio $\lambda_{\rm ion}$ with $\theta=30\deg, p=1.3, f_D=1, \Gamma=2$ and $\alpha_{\rm OX}=1.5$. }
\label{fig:lambda2}
\end{figure}

\begin{figure}[t]% ------------------------------------- Figure~15
\begin{center}
\includegraphics[trim=0in 0in 0in
0in,keepaspectratio=false,width=3.3in,angle=-0,clip=false]{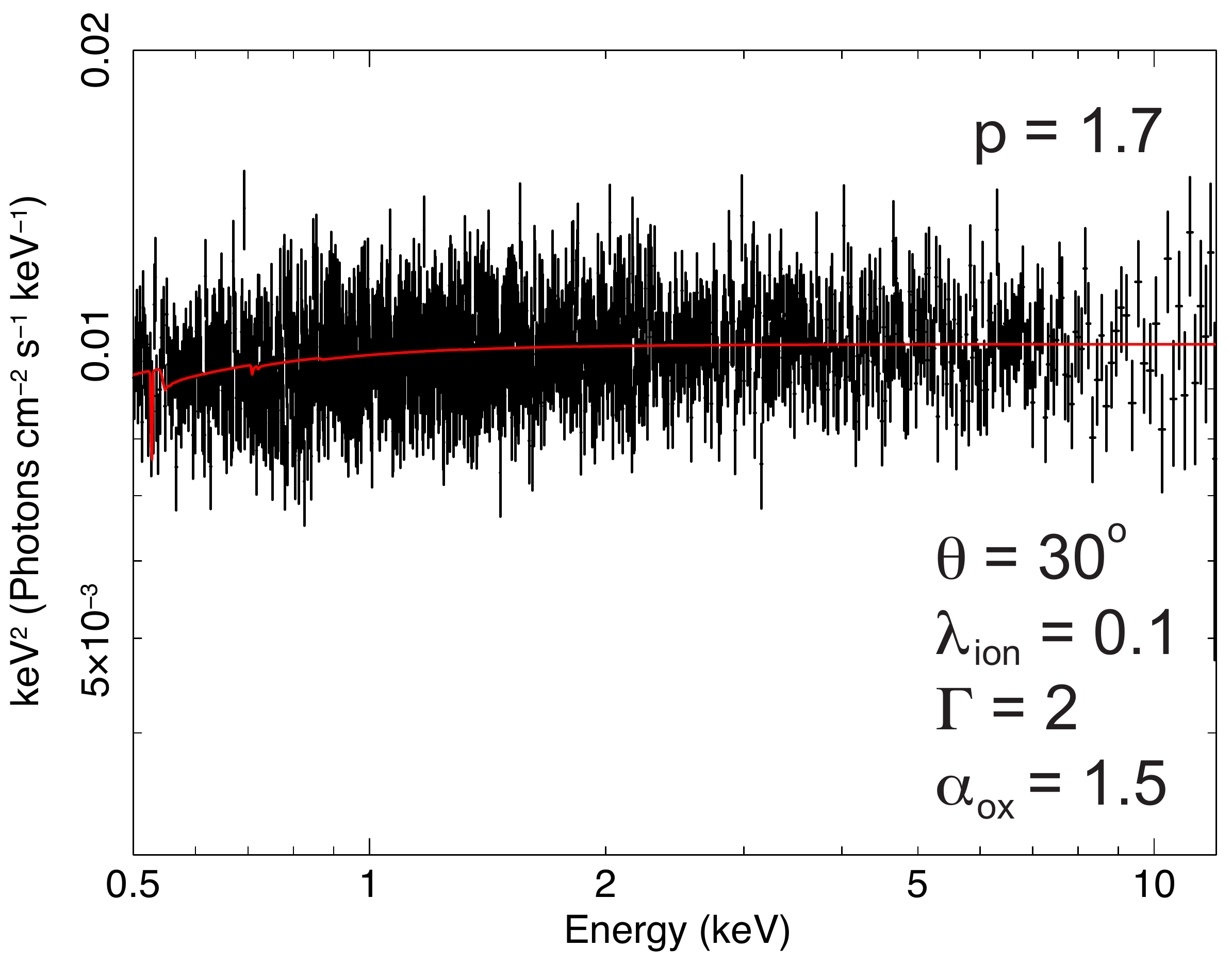}\includegraphics[trim=0in 0in 0in
0in,keepaspectratio=false,width=3.3in,angle=-0,clip=false]{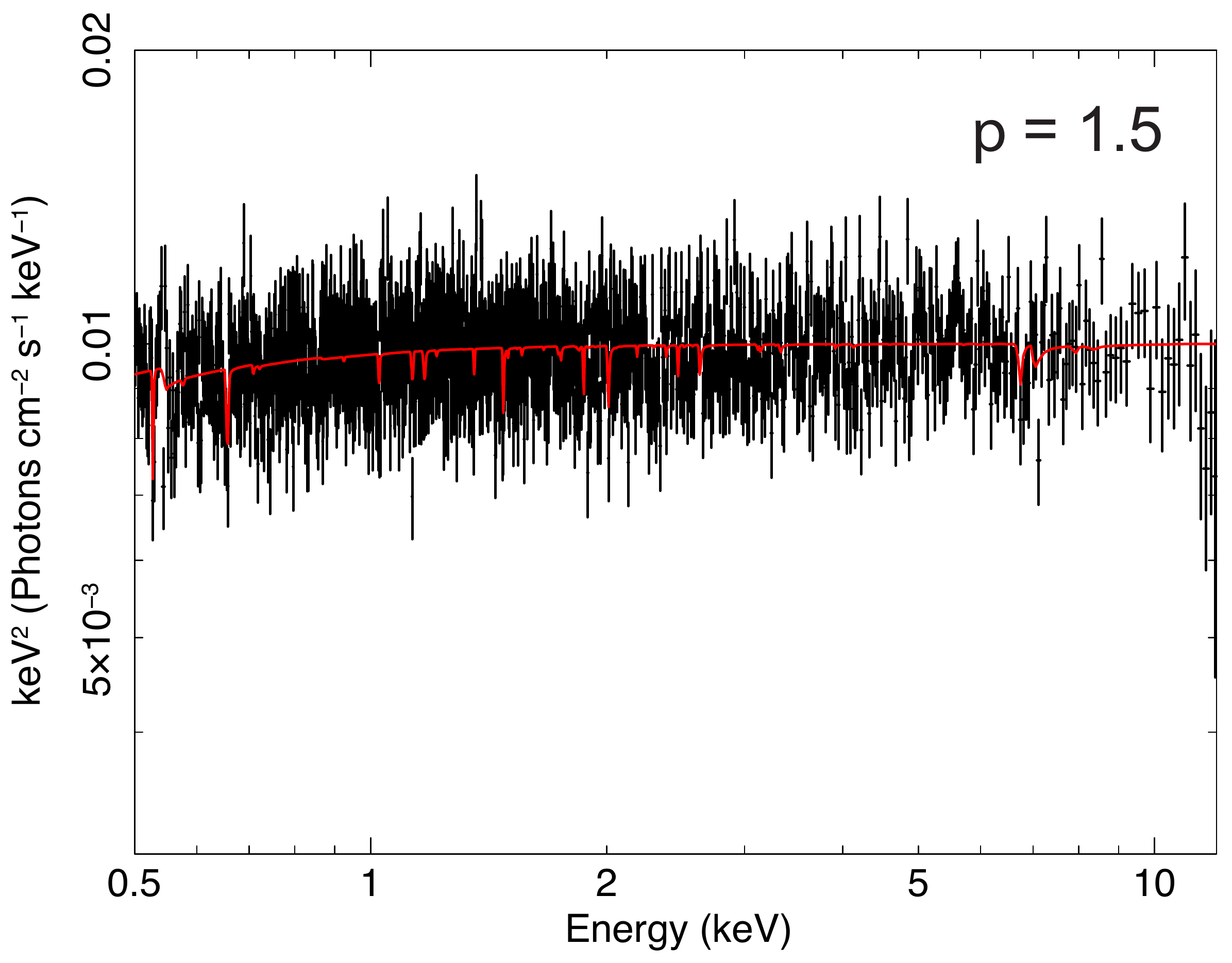}
\includegraphics[trim=0in 0in 0in
0in,keepaspectratio=false,width=3.3in,angle=-0,clip=false]{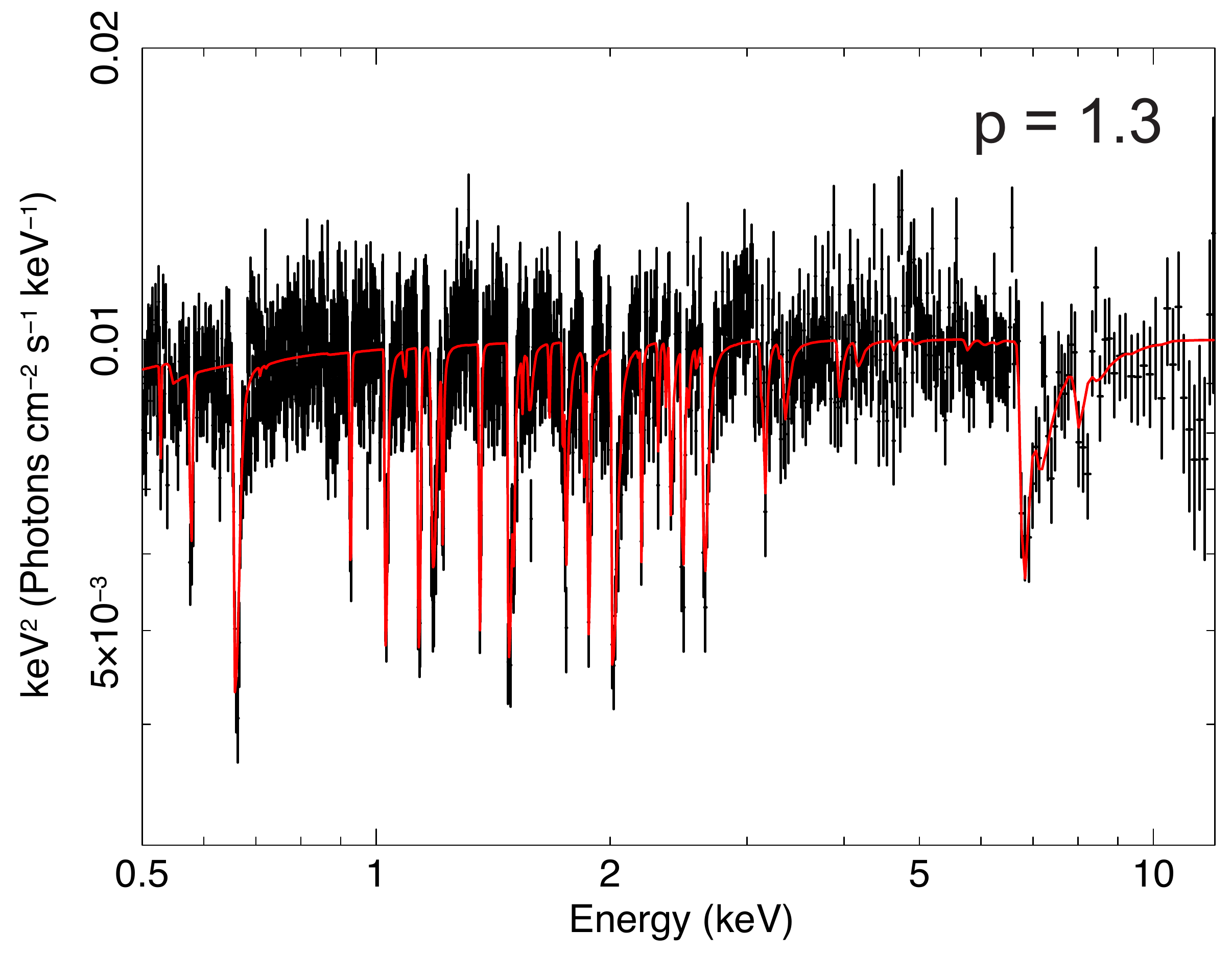}\includegraphics[trim=0in 0in 0in
0in,keepaspectratio=false,width=3.3in,angle=-0,clip=false]{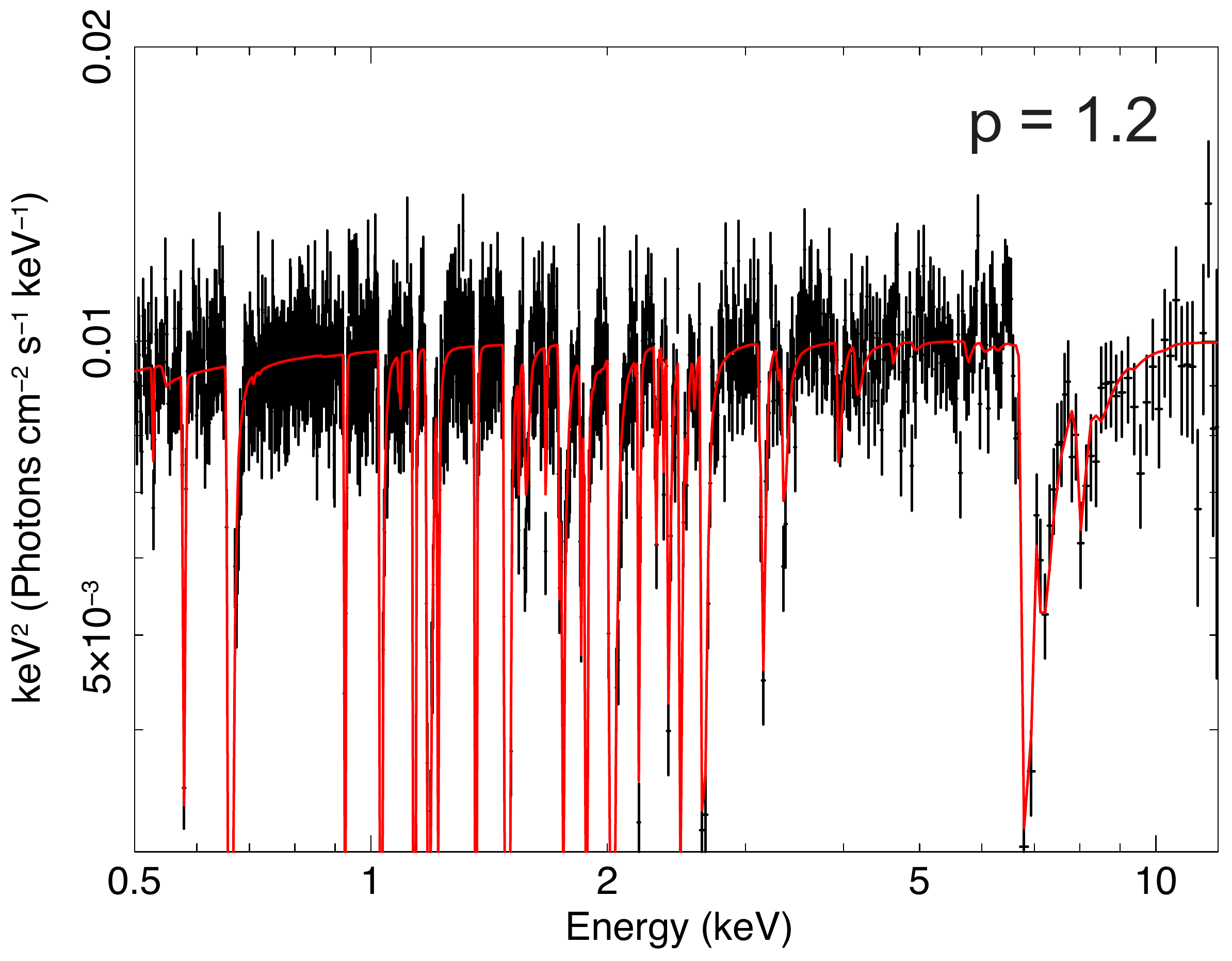}
\end{center}
\caption{Simulated 100ks {\it XRISM}/Resolve spectra for different density slope $p$ with $\theta=30\deg, \Gamma=2, \alpha_{\rm OX}=1.5, f_D=1$ and $\lambda_{\rm ion}=0.1$.}
\label{fig:slope2}
\end{figure}

\begin{figure}[t]% ------------------------------------- Figure~16
\begin{center}
\includegraphics[trim=0in 0in 0in
0in,keepaspectratio=false,width=3.3in,angle=-0,clip=false]{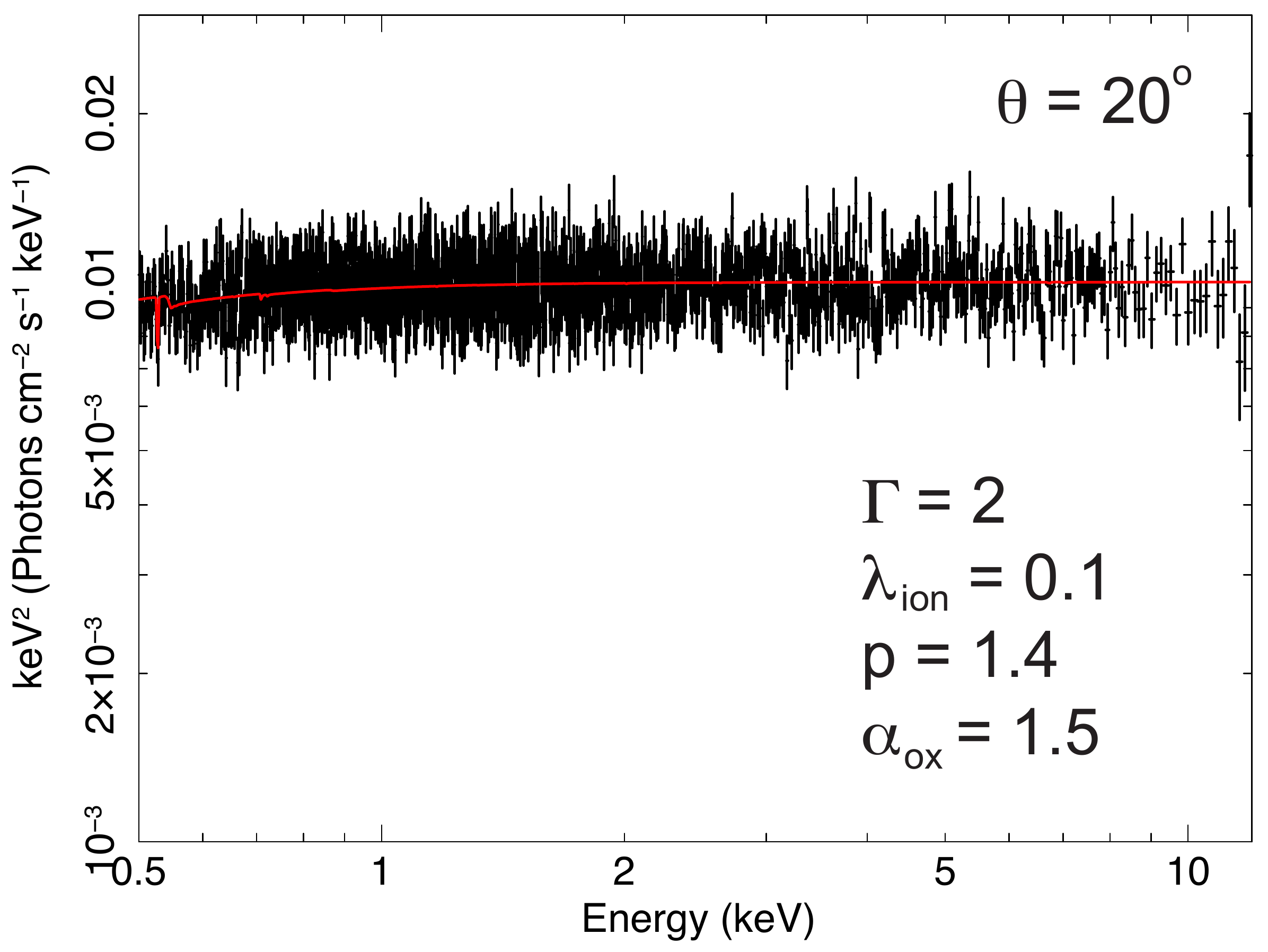}\includegraphics[trim=0in 0in 0in
0in,keepaspectratio=false,width=3.3in,angle=-0,clip=false]{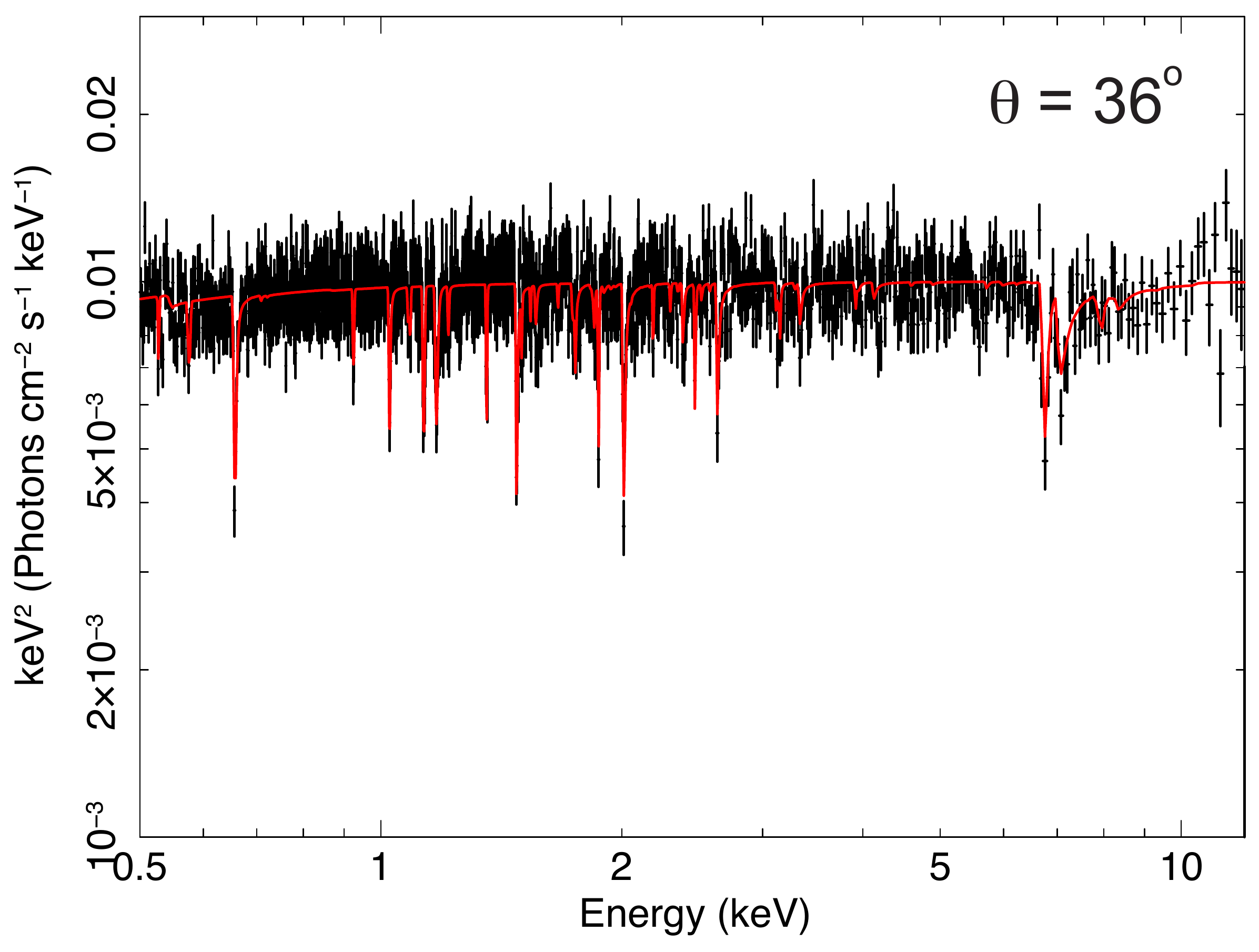}
\includegraphics[trim=0in 0in 0in
0in,keepaspectratio=false,width=3.3in,angle=-0,clip=false]{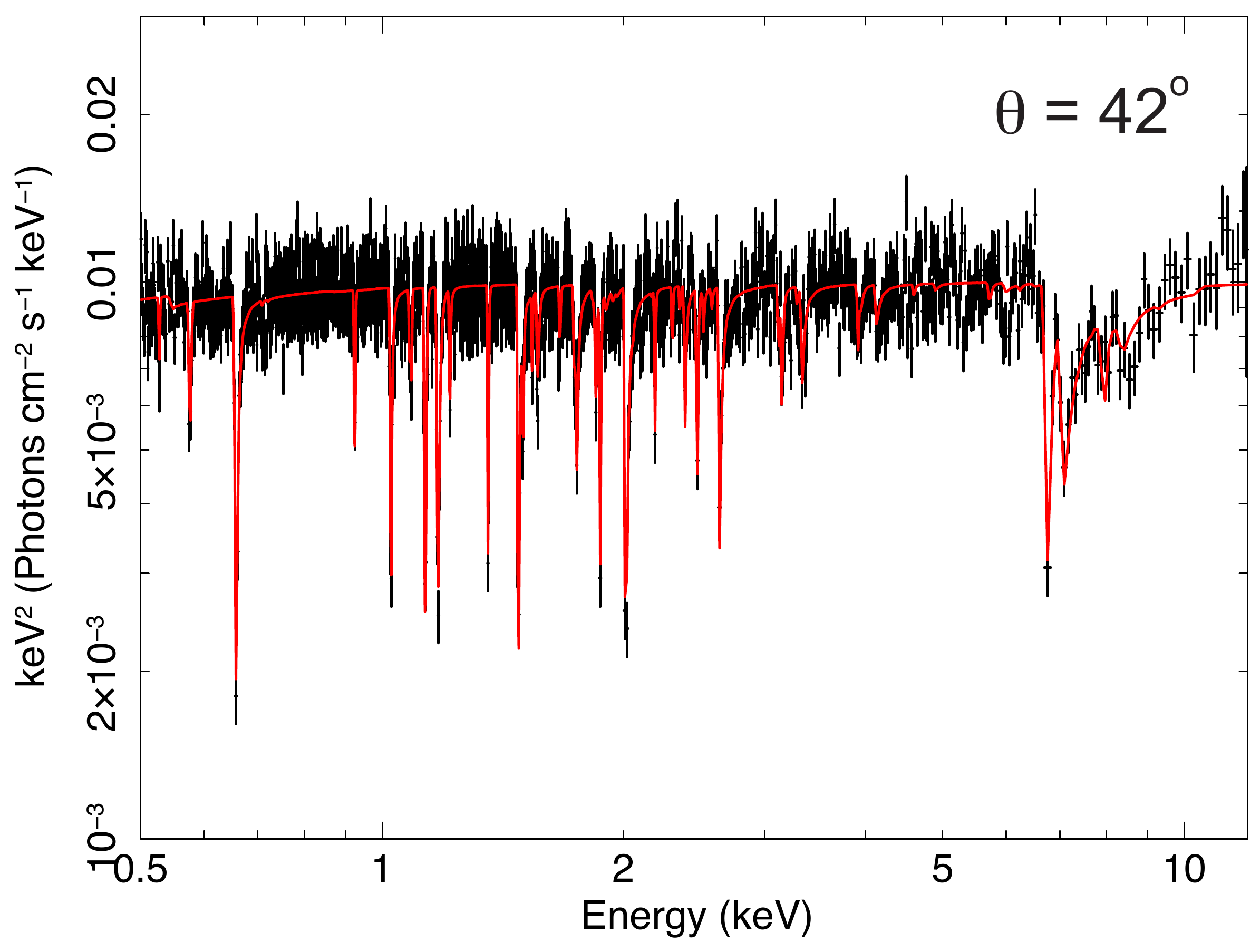}\includegraphics[trim=0in 0in 0in
0in,keepaspectratio=false,width=3.3in,angle=-0,clip=false]{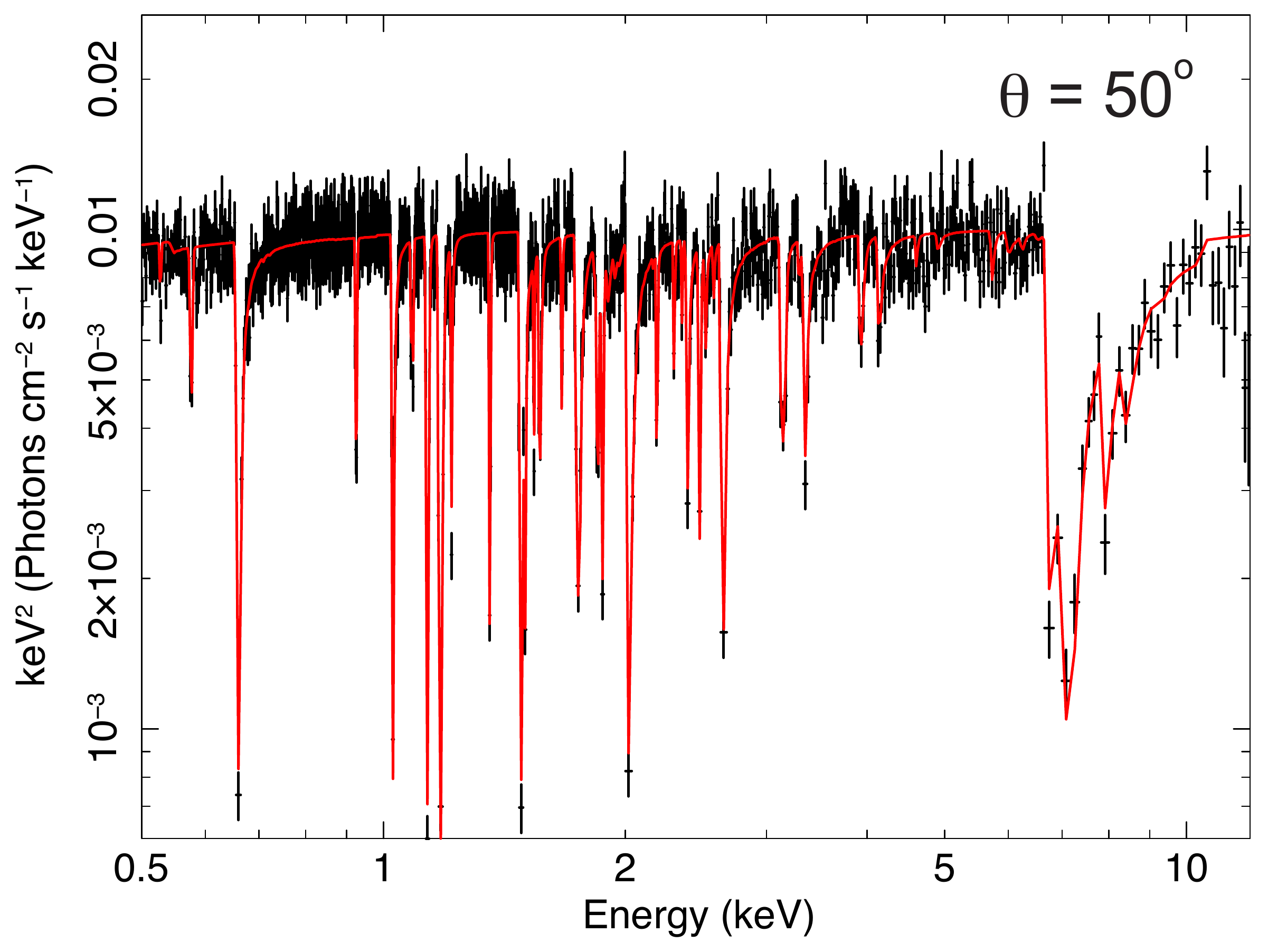}
\end{center}
\caption{Simulated 100ks {\it XRISM}/Resolve spectra for different inclination $\theta$ with $\Gamma=2, p=1.4, \alpha_{\rm OX}=1.5, f_D=1$ and $\lambda_{\rm ion}=0.1$.}
\label{fig:theta2}
\end{figure}

To further follow up the \xrism/Resolve simulations of our MHD UFO models discussed in \S 3.3.1, we present here in {\bf Figures~\ref{fig:alphaox2}-\ref{fig:theta2}}  the other simulated spectra for various model parameters listed in {\bf Table~1}. 

In {\bf Figure~\ref{fig:alphaox2}}, the well-defined blueshifted UFO features of asymmetry is similarly seen  in which X-ray-weak AGNs (e.g. $\alpha_{\rm OX}=1.8$) can clearly imprint such an asymmetric blueshift structure in the spectrum, most notably seen in  Fe K lines. Note that \fexxv\ and \fexxvi\ lines are made so broad that they are blended together to form a single feature, for example, for $\Gamma=2.5$ and $\alpha_{\rm OX}=1.8$. As expected from the synthetic spectra in {\bf Figure~\ref{fig:lambda}}, simulated data for different ionizing luminosity $\lambda_{\rm ion}$ in {\bf Figure~\ref{fig:lambda2}} do not seem to change much with $\lambda_{\rm ion}$, while the suppression of atomic lines, especially in the soft X-ray band, are found.  
Since the wind density slope and inclination can probably provide the most direct effect on the absorption spectra, dramatic changes in UFO signatures are indeed coded in {\it XRISM}/Resolve data as demonstrated in {\bf Figures~\ref{fig:slope2} and \ref{fig:theta2}}. For example, it is found that microcalorimeters are capable of detecting a series of blueshifted  lines of unique asymmetry in the soft X-ray UFOs as well as a pronounced Fe K UFOs for $p=1.2$ winds. For a fixed wind density factor $f_D =1$, the simulated spectrum would exhibit little wind features for $p=1.5-1.7$, as is previously shown from the model with $p=1.4$ in {\bf Figure~\ref{fig:slope}}. Similarly,  very weak line features expected from a polar wind with $\theta=20\deg$ would not be visible even with 80ks microcalorimeter observation under this parameter set, as indicated earlier in {\bf Figure~\ref{fig:theta}}. 
On the other hand, low inclination winds of $\theta=30\deg-40\deg$, being relevant for canonical Seyfert 1 AGNs, would be securely resolved.  
Moderate inclination winds of $\theta=50\deg$ would prominently exhibit stronger UFO features.

\section*{Appendix C: Impact of Reflection Component on Simulated Multi-Ion Broadband UFO Spectra}

One of the characteristic spectral features often identified in bright Seyfert AGNs is reflection features from disks; i.e. Compton hump and various atomic lines including emission and absorption/edges due to reprocessed photons  \citep[e.g.][]{MagdziarzZdziarski95,RossFabian05}. To follow up our  investigations on a persistent presence of MHD signatures in UFO spectra, we further examine the effects of such reflection component and see if the expected characteristics of asymmetric line profiles might be contaminated or distorted to the extent that identification of the MHD-driven UFO features may be difficult as a consequence. 

\begin{figure}[t]% ------------------------------------- Figure~17
\begin{center}
\includegraphics[trim=0in 0in 0in
0in,keepaspectratio=false,width=5.3in,angle=-0,clip=false]{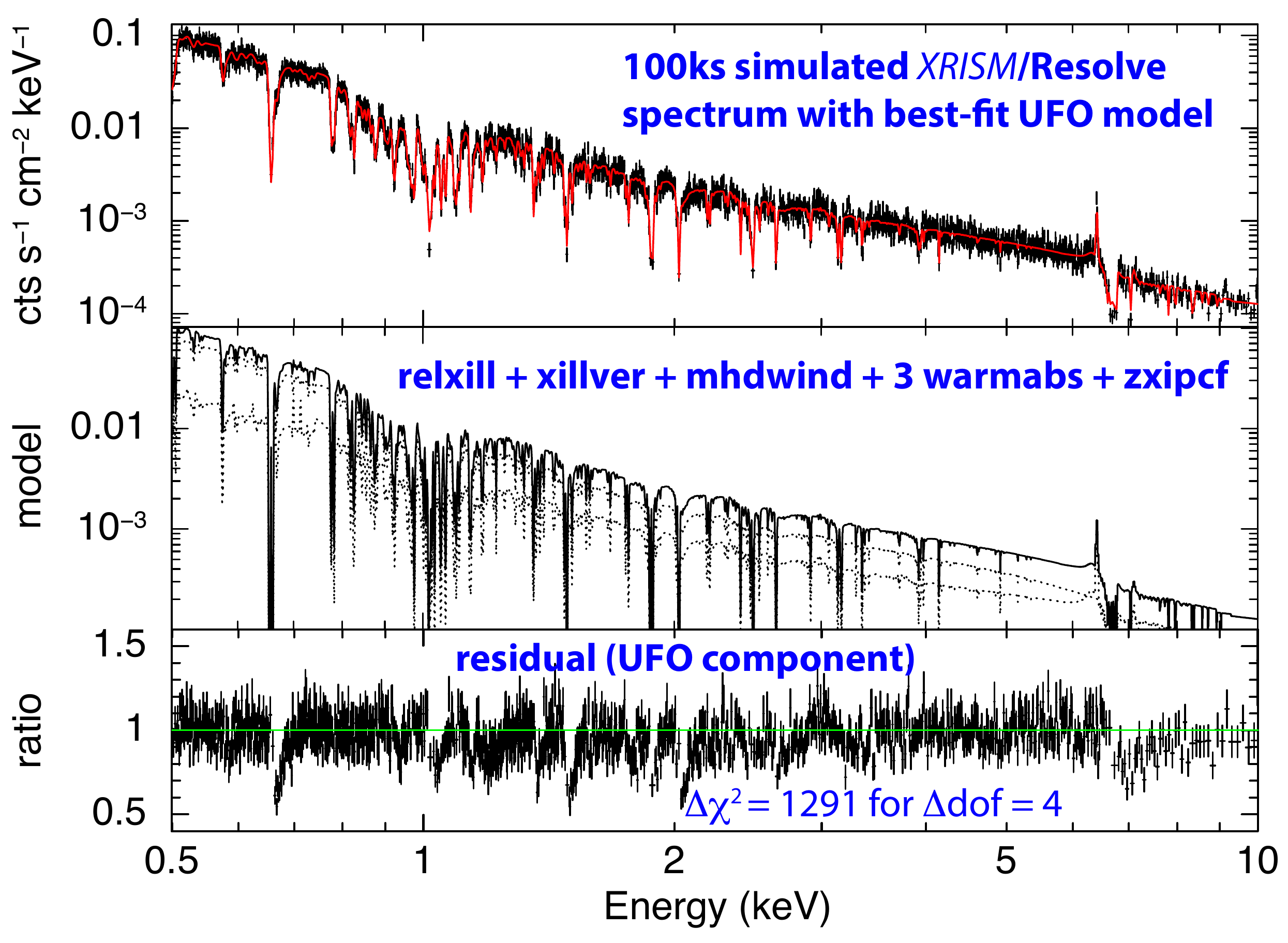}
\end{center}
\caption{Similar to Figure~\ref{fig:composite} but in the presence of reflection component. }
\label{fig:reflection}
\end{figure}

To this end, we replace a single power-law ({\tt po}) by  adding both {\tt xillverCp} (for distant reflection) and {\tt relxillCp} (for relativistic reflection) models in the original composite model (as shown in Table~2 and Fig.~\ref{fig:composite}) for \xrism/Resolve simulations; i.e. a new {\tt composite} model is expressed symbolically as 
\begin{eqnarray}
{\tt  tbabs \times (relxillCp + xillverCp) \times mhdwind \times warmabs_{\tt C} \times warmabs_{\tt W} \times warmabs_{\tt H} \times zxipcf} 
\end{eqnarray}
where we have fixed a number of baseline parameter values  in the reflection components to minimize the extra degrees of freedom, while others are to be varied to find the bestfit solution. 
Figure~\ref{fig:reflection} demonstrates a feasibility of extracting the MHD signatures of the expected UFO spectrum lines in such a case. We set the normalizations of the reflection components such that the 2-10 keV flux remains of the order of $10^{-11}$ erg~cm$^{-2}$~s$^{-1}$  as  considered earlier in \S 3.3.2 where reflection feature is absent. 
In this simulation, we assume a relatively strong reflection contribution to the underlying continuum; e.g. refl\_frac = 2 and  $A_{\rm Fe}/A_{\rm Fe,\odot}=5$. In order to reduce the extra degrees of freedom in the model, we fix the emissivity profile of the disk emission in such a way that $\epsilon(r) \propto r^{-6}$ for $r/R_g \le 10$ (the innermost) and $\propto r^{-3}$ for $r/R_g > 10$ (the outer part) to take into account relativistic effects under strong gravity (these parameters are thus frozen throughout fitting) for the dimensionless BH spin of $a=0.98$. The assumed parameter values and the bestfit model parameters are listed in Table 3. 

It is clearly demonstrated that the spectral signatures of UFOs due to the MHD wind component are still unambiguously imprinted in the broadband spectrum (i.e. extended asymmetric blue tails) despite the complex spectral curvature of relatively strong reflection flux including the extra absorption lines. 
Identification of the individual MHD-driven UFO lines in the bestfit model is equally robust as that in the power-law continuum simulation (in the absence of reflection; see Fig.~\ref{fig:composite}) indicating that the reflection features interfere very little with the UFO lines to the extent that the predicted tell-tale MHD signatures are less likely to be washed out. The adopted parameter values of MHD wind component ({\tt mhdwind}) are almost all well retained by fitting the simulated spectra in both cases (see Tables~2 and 3). 
Note that even strong Fe K emission lines (both narrow and broad) in this simulation does not significantly distort  the original Fe K UFO absorption feature.

\clearpage

%------------------------------- Table~3
\begin{deluxetable}{ll|cc}
\tabletypesize{\small} \tablecaption{Best-Fit {\tt Composite} Model for 100 ks Simulated {\it XRISM}/Resolve Spectrum with Reflection} \tablewidth{0pt}
\tablehead{Component & Parameter & Best-Fit Value & Model Value } 
\startdata
%Wind density slope $p$ & $0.7, 0.8, 0.9, 1.0, 1.15, 1.29$ \\
{\tt tbabs} & $N_H^{\rm Gal}$ [cm$^{-2}$] & $10^{20}$$^\dagger$ & $10^{20}$$^\dagger$  \\ \hline
{\tt xillverCp} & $\Gamma$  & $2.01^{+0.01}_{-0.01}$ & 2 $^\flat$  \\
                & $\theta$ [deg] & $36.8_{-1.9}^{+2.5}$  &  $35$ $^\flat$ \\ 
                & A$_{\rm Fe}$ [A$_{\rm Fe,\odot}$] & $5.04_{-0.61}^{+0.98}$  &  5 $^\flat$ \\ 
                & $kT_{e}$ [keV] & $163$ $^\triangle$ &  150 $^\flat$ \\
               & refl\_frac & $1.82_{-0.37}^{+0.44}$ &  2 $^\flat$ \\
               & $\log n_D \rm{[cm^{-3}]}$ & $18.9_{-0.11}^{+0.08}$ & 19 $^\flat$ \\
               & $\log \xi$ & $0.96_{-0.09}^{+0.09}$ & 1  \\
               & K$_{\rm ref}$ [$10^{-4}$] & $0.73_{-0.11}^{+0.17}$ &  1  $^\flat$ \\             
{\tt relxillCp}$^\sharp$ & $a/M$  & $0.979_{-0.013}^{+0.010}$ & 0.98   \\ 
              & $\log \xi$ & $2.96_{-0.10}^{+0.09}$ & 3  \\ \hline
% & K$_{\rm xillver}$ &  &  $10^{-4}$ \\ 
%Inclination angle $\theta$ [degrees] & $71.3^{+0.79}_{-0.69}$$^\sharp$ ~ ($67.0^{+1.07}_{-1.01}$)$^\diamond$ \\
{\tt mhdwind} & $\alpha_{\rm OX}$  & $1.50^{+0.01}_{p}$ & 1.5   \\
& $\Gamma$  & $2.01^{+0.01}_{-0.01}$ & 2 $^\flat$  \\
& $\lambda_{\rm ion}$   & $0.084_{-0.003}^{+0.003}$  & 0.08 \\
& $\theta$ [deg] & $36.8_{-1.9}^{+2.5}$  &  $35$ $^\flat$ \\ 
& $p$  & $1.26_{-0.03}^{+0.03}$ & 1.25 \\ 
& $f_D$ & $0.093_{-0.01}^{+0.02}$ & 0.1 \\
\hline
{\tt warmabs} & $N_{\rm H,Cold}$  & $1.01_{-1.01}^{+0.33} \times 10^{21}$ & $10^{21}$ \\
& $\log \xi_{\rm Cold}$  & $2.05_{-0.05}^{+0.05}$  & $2$ \\
& $v_{\rm out,Cold}/c$ & $-8.74_{-4.5}^{+3.3}  \times 10^{-4}$ & $10^{-3}$ \\
& $N_{\rm H,Warm}$  & $1.00_{-0.06}^{+0.29} \times 10^{22}$  & $10^{22}$ \\
& $\log \xi_{\rm Warm}$  & $2.97_{-0.02}^{+0.03}$  & $3$ \\
& $v_{\rm out,Warm}/c$ & $-2.96_{-0.14}^{+0.12} \times 10^{-3}$ & $3 \times 10^{-3}$ \\
& $N_{\rm H,Hot}$  & $1.0_{-0.02}^{p} \times 10^{23}$ & $10^{23}$ \\
& $\log \xi_{\rm Hot}$  & $4.00_{-0.01}^{+0.01}$ & $4$ \\ 
& $v_{\rm out,Hot}/c$ & $-1.0_{-0.07}^{+0.08} \times 10^{-2}$  & $10^{-2}$ \\ \hline
{\tt zxipcf} & $N_{\rm H,zxipcf}$  & $4.05_{-0.9}^{+1.0} \times 10^{23}$ & $5 \times 10^{23}$ \\
& $\log \xi_{\rm zxipcf}$  & $3.95_{-0.07}^{+0.06}$ & $4$ \\ 
& $v_{\rm out,zxipcf}/c$ & $-1.00_{-0.09}^{+0.01} \times 10^{-2}$ & $10^{-2}$ \\ \hline
\hline
%$r(\textmd{\fexxvi})/r_g$ & $\lesssim 25$  \\
& $\chi^2/$dof & $19681.51/18975$  ~ (20972.51/18979)$^\ddagger$ \\
%Downstream Elecetron Energy $kT_e$ & See \S 3.3 \\
%Thickness  $H \equiv h/r$  & $0.1, 0.5, 1$   \\
%Mass-Accretion Rate $\dot{m} \equiv \dot{M}/\dot{M}_{E}$ & $0.1, 0.5, 1$  \\
%Accreting Plasma  $r_{\rm sh}/r_g$  & $2?, 3?$   \\
\enddata
\vspace{0.05in}
%\begin{flushleft}
$^\sharp$ Disk emissivity law is fixed as $\epsilon \propto r^{-6}$ and $r^{-3}$ across $R_{\rm break}/R_g=10$ in {\tt relxill}.
\\
$^\triangle$ Unconstrained. $^\flat$ Tied among models. ~~ $^\ddagger$ In the absence of the UFO component ({\tt mhdwind}).
\\
$^p$ Pegged values.
%$^\sharp$ Treated as a free parameter. \\
%$^\diamond$ A fixed value obtained from N15.
%\end{flushleft}
\label{tab:reflection}
\end{deluxetable}

\end{document}